\documentclass[twoside]{article}

\usepackage{amsmath,amsfonts}
\usepackage{amssymb,latexsym}
\usepackage{amsxtra}
\usepackage{upref}
\usepackage{exscale}
\usepackage[mathscr]{eucal}

\newcommand{\be}{\begin{equation}}
\newcommand{\ee}{\end{equation}}
\newcommand{\beq}{\begin{eqnarray}}
\newcommand{\eeq}{\end{eqnarray}}
\newcommand{\no}{\nonumber}
\newcommand{\bea}{\begin{array}}
\newcommand{\eea}{\end{array}}
\newcommand{\lb}{\label}
\newcommand{\mcal}{\mathcal}
\newcommand{\mscr}{\mathscr}
\newcommand{\mfrak}{\mathfrak}
\newcommand{\ve}{\varepsilon}
\newcommand{\ds}{\displaystyle}
\newcommand{\ts}{\textstyle}
\newcommand{\pp}{\partial}
\newcommand{\im}{\imath}
\newcommand{\ppr}{^{\boldsymbol{\prime}}}
\newcommand{\bprime}{\boldsymbol{\prime}}

\newcommand{\pdag}{^{\dagger}}
\newcommand{\wt}{\widetilde}
\newcommand{\ovv}{\overline}
\newcommand{\trs}{\raisebox{-4pt}{$\mbox{\large tr}\atop {\scriptstyle s,s^{\boldsymbol{\prime}}=\uparrow,\downarrow}$}}
\newcommand{\trxs}{\raisebox{-4pt}{$\mbox{\large tr}\atop {\vec{x},\scriptstyle s}$}}
\newcommand{\TRS}{\raisebox{-4pt}{$\mbox{\large Tr}\atop {\scriptstyle a,s;b,s^{\boldsymbol{\prime}}}$}}

\newcommand{\ph}{\phantom}
\newcommand{\scr}{\scriptstyle}
\newcommand{\scrscr}{\scriptscriptstyle}
\newcommand{\scz}{\scriptsize}

\numberwithin{equation}{section}
\allowdisplaybreaks[4]

\begin{document}

\begin{center}
{\large\bf Field theory with coherent states for many-body problems} \vspace*{0.1cm}\\
{\large\bf with specified particle- and symmetry- quantum numbers} \vspace*{0.3cm}\\
{\bf (Non-relativistic electrons in a central potential and an external magnetic field.)} \vspace*{0.3cm}\\
{\bf Bernhard Mieck}\footnote{e-mail: "bjmeppstein@arcor.de";\newline freelance activity during 2007-2009;
current location : Zum Kohlwaldfeld 16, D-65817 Eppstein, Germany.}
\end{center}

\begin{abstract}
Coherent states and coherent state path integrals are applied to a many-body problem
for non-relativistic electrons in a central potential and an external magnetic field;
however, in addition to previous problems of coherent state path integrals, we definitely
fix the particle number and the other conserved symmetry quantum numbers to specific
values. We determine the maximal commuting set of symmetry quantum numbers in terms
of second quantized field operators which are restricted by corresponding delta functions
to fixed, specific values in a trace representation of anti-commuting coherent states.
One has to distinguish between one-particle and two-particle operators which are all
transformed to a normal ordering. After calculation of the delta functions for one-particle
operators with exponential integrals from the Dirac identity, we perform an anomalous
doubling within the coherent state representation for the delta functions of the two-particle
parts so that a Hubbard-Stratonovich transformation (HST) with 'hinge' fields can be taken
for corresponding self-energies with a subsequent coset decomposition. Therefore, the
remaining field theory of the Coulomb problem only consists of scalar self-energy densities
and coset matrices with the anomalous field degrees of freedom. This construction
allows for the obvious separation to self-energy densities with a saddle point approximation
which are to be inserted as definite values into the remaining field theory of coset matrices.
The extension of the given field theory is also briefly described for an ensemble average over the
external magnetic field so that mean eigenvalue densities and eigenvalue correlations
can be obtained from coherent state representations of delta functions with fixed,
maximal commuting sets of symmetry quantum numbers in a trace of anti-commuting fields
for further HST's and coset decompositions of self-energies.\newline

\vspace*{0.1cm}

\end{abstract}

\tableofcontents

\newpage

\section{Introduction} \lb{s1}

\subsection{Traces of delta functions of maximal-commuting, second-quantized operator-sets}\lb{s11}

Many phenomena of many-body physics can conveniently be represented by coherent states and their
corresponding coherent state path integrals \cite{Negele}-\cite{Nagaosa2}.
Coherent state path integrals have been mainly applied
for problems with varying, unspecified particle numbers where a Hamiltonian of second quantized operators
in the '$d$' dimensional case can be transferred to a time-ordered path integral for a '$(d+1)$'
dimensional, classical system. These coherent state path integrals therefore allow various classical approximations,
both in the bosonic and fermionic case, and can be further changed by a Hubbard-Stratonovich transformation
(HST) into self-energy matrices with anomalous doubled pairing of fields so that one can perform a
coset decomposition into density related and 'Nambu-doubled' parts \cite{Nambu}-\cite{pop2}.

However, in the present article we also use coherent states for a many electron atom with a fixed number of
electrons and specifically fixed symmetry quantum numbers. This is accomplished by a coherent state
representation for the trace over delta functions for a definite number of electrons and for specific
symmetry quantum numbers which are originally given in terms of second quantized field operators.
We calculate the density of states \(\varrho(E,l,s;n_{0},l_{z},s_{z})\) (\ref{s1_1}) for a many electron atom in
the infinite nuclear mass approximation and take the maximal commuting set of symmetry operators for the problem
\beq\lb{s1_1}
\lefteqn{\varrho(E,l,s;n_{0},l_{z},s_{z}) = \mbox{Coherent state path integral of the trace of}} \\ \no
&:=&\mbox{Tr}\bigg[\delta\Big(\hbar\,s_{z}-\boldsymbol{\hat{S}_{z}}(\hat{\psi}\pdag,\hat{\psi})\,\Big)\;
\delta\Big(\hbar\,l_{z}-\boldsymbol{\hat{L}_{z}}(\hat{\psi}\pdag,\hat{\psi})\,\Big)\;
\delta\Big(n_{0}-\boldsymbol{\hat{N}}(\hat{\psi}\pdag,\hat{\psi})\,\Big)\;\times \\ \no &\times&
\delta\Big(\hbar^{2}\,s(s+1)-\boldsymbol{\vec{S}}(\hat{\psi}\pdag,\hat{\psi})\cdot
\boldsymbol{\vec{S}}(\hat{\psi}\pdag,\hat{\psi})\,\Big)\;
\delta\Big(\hbar^{2}\,l(l+1)-\boldsymbol{\vec{L}}(\hat{\psi}\pdag,\hat{\psi})\cdot
\boldsymbol{\vec{L}}(\hat{\psi}\pdag,\hat{\psi})\,\Big)\;
\delta\Big(E-\boldsymbol{\hat{H}}(\hat{\psi}\pdag,\hat{\psi};B_{z})\,\Big)\bigg] \;.
\eeq
In later sections we have to distinguish between delta functions of one-particle operators
\(\boldsymbol{\hat{N}}(\hat{\psi}\pdag,\hat{\psi})\), \(\boldsymbol{\hat{L}_{z}}(\hat{\psi}\pdag,\hat{\psi})\),
\(\boldsymbol{\hat{S}_{z}}(\hat{\psi}\pdag,\hat{\psi})\) and two-particle operators
\(\boldsymbol{\hat{H}}(\hat{\psi}\pdag,\hat{\psi};B_{z})\),
\(\boldsymbol{\vec{L}}(\hat{\psi}\pdag,\hat{\psi})\boldsymbol{\cdot}
\boldsymbol{\vec{L}}(\hat{\psi}\pdag,\hat{\psi})\),
\(\boldsymbol{\vec{S}}(\hat{\psi}\pdag,\hat{\psi})\boldsymbol{\cdot}
\boldsymbol{\vec{S}}(\hat{\psi}\pdag,\hat{\psi})\) with the quantum numbers
\(n_{0}\), \(\hbar\,l_{z}\), \(\hbar\,s_{z}\) and \(E\), \(\hbar^{2}\,l(l+1)\), \(\hbar^{2}\,s(s+1)\), respectively.
We note that relation (\ref{s1_1}) defines a quantum field theory of second quantized operators where
the delta functions constrain the maximal commuting set of symmetry operators to those numbers which are
anticipated by ordinary quantum mechanics with a Fock space. It is furthermore possible to determine with coherent state path integrals a mean eigenvalue density (\ref{s1_2}) of a disordered system for fixed symmetry quantum numbers within an ensemble average, e.\ g.\ with a Gaussian distribution
for an external magnetic field $B_{z}$.
Moreover, one can extend the coherent state representation for specific symmetry numbers to the computation
of eigenvalue correlations (\ref{s1_3}) of a disordered system with an ensemble average, as e.\ g.\
also with a Gaussian distribution of an external magnetic field
\beq\lb{s1_2}
\lefteqn{\ovv{\varrho(E,l,s;n_{0},l_{z},s_{z})} =
\mbox{Tr}\bigg[\delta\Big(\hbar\,s_{z}-\boldsymbol{\hat{S}_{z}}(\hat{\psi}\pdag,\hat{\psi})\,\Big)\;
\delta\Big(\hbar\,l_{z}-\boldsymbol{\hat{L}_{z}}(\hat{\psi}\pdag,\hat{\psi})\,\Big)\;
\delta\Big(n_{0}-\boldsymbol{\hat{N}}(\hat{\psi}\pdag,\hat{\psi})\,\Big)\;\times }  \\ \no &\times&
\delta\Big(\hbar^{2}\,s(s+1)-\boldsymbol{\vec{S}}(\hat{\psi}\pdag,\hat{\psi})\boldsymbol{\cdot}
\boldsymbol{\vec{S}}(\hat{\psi}\pdag,\hat{\psi})\,\Big)\;
\delta\Big(\hbar^{2}\,l(l+1)-\boldsymbol{\vec{L}}(\hat{\psi}\pdag,\hat{\psi})\boldsymbol{\cdot}
\boldsymbol{\vec{L}}(\hat{\psi}\pdag,\hat{\psi})\,\Big)\;\times\;
\ovv{\delta\Big(E-\boldsymbol{\hat{H}}(\hat{\psi}\pdag,\hat{\psi};B_{z})\,\Big)}\bigg] \;; \\  \lb{s1_3}
\lefteqn{\ovv{\varrho(E_{1},E_{2},l,s;n_{0},l_{z},s_{z})} =
\mbox{Tr}\bigg[\delta\Big(\hbar\,s_{z}-\boldsymbol{\hat{S}_{z}}(\hat{\psi}\pdag,\hat{\psi})\,\Big)\;
\delta\Big(\hbar\,l_{z}-\boldsymbol{\hat{L}_{z}}(\hat{\psi}\pdag,\hat{\psi})\,\Big)\;
\delta\Big(n_{0}-\boldsymbol{\hat{N}}(\hat{\psi}\pdag,\hat{\psi})\,\Big)\;\times}   \\ \no &\times&
\delta\Big(\hbar^{2}\,s(s+1)-\boldsymbol{\vec{S}}(\hat{\psi}\pdag,\hat{\psi})\cdot
\boldsymbol{\vec{S}}(\hat{\psi}\pdag,\hat{\psi})\,\Big)\;
\delta\Big(\hbar^{2}\,l(l+1)-\boldsymbol{\vec{L}}(\hat{\psi}\pdag,\hat{\psi})\cdot
\boldsymbol{\vec{L}}(\hat{\psi}\pdag,\hat{\psi})\,\Big)\;\times \\  \no &\times&
\ovv{\delta\Big(E_{2}-\boldsymbol{\hat{H}}(\hat{\psi}\pdag,\hat{\psi};B_{z})\,\Big)\;
\delta\Big(E_{1}-\boldsymbol{\hat{H}}(\hat{\psi}\pdag,\hat{\psi};B_{z})\,\Big)}\bigg] \;.
\eeq
In the following we describe the considered Hamiltonian for a many electron atom with a rigid position of
the nucleus at the coordinate origin in terms of second quantized Fermi operators (\ref{s1_4}), labelled by
the coordinate vector $\vec{x}$ and the electron spin \(s=\uparrow,\downarrow\) which are also abbreviated
into the common labels of vectors \(\vec{y}=(\vec{x},\,s=\uparrow,\downarrow)\),
\(\vec{y}\ppr=(\vec{x},\,s\ppr=\uparrow,\downarrow)\),
\(\vec{y}_{1}=(\vec{x}_{1},\,s_{1}=\uparrow,\downarrow)\),etc.\ . The sum (\ref{s1_5}) over
the field operators is normalized by a volume \(V^{(d)}\) which gives rise to a total number of discrete,
spatial points \(\mcal{N}_{x}\) from the underlying, smallest volume element \(({\scrscr \Delta}x)^{d}\).
In the remainder the usual summation convention with normalization
\(1/\mcal{N}_{x}\) according to (\ref{s1_5}) is always
implied for the repeated occurrence of indices \(\vec{x}\), \(s=\uparrow,\downarrow\) or the combined form
\(\vec{y}\) in order to attain an abbreviated kind of equations and relations
\beq \lb{s1_4}
\hat{\psi}_{\vec{x},s}\;,\;\hat{\psi}_{\vec{x},s}\pdag &;& s=\uparrow,\downarrow\;;\;\;\;
\vec{x}=\{x^{(1)},x^{(2)},x^{(3)}\} \;;\;\;\;(d=3)\;; \\  \no
\vec{y} &=& \big(\vec{x}=\{x^{(1)},x^{(2)},x^{(3)}\}\:,\:s=\{\uparrow,\downarrow\}\big)\;;  \\  \lb{s1_5}
\sum_{\vec{x}}\ldots  &=& \int_{V^{(d)}}\frac{\big(\Delta x\big)^{d}}{V^{(d)}}\ldots =
\sum_{\{\vec{x}_{i}\}}\frac{1}{\mcal{N}_{x}}\ldots \;\;;\;\;\;\mcal{N}_{x}=
\frac{V^{(d)}}{\big(\Delta x\big)^{d}}\;;\\ \no
\sum_{\vec{y}}\ldots &=&\sum_{\vec{x},s=\uparrow,\downarrow}\ldots
=\sum_{s=\uparrow,\downarrow}
\int_{V^{(d)}}\frac{\big(\Delta x\big)^{d}}{V^{(d)}}\ldots
=\sum_{s=\uparrow,\downarrow}
\sum_{\{\vec{x}_{i}\}}\frac{1}{\mcal{N}_{x}}\ldots\;; \\ \no
m_{e}/M_{nucl.} &\rightarrow& 0 \;.
\eeq
The Hamiltonian \(\boldsymbol{\hat{H}}(\hat{\psi}\pdag,\hat{\psi};B_{z})\) (\ref{s1_6}) consists of the
one-particle part \(\boldsymbol{\hat{H}}^{(1)}(\hat{\psi}\pdag,\hat{\psi};B_{z})\) (\ref{s1_7}) with
an external magnetic field $B_{z}$ and the repulsive
Coulomb interaction \(\boldsymbol{\hat{V}_{\!c}}(\hat{\psi}\pdag,\hat{\psi})\) among the electrons.
The standard one-particle Hamiltonian has density related terms of bilinear, second quantized fields
\(\hat{\psi}_{\vec{x}\ppr,s\ppr}\pdag\ldots\hat{\psi}_{\vec{x},s}\) with one-particle matrix
elements \(\hat{H}^{(1)}_{\vec{x}\ppr,s\ppr;\vec{x},s}(B_{z})\) (\ref{s1_8}). The latter one-particle
matrix elements (\ref{s1_8}) contain the kinetic energy with an external magnetic field $B_{z}$ (\ref{s1_9})
in symmetric gauge and the Coulomb attraction of the nucleus with total charge '\(z_{0}\;e\)' at the coordinate
origin. Furthermore, the coupling of the electron's magnetic moment to the external magnetic field is added
with inclusion of the Pauli matrix \((\hat{\sigma}_{z})_{s\ppr s}=(\hat{\sigma}_{3})_{s\ppr s}\) for the
external magnetic field $B_{z}$ in z-direction
\beq \lb{s1_6}
\boldsymbol{\hat{H}}(\hat{\psi}\pdag,\hat{\psi};B_{z}) &=&
\boldsymbol{\hat{H}}^{(1)}(\hat{\psi}\pdag,\hat{\psi};B_{z}) +
\boldsymbol{\hat{V}_{\!c}}(\hat{\psi}\pdag,\hat{\psi}) \;;  \\  \lb{s1_7}
\boldsymbol{\hat{H}}^{(1)}(\hat{\psi}\pdag,\hat{\psi};B_{z}) &=&
\hspace*{-0.3cm}\sum_{\vec{x};s,s\ppr}\hat{\psi}_{\vec{x},s\ppr}\pdag\bigg[
\bigg(\frac{(\hat{\vec{p}}+e\,\vec{A}(\vec{x})\,)^{2}}{2\,m_{e}}
-\frac{1}{4\pi\,\ve_{0}}\,
\frac{z_{0}\:e^{2}}{|\vec{x}|+k_{\mbox{\scz nucl.}}}\bigg)\delta_{s\ppr s}+\mu_{B}\; B_{z}
\left(\bea{cc} 1 & \\ & -1\eea\right)_{s\ppr s}\bigg]\,\hat{\psi}_{\vec{x},s} \\ \no
&=&\hspace*{-0.3cm}
\sum_{\vec{x};s,s\ppr}\hat{\psi}_{\vec{x},s\ppr}\pdag\bigg[\bigg(-\frac{\hbar^{2}}{2\,m_{e}}\;
\frac{\pp}{\pp\vec{x}}\cdot\frac{\pp}{\pp\vec{x}}+\frac{e\,B_{z}}{2\,m_{e}}\;\hat{L}_{z}+
\frac{e^{2}\,B_{z}^{2}}{8\,m_{e}}\;\big(x^{2}+y^{2}\big)+  \\ \no &-& \frac{1}{4\pi\,\ve_{0}}\,
\frac{z_{0}\:e^{2}}{|\vec{x}|+k_{\mbox{\scz nucl.}}}\bigg)\delta_{s\ppr s}+\mu_{B}\; B_{z}
\left(\bea{cc} 1 & \\ & -1\eea\right)_{s\ppr s}\bigg]\,\hat{\psi}_{\vec{x},s} =
\sum_{\vec{x},s;\vec{x}\ppr,s\ppr}\hat{\psi}_{\vec{x}\ppr,s\ppr}\pdag\;
\hat{H}^{(1)}_{\vec{x}\ppr,s\ppr;\vec{x},s}(B_{z})\;\;\hat{\psi}_{\vec{x},s} \;;  \\  \lb{s1_8}
\hat{H}^{(1)}_{\vec{x}\ppr,s\ppr;\vec{x},s}(B_{z}) &=&
\bigg[\bigg(-\frac{\hbar^{2}}{2\,m_{e}}\;
\frac{\pp}{\pp\vec{x}\ppr}\cdot\frac{\pp}{\pp\vec{x}\ppr}+\frac{e\,B_{z}}{2\,m_{e}}\;
\big(\vec{x}\ppr\times{\ts\frac{\hbar}{\im}\frac{\pp}{\pp\vec{x}\ppr}}\big)_{z}+
\frac{e^{2}\,B_{z}^{2}}{8\,m_{e}}\;\big(x^{\bprime 2}+y^{\bprime 2}\big) + \\ \no &-& \frac{1}{4\pi\,\ve_{0}}\,
\frac{z_{0}\:e^{2}}{|\vec{x}\ppr|+k_{\mbox{\scz nucl.}}}\bigg)\delta_{s\ppr s}+\mu_{B}\; B_{z}
\left(\bea{cc} 1 & \\ & -1\eea\right)_{s\ppr s}\bigg]\;\delta_{\vec{x}\ppr,\vec{x}}\:\mcal{N}_{x} \;; \\  \lb{s1_9}
\vec{A}(\vec{x}) &=& (B_{z}/2)\;\big(-y\,,\,x\,,\,0\big)^{T}\;;\;\;\;\;
\mu_{B}=e\hbar/(2m_{e})\;;\;\;\;(e>0)\;.
\eeq
The Coulomb repulsion \(\boldsymbol{\hat{V}_{\!c}}(\hat{\psi}\pdag,\hat{\psi})\) of the electrons is described in
relation (\ref{s1_10}) where one can also introduce an inverse screening length \(k_{\mbox{\scz e}}\) as for the nuclear screening
with \(k_{\mbox{\scz nucl.}}\) of (\ref{s1_7},\ref{s1_8}) in order to regularize the Coulomb potential
\beq  \lb{s1_10}
\boldsymbol{\hat{V}_{\!c}}(\hat{\psi}\pdag,\hat{\psi}) &=& \sum_{\vec{x},s;\vec{x}\ppr,s\ppr}
\hat{\psi}_{\vec{x},s}\pdag\;\hat{\psi}_{\vec{x}\ppr,s\ppr}\pdag\;\frac{e^{2}}{4\pi\,\ve_{0}}\;
\frac{1}{|\vec{x}-\vec{x}\ppr|+k_{\mbox{\scz e}}}\;\hat{\psi}_{\vec{x}\ppr,s\ppr}\;
\hat{\psi}_{\vec{x},s}\;.
\eeq
The rotational symmetry of the electrons in the central potential and the interaction with an external
magnetic field in absence of a spin-orbit coupling imply the following, maximal commuting set of symmetry operators in
second quantized form
\beq  \lb{s1_11} &&\hspace*{-1.3cm}
\boldsymbol{\hat{H}}(\hat{\psi}\pdag,\hat{\psi};B_{z}) \;,\;
\boldsymbol{\vec{L}}(\hat{\psi}\pdag,\hat{\psi})\cdot
\boldsymbol{\vec{L}}(\hat{\psi}\pdag,\hat{\psi})\; ,\;
\boldsymbol{\vec{S}}(\hat{\psi}\pdag,\hat{\psi})\cdot
\boldsymbol{\vec{S}}(\hat{\psi}\pdag,\hat{\psi})\;,\;
\boldsymbol{\hat{N}}(\hat{\psi}\pdag,\hat{\psi})\;,\;
\boldsymbol{\hat{L}_{z}}(\hat{\psi}\pdag,\hat{\psi})\;,\;
\boldsymbol{\hat{S}_{z}}(\hat{\psi}\pdag,\hat{\psi})\;\longrightarrow \\ \no & \rightarrow &
\mbox{maximal commuting set of second quantized operators according to the symmetries}\;.
\eeq
In the remaining of this section the various commutator relations are computed in order to verify
the given set (\ref{s1_11}) of maximal commuting symmetry operators. Aside from the Hamiltonian
as a two-particle operator, there appear the one-particle operators
\(\boldsymbol{\hat{\mfrak{O}}^{(k=1,2,3)}}(\hat{\psi}\pdag,\hat{\psi})\),
\(\boldsymbol{\hat{\mfrak{J}}_{i}^{(k=1,2)}}(\hat{\psi}\pdag,\hat{\psi})\) (\ref{s1_12})
with matrix elements \(\hat{\mfrak{O}}^{(k)}_{\vec{x}\ppr,s\ppr;\vec{x},s}\) (\ref{s1_13}),
\(\hat{\mfrak{J}}^{(k)}_{i;\vec{x}\ppr,s\ppr;\vec{x},s}\) (\ref{s1_18}), which are introduced
for the total number \(\boldsymbol{\hat{N}}(\hat{\psi}\pdag,\hat{\psi})\) (\ref{s1_14})
of electrons and the total orbital angular momentum \(\boldsymbol{\hat{L}_{z}}(\hat{\psi}\pdag,\hat{\psi})\)
(\ref{s1_15}) and spin momentum \(\boldsymbol{\hat{S}_{z}}(\hat{\psi}\pdag,\hat{\psi})\) (\ref{s1_16})
in z-direction. However, the squares of absolute values
\(\boldsymbol{\hat{\mfrak{V}}^{(\kappa)}}(\hat{\psi}\pdag,\hat{\psi})\)
(\ref{s1_19}) of orbital \(\boldsymbol{\vec{L}}(\hat{\psi}\pdag,\hat{\psi})\boldsymbol{\cdot}
\boldsymbol{\vec{L}}(\hat{\psi}\pdag,\hat{\psi})\) and spin angular momentum
\(\boldsymbol{\vec{S}}(\hat{\psi}\pdag,\hat{\psi})\boldsymbol{\cdot}
\boldsymbol{\vec{S}}(\hat{\psi}\pdag,\hat{\psi})\) contain two-particle parts which therefore
require transformations analogous to the Hamiltonian
\beq\lb{s1_12}
\mbox{one-particle operators} &:& \boldsymbol{\hat{\mfrak{O}}^{(k=1,2,3)}}(\hat{\psi}\pdag,\hat{\psi})\;;\;
\boldsymbol{\hat{\mfrak{J}}_{i}^{(k=1,2)}}(\hat{\psi}\pdag,\hat{\psi})\;;  \\ \lb{s1_13}
\boldsymbol{\hat{\mfrak{O}}^{(k)}}(\hat{\psi}\pdag,\hat{\psi}) &=&\sum_{\vec{x},s;\vec{x}\ppr,s\ppr}
\hat{\psi}_{\vec{x}\ppr,s\ppr}\pdag\;\hat{\mfrak{O}}^{(k)}_{\vec{x}\ppr,s\ppr;\vec{x},s}\;\hat{\psi}_{\vec{x},s}
\;\;\;;\;(k=1,2,3)\;; \\  \lb{s1_14}
\boldsymbol{\hat{\mfrak{O}}^{(k=1)}}(\hat{\psi}\pdag,\hat{\psi}) &=&
\boldsymbol{\hat{N}}(\hat{\psi}\pdag,\hat{\psi})=\sum_{\vec{x},s}
\hat{\psi}_{\vec{x},s}\pdag\;\hat{\psi}_{\vec{x},s} \;;\;\rightarrow
\hat{\mfrak{O}}^{(k=1)}_{\vec{x}\ppr,s\ppr;\vec{x},s} =
\mcal{N}_{x}\;\delta_{\vec{x}\ppr,\vec{x}}\;\delta_{s\ppr,s}\;;    \\  \lb{s1_15}
\boldsymbol{\hat{\mfrak{O}}^{(k=2)}}(\hat{\psi}\pdag,\hat{\psi}) &=&\hspace*{-0.2cm}
\boldsymbol{\hat{L}_{z}}(\hat{\psi}\pdag,\hat{\psi})=\sum_{\vec{x},s}
\hat{\psi}_{\vec{x},s}\pdag\;\big(\vec{x}\times\hat{\vec{p}}\big)_{z}\hat{\psi}_{\vec{x},s} \;;\rightarrow
\hat{\mfrak{O}}^{(k=2)}_{\vec{x}\ppr,s\ppr;\vec{x},s} =
\big(\vec{x}\ppr\times\vec{p}\ppr\big)_{z}\delta_{\vec{x}\ppr,\vec{x}}\mcal{N}_{x}\,\delta_{s\ppr,s}\;;  \\   \lb{s1_16}
\boldsymbol{\hat{\mfrak{O}}^{(k=3)}}(\hat{\psi}\pdag,\hat{\psi}) &=&
\boldsymbol{\hat{S}_{z}}(\hat{\psi}\pdag,\hat{\psi})=\sum_{\vec{x};s,s\ppr}
\hat{\psi}_{\vec{x},s\ppr}\pdag\frac{\hbar}{2}\bigg(\bea{cc} 1 & \\ &\hspace*{-0.2cm} -1\eea\bigg)_{s\ppr s}
\hspace*{-0.1cm}\hat{\psi}_{\vec{x},s}\;;\;\rightarrow
\hat{\mfrak{O}}^{(k=3)}_{\vec{x}\ppr,s\ppr;\vec{x},s} =\frac{\hbar}{2}
\bigg(\bea{cc} 1 & \\ & \hspace*{-0.2cm}-1 \eea\bigg)_{s\ppr s}\hspace*{-0.2cm}\mcal{N}_{x}\,
\delta_{\vec{x}\ppr,\vec{x}}\;;  \\  \lb{s1_17}
\boldsymbol{\hat{\mfrak{J}}_{i}^{(k)}}(\hat{\psi}\pdag,\hat{\psi}) &=&
\underbrace{\boldsymbol{\hat{L}_{i}}(\hat{\psi}\pdag,\hat{\psi})}_{k=1}\;,\,
\underbrace{\boldsymbol{\hat{S}_{i}}(\hat{\psi}\pdag,\hat{\psi})}_{k=2} \;; \;(k=1,2)\;;  \\  \lb{s1_18}
\boldsymbol{\hat{\mfrak{J}}_{i}^{(k)}}(\hat{\psi}\pdag,\hat{\psi}) &=& \sum_{\vec{x},s;\vec{x}\ppr,s\ppr}
\hat{\psi}_{\vec{x}\ppr,s\ppr}\pdag\;\hat{\mfrak{J}}^{(k)}_{i;\vec{x}\ppr,s\ppr;\vec{x},s}\;\hat{\psi}_{\vec{x},s}
\;\;\;;\;(k=1,2)\;;   \\   \lb{s1_19}
\mbox{two-particle operators} &:& \boldsymbol{\hat{\mfrak{V}}^{(\kappa)}}(\hat{\psi}\pdag,\hat{\psi})
=\boldsymbol{\hat{\mfrak{J}}_{i}^{(k)}}(\hat{\psi}\pdag,\hat{\psi})\;\cdot\;
\boldsymbol{\hat{\mfrak{J}}_{i}^{(k)}}(\hat{\psi}\pdag,\hat{\psi})\;\;\;;\;(k=1,2)\;.
\eeq
Concerning the property of a maximal commuting set of symmetry operators, we attain relations (\ref{s1_20}-\ref{s1_29})
for one-particle and two-particle operators (including the Hamiltonian) and therefore attest through the given
commutators the validity of maximal chosen symmetries
\beq \lb{s1_20}
\boldsymbol{\Big[}\boldsymbol{\hat{\mfrak{O}}^{(k)}}(\hat{\psi}\pdag,\hat{\psi})\;\boldsymbol{,}\;
\boldsymbol{\hat{\mfrak{O}}^{(\kappa)}}(\hat{\psi}\pdag,\hat{\psi}) \boldsymbol{\Big]_{-}} &=&
\sum_{\vec{x}\ppr,s\ppr;\vec{x},s}\hat{\psi}_{\vec{x}\ppr,s\ppr}\pdag\;
\bigg(\boldsymbol{\Big[}\hat{\mfrak{O}}_{\vec{x}_{1}\ppr,s_{1}\ppr;\vec{x}_{1},s_{1}}^{(k)}\;\boldsymbol{,}\;
\hat{\mfrak{O}}_{\vec{x}_{2}\ppr,s_{2}\ppr;\vec{x}_{2},s_{2}}^{(\kappa)} \boldsymbol{\Big]_{-}}
\bigg)_{\vec{x}\ppr,s\ppr;\vec{x},s}\;\hat{\psi}_{\vec{x},s} \;; \\ \lb{s1_21}
\boldsymbol{\big\{}\hat{\psi}_{\vec{x},s}\;\boldsymbol{,}\;\hat{\psi}_{\vec{x}\ppr,s\ppr}\pdag\boldsymbol{\big\}_{+}}
&=&\mcal{N}_{x}\:\delta_{\vec{x},\vec{x}\ppr}\:\delta_{ss\ppr}\;;\;\;\;\;
\boldsymbol{\big\{}\hat{\psi}_{\vec{x},s}\;\boldsymbol{,}\;\hat{\psi}_{\vec{x}\ppr,s\ppr}\boldsymbol{\big\}_{+}}=
\boldsymbol{\big\{}\hat{\psi}_{\vec{x},s}\pdag\;\boldsymbol{,}\;
\hat{\psi}_{\vec{x}\ppr,s\ppr}\pdag\boldsymbol{\big\}_{+}}=0\;;  \\  \lb{s1_22}
\boldsymbol{\hat{L}_{i}}(\hat{\psi}\pdag,\hat{\psi}) &=&\sum_{\vec{x},s}
\hat{\psi}_{\vec{x},s}\pdag\;\Big(\vec{x}\times\hat{\vec{p}}\Big)_{i}\,\hat{\psi}_{\vec{x},s}  \;; \\ \lb{s1_23}
\boldsymbol{\Big[}\boldsymbol{\hat{L}_{i}}(\hat{\psi}\pdag,\hat{\psi})\;\boldsymbol{,}\;
\boldsymbol{\hat{L}_{j}}(\hat{\psi}\pdag,\hat{\psi}) \boldsymbol{\Big]_{-}} &=&\im\;\ve_{ijk}\;\hbar\;
\boldsymbol{\hat{L}_{k}}(\hat{\psi}\pdag,\hat{\psi}) \;;\;\;\;
\boldsymbol{\Big[}\boldsymbol{\hat{L}_{i}}(\hat{\psi}\pdag,\hat{\psi})\;\boldsymbol{,}\;
\boldsymbol{\vec{L}}(\hat{\psi}\pdag,\hat{\psi})\cdot
\boldsymbol{\vec{L}}(\hat{\psi}\pdag,\hat{\psi})\boldsymbol{\Big]_{-}} = 0 \;; \\  \lb{s1_24}
\boldsymbol{\hat{S}_{i}}(\hat{\psi}\pdag,\hat{\psi}) &=&\sum_{\vec{x};s\ppr,s}
\hat{\psi}_{\vec{x},s\ppr}\pdag\;\frac{\hbar}{2}
\Big(\hat{\sigma}_{i}\Big)_{s\ppr s}\,\hat{\psi}_{\vec{x},s}  \;; \\ \lb{s1_25}
\boldsymbol{\Big[}\boldsymbol{\hat{S}_{i}}(\hat{\psi}\pdag,\hat{\psi})\;\boldsymbol{,}\;
\boldsymbol{\hat{S}_{j}}(\hat{\psi}\pdag,\hat{\psi}) \boldsymbol{\Big]_{-}} &=&\im\;\ve_{ijk}\;\hbar\;
\boldsymbol{\hat{S}_{k}}(\hat{\psi}\pdag,\hat{\psi}) \;;\;\;\;
\boldsymbol{\Big[}\boldsymbol{\hat{S}_{i}}(\hat{\psi}\pdag,\hat{\psi})\;\boldsymbol{,}\;
\boldsymbol{\vec{S}}(\hat{\psi}\pdag,\hat{\psi})\cdot
\boldsymbol{\vec{S}}(\hat{\psi}\pdag,\hat{\psi})\boldsymbol{\Big]_{-}} = 0 \;;  \\  \lb{s1_26}
\boldsymbol{\Big[}\boldsymbol{\hat{L}_{i}}(\hat{\psi}\pdag,\hat{\psi})\;\boldsymbol{,}\;
\boldsymbol{\hat{S}_{j}}(\hat{\psi}\pdag,\hat{\psi}) \boldsymbol{\Big]_{-}} &=&0 \;;  \\  \lb{s1_27}
\boldsymbol{\Big[}\boldsymbol{\hat{N}}(\hat{\psi}\pdag,\hat{\psi})\;\boldsymbol{,}\;
\boldsymbol{\hat{L}_{i}}(\hat{\psi}\pdag,\hat{\psi}) \boldsymbol{\Big]_{-}} &=&
\boldsymbol{\Big[}\boldsymbol{\hat{N}}(\hat{\psi}\pdag,\hat{\psi})\;\boldsymbol{,}\;
\boldsymbol{\hat{S}_{i}}(\hat{\psi}\pdag,\hat{\psi}) \boldsymbol{\Big]_{-}} =
\boldsymbol{\Big[}\boldsymbol{\hat{N}}(\hat{\psi}\pdag,\hat{\psi})\;\boldsymbol{,}\;
\boldsymbol{\hat{H}}(\hat{\psi}\pdag,\hat{\psi};B_{z}) \boldsymbol{\Big]_{-}} = 0 \;; \\ \lb{s1_28}
\boldsymbol{\Big[}\boldsymbol{\hat{L}_{z}}(\hat{\psi}\pdag,\hat{\psi})\;\boldsymbol{,}\;
\boldsymbol{\hat{H}}(\hat{\psi}\pdag,\hat{\psi};B_{z}) \boldsymbol{\Big]_{-}} &=&
\boldsymbol{\Big[}\boldsymbol{\hat{S}_{z}}(\hat{\psi}\pdag,\hat{\psi})\;\boldsymbol{,}\;
\boldsymbol{\hat{H}}(\hat{\psi}\pdag,\hat{\psi};B_{z}) \boldsymbol{\Big]_{-}} =  0 \;; \\  \lb{s1_29}
\boldsymbol{\Big[}\boldsymbol{\vec{L}}(\hat{\psi}\pdag,\hat{\psi})\cdot
\boldsymbol{\vec{L}}(\hat{\psi}\pdag,\hat{\psi})\;\boldsymbol{,}\;
\boldsymbol{\hat{H}}(\hat{\psi}\pdag,\hat{\psi};B_{z}) \boldsymbol{\Big]_{-}} &=&
\boldsymbol{\Big[}\boldsymbol{\vec{S}}(\hat{\psi}\pdag,\hat{\psi})\cdot
\boldsymbol{\vec{S}}(\hat{\psi}\pdag,\hat{\psi})\;\boldsymbol{,}\;
\boldsymbol{\hat{H}}(\hat{\psi}\pdag,\hat{\psi};B_{z}) \boldsymbol{\Big]_{-}} =  0 \;.
\eeq
The commutators (\ref{s1_20}-\ref{s1_29}) result from the fundamental anti-commutators (\ref{s1_21},\ref{s1_30})
of second quantization which we further apply to determine the two-particle content (\ref{s1_31}-\ref{s1_34})
of the absolute values of the orbital \(\boldsymbol{\hat{\mfrak{V}}^{(\kappa=1)}}(\hat{\psi}\pdag,\hat{\psi})=
\boldsymbol{\vec{L}}(\hat{\psi}\pdag,\hat{\psi})\boldsymbol{\cdot}
\boldsymbol{\vec{L}}(\hat{\psi}\pdag,\hat{\psi})\) (\ref{s1_31},\ref{s1_32}) and spin angular momentum
\(\boldsymbol{\hat{\mfrak{V}}^{(\kappa=2)}}(\hat{\psi}\pdag,\hat{\psi})=
\boldsymbol{\vec{S}}(\hat{\psi}\pdag,\hat{\psi})\boldsymbol{\cdot}
\boldsymbol{\vec{S}}(\hat{\psi}\pdag,\hat{\psi})\) (\ref{s1_33},\ref{s1_34}). Apart from the matrix elements
\((\vec{L}\cdot\vec{L})_{\vec{x}_{2},\vec{x}_{1}}\), \((\vec{S}\cdot\vec{S})_{s\ppr s}\) of the one-particle ingredients,
we point out the two-particle parts of the absolute values of the two angular momentum types, similar and analogous to the
total Hamiltonian (\ref{s1_40})
\(\boldsymbol{\hat{\mfrak{V}}^{(\kappa=0)}}(\hat{\psi}\pdag,\hat{\psi})=
\boldsymbol{\hat{H}}(\hat{\psi}\pdag,\hat{\psi};B_{z})\) with one-particle elements
\(\hat{v}_{\vec{x}\ppr,s\ppr;\vec{x},s}^{(\kappa=0)}=\hat{H}^{(1)}_{\vec{x}\ppr,s\ppr;\vec{x},s}(B_{z})\) (\ref{s1_41}) and
two-particle matrix elements (\ref{s1_42}). Although the two-particle parts of the absolute values of orbital and spin
momentum can be summarized by two-particle matrix elements (\ref{s1_37},\ref{s1_39}) in correspondence to the two-particle,
Coulomb interaction matrix elements (\ref{s1_42}), one has to regard the following, crucial difference of
elements \(\hat{\mscr{V}}_{\vec{y}_{2},\vec{y}_{2}\ppr;\vec{y}_{1}\ppr,\vec{y}_{1}}^{(\kappa=0,1,2)}\)
(\ref{s1_35},\ref{s1_37},\ref{s1_39},\ref{s1_42}) : \newline
The interaction matrix elements (\ref{s1_37},\ref{s1_39}) of the absolute values of orbital and spin angular momentum separate
into a sum over three components \(i=1,2,3\) where each summand of the three components results into a factorization of
two independent factors.
This is in contrast to the Coulomb matrix element
\(\hat{\mscr{V}}_{\vec{y}_{2},\vec{y}_{2}\ppr;\vec{y}_{1}\ppr,\vec{y}_{1}}^{(\kappa=0)}\)
where such a factorization does not occur in the interaction. Therefore, the HST transformations of the various two-particle
parts within \(\boldsymbol{\hat{\mfrak{V}}^{(\kappa=1,2)}}(\hat{\psi}\pdag,\hat{\psi})\) have to be distinguished from the
Hamiltonian part \(\boldsymbol{\hat{\mfrak{V}}^{(\kappa=0)}}(\hat{\psi}\pdag,\hat{\psi})=
\boldsymbol{\hat{H}}(\hat{\psi}\pdag,\hat{\psi};B_{z})\) with a 'true' Coulomb 'interaction'
\(\boldsymbol{\hat{V}_{\!c}}(\hat{\psi}\pdag,\hat{\psi})\) and non-vanishing matrix elements (\ref{s1_42}) between
primed and unprimed coordinate labels so that a corresponding factorization of the interaction is not possible
\beq\lb{s1_30}
\boldsymbol{\big\{}\hat{\psi}_{\vec{x},s}\;\boldsymbol{,}\;\hat{\psi}_{\vec{x}\ppr,s\ppr}\pdag\boldsymbol{\big\}_{+}}
&=&\mcal{N}_{x}\:\delta_{\vec{x},\vec{x}\ppr}\:\delta_{ss\ppr}\;;\;\;\;\;
\boldsymbol{\big\{}\hat{\psi}_{\vec{x},s}\;\boldsymbol{,}\;\hat{\psi}_{\vec{x}\ppr,s\ppr}\boldsymbol{\big\}_{+}}=
\boldsymbol{\big\{}\hat{\psi}_{\vec{x},s}\pdag\;\boldsymbol{,}\;\hat{\psi}_{\vec{x}\ppr,s\ppr}\pdag\boldsymbol{\big\}_{+}}=0\;;
\\  \lb{s1_31} \boldsymbol{\hat{\mfrak{V}}^{(\kappa=1)}}(\hat{\psi}\pdag,\hat{\psi}) &=&
\boldsymbol{\vec{L}}(\hat{\psi}\pdag,\hat{\psi})\boldsymbol{\cdot}
\boldsymbol{\vec{L}}(\hat{\psi}\pdag,\hat{\psi}) =
\sum_{\vec{x}_{1},\vec{x}_{2},s;\vec{x}_{1}\ppr,\vec{x}_{2}\ppr,s\ppr}
\hat{\psi}_{\vec{x}_{2}\ppr,s\ppr}\pdag\;\big(\hat{L}_{i}\big)_{\vec{x}_{2}\ppr,\vec{x}_{1}\ppr}\;
\hat{\psi}_{\vec{x}_{1}\ppr,s\ppr}\;\;
\hat{\psi}_{\vec{x}_{2},s}\pdag\;\big(\hat{L}_{i}\big)_{\vec{x}_{2},\vec{x}_{1}}\;
\hat{\psi}_{\vec{x}_{1},s}  \\ \no &=&
\sum_{\vec{x}_{1},\vec{x}_{2},s}\hat{\psi}_{\vec{x}_{2},s}\pdag\;
\big(\vec{L}\cdot\vec{L}\big)_{\vec{x}_{2},\vec{x}_{1}}\;\hat{\psi}_{\vec{x}_{1},s} +
\sum_{\vec{x}_{1},\vec{x}_{2},s;\vec{x}_{1}\ppr,\vec{x}_{2}\ppr,s\ppr}
\hat{\psi}_{\vec{x}_{2},s}\pdag\;\hat{\psi}_{\vec{x}_{2}\ppr,s\ppr}\pdag\;
\big(\hat{L}_{i}\big)_{\vec{x}_{2}\ppr,\vec{x}_{1}\ppr}\;
\hat{\psi}_{\vec{x}_{1}\ppr,s\ppr}\;\;
\big(\hat{L}_{i}\big)_{\vec{x}_{2},\vec{x}_{1}}\;
\hat{\psi}_{\vec{x}_{1},s} \;;  \\  \lb{s1_32}
\big(\hat{L}_{i}\big)_{\vec{x}_{2},\vec{x}_{1}} &=&
\big(\vec{x}_{2}\times\hat{\vec{p}}_{2}\big)_{i}\;\delta_{\vec{x}_{2},\vec{x}_{1}}\;\mcal{N}_{x}\;;\;\;\;
\big(\vec{L}\cdot\vec{L}\big)_{\vec{x}_{2},\vec{x}_{1}} =
\Big(\big(\vec{x}_{2}\times\vec{p}_{2}\big)\cdot\big(\vec{x}_{2}\times\vec{p}_{2}\big)\Big)
\delta_{\vec{x}_{2},\vec{x}_{1}}\mcal{N}_{x}\;;   \\   \lb{s1_33}
\boldsymbol{\hat{\mfrak{V}}^{(\kappa=2)}}(\hat{\psi}\pdag,\hat{\psi}) &=&
\boldsymbol{\vec{S}}(\hat{\psi}\pdag,\hat{\psi})\boldsymbol{\cdot}
\boldsymbol{\vec{S}}(\hat{\psi}\pdag,\hat{\psi}) =
\sum_{\vec{x}_{1},s_{1},s_{1}\ppr;\vec{x}_{2},s_{2},s_{2}\ppr}
\hat{\psi}_{\vec{x}_{2},s_{2}\ppr}\pdag\;\big(\hat{S}_{i}\big)_{s_{2}\ppr s_{2}}\;
\hat{\psi}_{\vec{x}_{2},s_{2}}\;\;
\hat{\psi}_{\vec{x}_{1},s_{1}\ppr}\pdag\;\big(\hat{S}_{i}\big)_{s_{1}\ppr s_{1}}\;
\hat{\psi}_{\vec{x}_{1},s_{1}}  \\ \no &=&
\sum_{\vec{x},s,s\ppr}\hat{\psi}_{\vec{x},s\ppr}\pdag\;
\big(\vec{S}\cdot\vec{S}\big)_{s\ppr s}\;\hat{\psi}_{\vec{x},s} +
\sum_{\vec{x}_{1},s_{1},s_{1}\ppr;\vec{x}_{2},s_{2},s_{2}\ppr}
\hat{\psi}_{\vec{x}_{1},s_{1}\ppr}\pdag\;\hat{\psi}_{\vec{x}_{2},s_{2}\ppr}\pdag\;
\big(\hat{S}_{i}\big)_{s_{2}\ppr s_{2}}\;
\hat{\psi}_{\vec{x}_{2},s_{2}}\;\;
\big(\hat{S}_{i}\big)_{s_{1}\ppr s_{1}}\;
\hat{\psi}_{\vec{x}_{1},s_{1}} \;;  \\  \lb{s1_34}
\big(\hat{S}_{i}\big)_{s_{1}\ppr s_{1}} &=&\frac{\hbar}{2}
\big(\hat{\sigma}_{i}\big)_{s_{1}\ppr s_{1}}\;;\;\;\;
\big(\vec{S}\cdot\vec{S}\big)_{s\ppr s}=\frac{3}{4}\,\hbar^{2}\;\delta_{s\ppr s}\;;  \\  \lb{s1_35}
\boldsymbol{\hat{\mfrak{V}}^{(\kappa)}}(\hat{\psi}\pdag,\hat{\psi}) &=&
\sum_{\vec{y};\vec{y}\ppr}\hat{\psi}_{\vec{y}\ppr}\pdag\:
\hat{v}_{\vec{y}\ppr;\vec{y}}^{(\kappa)}\:\hat{\psi}_{\vec{y}} +
\sum_{\vec{y}_{1/2};\vec{y}_{1/2}\ppr}
\hat{\psi}_{\vec{y}_{2}}\pdag\:\hat{\psi}_{\vec{y}_{2}\ppr}\pdag\;
\hat{\mscr{V}}_{\vec{y}_{2},\vec{y}_{2}\ppr;\vec{y}_{1}\ppr,\vec{y}_{1}}^{(\kappa)}\;
\hat{\psi}_{\vec{y}_{1}\ppr}\:\hat{\psi}_{\vec{y}_{1}}  \;;  \\  \lb{s1_36}
\hat{v}_{\vec{y}\ppr;\vec{y}}^{(\kappa=1)} &=& \hat{v}_{\vec{x}\ppr,s\ppr;\vec{x},s}^{(\kappa=1)} =
\Big(\big(\vec{x}\ppr\times\vec{p}\ppr\big)\cdot
\big(\vec{x}\ppr\times\vec{p}\ppr\big)\delta_{\vec{x}\ppr,\vec{x}}\:\mcal{N}_{x}\:
\delta_{s\ppr s} \Big)  \;;   \\  \lb{s1_37}
\hat{\mscr{V}}_{\vec{y}_{2},\vec{y}_{2}\ppr;\vec{y}_{1}\ppr,\vec{y}_{1}}^{(\kappa=1)}&=&
\hat{\mscr{V}}_{\vec{x}_{2},s_{2},\vec{x}_{2}\ppr,s_{2}\ppr;
\vec{x}_{1}\ppr,s_{1}\ppr,\vec{x}_{1},s_{1}}^{(\kappa=1)} =
\Big(\big(\vec{x}_{2}\ppr\times\vec{p}_{2}\ppr\big)\:\delta_{\vec{x}_{2}\ppr,\vec{x}_{1}\ppr}\:\mcal{N}_{x}\:
\delta_{s_{2}\ppr s_{1}\ppr} \Big) \boldsymbol{\cdot}
\Big(\big(\vec{x}_{2}\times\vec{p}_{2}\big)\:\delta_{\vec{x}_{2},\vec{x}_{1}}\:\mcal{N}_{x}\:
\delta_{s_{2}s_{1}} \Big)  \\  \no &=&\sum_{i=1}^{3}
\Big(\big(\vec{x}_{2}\ppr\times\vec{p}_{2}\ppr\big)_{\boldsymbol{i}}\:
\delta_{\vec{x}_{2}\ppr,\vec{x}_{1}\ppr}\:\mcal{N}_{x}\:
\delta_{s_{2}\ppr s_{1}\ppr} \Big)
\Big(\big(\vec{x}_{2}\times\vec{p}_{2}\big)_{\boldsymbol{i}}\:\delta_{\vec{x}_{2},\vec{x}_{1}}\:\mcal{N}_{x}\:
\delta_{s_{2}s_{1}} \Big)    \;; \\  \lb{s1_38}
\hat{v}_{\vec{y}\ppr;\vec{y}}^{(\kappa=2)} &=& \hat{v}_{\vec{x}\ppr,s\ppr;\vec{x},s}^{(\kappa=2)} =
\frac{3}{4}\:\hbar^{2}\:\delta_{s\ppr s}\:\delta_{\vec{x}\ppr,\vec{x}}\:\mcal{N}_{x}\;;  \\  \lb{s1_39}
\hat{\mscr{V}}_{\vec{y}_{2},\vec{y}_{2}\ppr;\vec{y}_{1}\ppr,\vec{y}_{1}}^{(\kappa=2)} &=&
\hat{\mscr{V}}_{\vec{x}_{2},s_{2},\vec{x}_{2}\ppr,s_{2}\ppr;\vec{x}_{1}\ppr,s_{1}\ppr,\vec{x}_{1},s_{1}}^{(\kappa=2)} =
\sum_{i=1}^{3}\frac{\hbar}{2}\:\big(\hat{\sigma}_{i}\big)_{s_{2}\ppr s_{1}\ppr}\;
\delta_{\vec{x}_{2}\ppr,\vec{x}_{1}\ppr}\:\mcal{N}_{x}\;
\frac{\hbar}{2}\:\big(\hat{\sigma}_{i}\big)_{s_{2}s_{1}}\;
\delta_{\vec{x}_{2},\vec{x}_{1}}\:\mcal{N}_{x}  \;;  \\  \lb{s1_40}
\boldsymbol{\hat{\mfrak{V}}^{(\kappa=0)}}(\hat{\psi}\pdag,\hat{\psi}) &=&
\boldsymbol{\hat{H}}(\hat{\psi}\pdag,\hat{\psi};B_{z}) =
\boldsymbol{\hat{H}}^{(1)}(\hat{\psi}\pdag,\hat{\psi};B_{z}) +
\boldsymbol{\hat{V}_{\!c}}(\hat{\psi}\pdag,\hat{\psi}) \;;  \\  \lb{s1_41}
\hat{v}_{\vec{y}\ppr;\vec{y}}^{(\kappa=0)}  &=&
\hat{v}_{\vec{x}\ppr,s\ppr;\vec{x},s}^{(\kappa=0)} = \hat{H}^{(1)}_{\vec{x}\ppr,s\ppr;\vec{x},s}(B_{z}) \;; \\  \lb{s1_42}
\hat{\mscr{V}}_{\vec{y}_{2},\vec{y}_{2}\ppr;\vec{y}_{1}\ppr,\vec{y}_{1}}^{(\kappa=0)} &=&
\hat{\mscr{V}}_{\vec{x}_{2},s_{2},\vec{x}_{2}\ppr,s_{2}\ppr;\vec{x}_{1}\ppr,s_{1}\ppr,\vec{x}_{1},s_{1}}^{(\kappa=0)} =
\frac{e^{2}}{4\pi\,\ve_{0}}\;
\frac{1}{|\vec{x}_{1}-\vec{x}_{1}\ppr|+k_{\mbox{\scz e}}}\;\delta_{\vec{x}_{2},\vec{x}_{1}}\,\mcal{N}_{x}\:\delta_{s_{2}s_{1}}\;
\delta_{\vec{x}_{2}\ppr,\vec{x}_{1}\ppr}\,\mcal{N}_{x}\:\delta_{s_{2}\ppr s_{1}\ppr} \;.
\eeq
In this introductory section we have described the maximal commuting set of symmetry operators in terms of their
commutators and have grouped the operator set into one-particle and two-particle classes so that the given definitions
allow collective, abbreviated treatment for just two types of operators instead of a total of six different operators.

\subsection{Coherent state representation of delta functions within trace operations} \lb{s12}

Aside from the appropriate choice of the maximal commuting set of symmetry operators, we use the Dirac identity (\ref{s1_43})
in order to represent the delta functions of second quantized Fermi operators as part of the inverse of the corresponding
symmetry operators. Since one has to distinguish between one-particle
\(\boldsymbol{\hat{\mfrak{O}}^{(k)}}(\hat{\psi}\pdag,\hat{\psi})\) and two-particle operators
\(\boldsymbol{\hat{\mfrak{V}}^{(\kappa)}}(\hat{\psi}\pdag,\hat{\psi})\) in later transformations, the analogous
Dirac identities are listed separately in relations (\ref{s1_43}-\ref{s1_45}) and (\ref{s1_46}-\ref{s1_49}), respectively.
Note that the non-hermitian, imaginary increments \(\mp\im\,\ve_{+}^{(k)}\) (\ref{s1_43}) and
\(\mp\im\,\mfrak{e}_{+}^{(\kappa)}\) (\ref{s1_46},\ref{s1_47}) have to take the physical dimensions as their one-particle
or two-particle operators or their chosen, corresponding quantum numbers (\ref{s1_45}) and (\ref{s1_49})
\beq\lb{s1_43}
\lim_{\ve_{+}^{(k)}\rightarrow0_{+}}^{k=1,2,3}\frac{1}{\mfrak{o}^{(k)}-
\boldsymbol{\hat{\mfrak{O}}^{(k)}}(\hat{\psi}\pdag,\hat{\psi})\mp\im\,\ve_{+}^{(k)}} &=&
\frac{\mbox{Principal value}}{\mfrak{o}^{(k)}-\boldsymbol{\hat{\mfrak{O}}^{(k)}}(\hat{\psi}\pdag,\hat{\psi})}\pm\im\,\pi\;
\delta\big(\mfrak{o}^{(k)}-\boldsymbol{\hat{\mfrak{O}}^{(k)}}(\hat{\psi}\pdag,\hat{\psi})\,\big) \;;  \\  \lb{s1_44}
\boldsymbol{\hat{\mfrak{O}}^{(k)}}(\hat{\psi}\pdag,\hat{\psi}) &=&
\underbrace{\boldsymbol{\hat{N}}(\hat{\psi}\pdag,\hat{\psi})}_{k=1}\;,\;
\underbrace{\boldsymbol{\hat{L}_{z}}(\hat{\psi}\pdag,\hat{\psi})}_{k=2}\;,\,
\underbrace{\boldsymbol{\hat{S}_{z}}(\hat{\psi}\pdag,\hat{\psi})}_{k=3}\;; \\  \lb{s1_45}
\mfrak{o}^{(k)} &=& \underbrace{n_{0}}_{k=1}\;,\;\underbrace{\hbar\,l_{z}}_{k=2}\;,\;
\underbrace{\hbar\,s_{z}}_{k=3}\;;    \\ \lb{s1_46}
\lim_{\mfrak{e}_{+}^{(E)}\rightarrow0_{+}}\frac{1}{E-\boldsymbol{\hat{H}}(\hat{\psi}\pdag,\hat{\psi};B_{z})
\mp\im\,\mfrak{e}_{+}^{(E)}} &=&
\frac{\mbox{Principal value}}{E-\boldsymbol{\hat{H}}(\hat{\psi}\pdag,\hat{\psi};B_{z})}\pm\im\,\pi\;
\delta\big(E-\boldsymbol{\hat{H}}(\hat{\psi}\pdag,\hat{\psi};B_{z})\,\big) \;;  \\  \lb{s1_47}
\lim_{\mfrak{e}_{+}^{(\kappa)}\rightarrow0_{+}}^{\kappa=0,1,2}\frac{1}{\mfrak{v}^{(\kappa)}-
\boldsymbol{\hat{\mfrak{V}}^{(\kappa)}}(\hat{\psi}\pdag,\hat{\psi})\mp\im\,\mfrak{e}_{+}^{(\kappa)}} &=&
\frac{\mbox{Principal value}}{\mfrak{v}^{(\kappa)}-\boldsymbol{\hat{\mfrak{V}}^{(\kappa)}}(\hat{\psi}\pdag,\hat{\psi})}\pm\im\,\pi\;
\delta\big(\mfrak{v}^{(\kappa)}-\boldsymbol{\hat{\mfrak{V}}^{(\kappa)}}(\hat{\psi}\pdag,\hat{\psi})\,\big) \;;  \\  \lb{s1_48}
\boldsymbol{\hat{\mfrak{V}}^{(\kappa)}}(\hat{\psi}\pdag,\hat{\psi}) &=&
\underbrace{\boldsymbol{\hat{H}}(\hat{\psi}\pdag,\hat{\psi};B_{z})}_{\kappa=0}\;,\;
\underbrace{\boldsymbol{\vec{L}}(\hat{\psi}\pdag,\hat{\psi})\cdot
\boldsymbol{\vec{L}}(\hat{\psi}\pdag,\hat{\psi})}_{\kappa=1}\;,\;
\underbrace{\boldsymbol{\vec{S}}(\hat{\psi}\pdag,\hat{\psi})\cdot
\boldsymbol{\vec{S}}(\hat{\psi}\pdag,\hat{\psi})}_{\kappa=2}    \;; \\   \lb{s1_49}
\mfrak{v}^{(\kappa)} &=& \underbrace{E}_{\kappa=0}\;,\;\underbrace{\hbar^{2}\,l(l+1)}_{\kappa=1}\;,\;
\underbrace{\hbar^{2}\,s(s+1)}_{\kappa=2}\;\;.
\eeq
Since the inverted operators of the Dirac identities (\ref{s1_43}-\ref{s1_49}) can be obtained from
integration of exponentials with generalized 'time' parameters \(t_{p_{k}}^{(k)}\,/\,\hbar\), \(\mfrak{t}_{q_{\kappa}}^{(\kappa)}\,/\,\hbar\)
having the inverted physical dimensions as their operators or quantum numbers, we can represent the delta functions
of one-particle and two-particle operators with relations (\ref{s1_50}) and (\ref{s1_52}) where we additionally list
the delta function (\ref{s1_51}) of the Hamilton operator for further reference and transformations of the other two
two-particle operators. In order to eliminate the principal values of the Dirac identities, one has to consider
two separate integration branches '\(p_{k}=\pm\)', '\(q_{\kappa}=\pm\)' under inclusion of metric signs
'\(\eta_{p_{k}=\pm}^{(k)}=\pm\)', '\(\zeta_{q_{\kappa}=\pm}^{(\kappa)}=\pm\)' for every, one-particle and two-particle observable
\(\boldsymbol{\hat{\mfrak{O}}^{(k)}}(\hat{\psi}\pdag,\hat{\psi})\),
\(\boldsymbol{\hat{\mfrak{V}}^{(\kappa)}}(\hat{\psi}\pdag,\hat{\psi})\) so that their delta functions are only left
in correspondence. However, we emphasize again the imaginary values '\(-\im\:\ve_{p_{k}}^{(k)}\)',
'\(-\im\:\mfrak{e}_{q_{\kappa}}^{(\kappa)}\)' of the limit process within the Dirac identities which have to be adapted
to their metric signs '\(\eta_{p_{k}}^{(k)}\)', '\(\zeta_{q_{\kappa}}^{(\kappa)}\)' in such a manner that convergent integrals
follow for infinitely increasing upper integration boundaries \(T^{(k)}\rightarrow+\infty\), \(\mcal{T}^{(\kappa)}\rightarrow+\infty\).
Furthermore, a kind of 'time'-ordering '\(\overleftarrow{\exp}\{\ldots\}\)' of the exponential step operators has to be
included where one starts out from smaller, generalized 'time'-parameters on the right-hand side to larger 'time'-values
on the left-hand side
\beq\lb{s1_50}
\underbrace{\delta\big(\mfrak{o}^{(k)}-
\boldsymbol{\hat{\mfrak{O}}^{(k)}}(\hat{\psi}\pdag,\hat{\psi})\,\big)}_{(k=1,2,3)} &=& \hspace*{-0.4cm}\sum_{p_{k}=\pm}\;
\lim_{|\ve_{p_{k}}^{(k)}|\rightarrow 0} \;\lim_{T^{(k)}\rightarrow+\infty}\int_{0}^{T^{(k)}}
\hspace*{-0.2cm}\frac{dt_{p_{k}}^{(k)}}{2\pi\,\hbar}\;\;
\overleftarrow{\exp}\bigg\{\!\!-\im\,\eta_{p_{k}}^{(k)}\:\frac{t_{p_{k}}^{(k)}}{\hbar}\;
\Big(\mfrak{o}^{(k)}-\boldsymbol{\hat{\mfrak{O}^{(k)}}}(\hat{\psi}\pdag,\hat{\psi})-
\im\:\ve_{p_{k}}^{(k)}\Big)\bigg\}\; ;  \\  \lb{s1_51}
\delta\big(E-\boldsymbol{\hat{H}}(\hat{\psi}\pdag,\hat{\psi};B_{z})\,\big)  &=& \hspace*{-0.4cm}\sum_{q_{E}=\pm}\;
\lim_{|\mfrak{e}_{q_{E}}^{(E)}|\rightarrow 0} \;\hspace*{-0.2cm}\;\lim_{\mcal{T}^{(E)}\rightarrow+\infty}
\int_{0}^{\mcal{T}^{(E)}}\hspace*{-0.2cm}
\frac{d\mfrak{t}_{q_{E}}^{(E)}}{2\pi\,\hbar}\;\;
\overleftarrow{\exp}\bigg\{\!\!-\im\,\zeta_{q_{E}}^{(E)}\:\frac{\mfrak{t}_{q_{E}}^{(E)}}{\hbar}\;
\Big(E-\boldsymbol{\hat{H}}(\hat{\psi}\pdag,\hat{\psi};B_{z})-\im\:\mfrak{e}_{q_{E}}^{(E)}\Big)\bigg\} ;  \\ \lb{s1_52}
\underbrace{\delta\big(\mfrak{v}^{(\kappa)}-
\boldsymbol{\hat{\mfrak{V}}^{(\kappa)}}(\hat{\psi}\pdag,\hat{\psi})\,\big)}_{(\kappa=0,1,2)}  &=& \hspace*{-0.4cm}\sum_{q_{\kappa}=\pm}\;
\lim_{|\mfrak{e}_{q_{\kappa}}^{(\kappa)}|\rightarrow 0} \lim_{\mcal{T}^{(\kappa)}\rightarrow+\infty}
\int_{0}^{\mcal{T}^{(\kappa)}}\hspace*{-0.2cm}
\frac{d\mfrak{t}_{q_{\kappa}}^{(\kappa)}}{2\pi\,\hbar}\;\;
\overleftarrow{\exp}\bigg\{\!\!-\im\,\zeta_{q_{\kappa}}^{(\kappa)}\:\frac{\mfrak{t}_{q_{\kappa}}^{(\kappa)}}{\hbar}
\Big(\mfrak{v}^{(\kappa)}-\boldsymbol{\hat{\mfrak{V}}^{(\kappa)}}(\hat{\psi}\pdag,\hat{\psi})-\im\:
\mfrak{e}_{q_{\kappa}}^{(\kappa)}\Big)\bigg\}_{\mbox{.}}
\eeq
The notation of the generalized time variables \(t_{p_{k}}^{(k)}/\hbar\), \(\mfrak{t}_{q_{\kappa}}^{(\kappa)}/\hbar\)
of one-particle \(\boldsymbol{\hat{\mfrak{O}}^{(k)}}(\hat{\psi}\pdag,\hat{\psi})\) and
two-particle operators \(\boldsymbol{\hat{\mfrak{V}}^{(\kappa)}}(\hat{\psi}\pdag,\hat{\psi})\) follows
according to the upper indices '\(^{(k=1,2,3)}\)' and '\(^{(\kappa=0,1,2)}\)' which specify the chosen symmetry
operators as \(\boldsymbol{\hat{N}}(\hat{\psi}\pdag,\hat{\psi})\) (\ref{s1_14}),
\(\boldsymbol{\hat{L}_{z}}(\hat{\psi}\pdag,\hat{\psi})\) (\ref{s1_15}),
\(\boldsymbol{\hat{S}_{z}}(\hat{\psi}\pdag,\hat{\psi})\) (\ref{s1_16}) or as two-particle operators
\(\boldsymbol{\hat{H}}(\hat{\psi}\pdag,\hat{\psi};B_{z})\) (\ref{s1_40}),
\(\boldsymbol{\vec{L}}(\hat{\psi}\pdag,\hat{\psi})\boldsymbol{\cdot}
\boldsymbol{\vec{L}}(\hat{\psi}\pdag,\hat{\psi})\) (\ref{s1_31}) and
\(\boldsymbol{\vec{S}}(\hat{\psi}\pdag,\hat{\psi})\boldsymbol{\cdot}
\boldsymbol{\vec{S}}(\hat{\psi}\pdag,\hat{\psi})\) (\ref{s1_33}), respectively. Since one has to remove
the principal values within the Dirac identity by subtraction for the remaining delta function of the
chosen symmetry operator, we have to introduce two different, separate integration branches for the generalized
time variables \(t_{p_{k}}^{(k)}/\hbar\), \(\mfrak{t}_{q_{\kappa}}^{(\kappa)}/\hbar\) which are denoted by the
additional lower indices \(p_{k}=\pm\), \(q_{\kappa}=\pm\) with respective metric signs '\(\eta_{p_{k}=\pm}^{(k)}=\pm\)',
'\(\zeta_{q_{\kappa}=\pm}^{(\kappa)}=\pm\)'.

It remains to transform the exponential integrand within the representation of the delta functions
of operators to coherent state path integrals. In order to abbreviate notations, we specify again
the total density of states (\ref{s1_1}) in terms of one-particle and two-particle parts
with additional separation of the Hamiltonian
\(\boldsymbol{\hat{\mfrak{V}}^{(\kappa=0=E)}}(\hat{\psi}\pdag,\hat{\psi})\) for further reference and transformations
\footnote{The index '\(\kappa=E\)' for the Hamiltonian case of particular consideration is used in parallel with the
equivalent index number '\(\kappa=0\)' apart from the other two indices '\(\kappa=1,2\)' in the class of operators
\(\boldsymbol{\hat{\mfrak{V}}^{(\kappa)}}(\hat{\psi}\pdag,\hat{\psi})\)
which specify our investigations for the absolute values of orbital '\(\kappa=1\)' and spin '\(\kappa=2\)'
angular momentum with their two-particle contents.}
\beq\lb{s1_53}\hspace*{-0.1cm}
\varrho(E,\mfrak{v}^{(\kappa=1,2)};\mfrak{o}^{(k=1,2,3)}) &\hspace*{-0.3cm}=&\hspace*{-0.3cm}
\mbox{Tr}\bigg[\bigg(\prod_{k=1}^{3}\delta\Big(\mfrak{o}^{(k)}-
\boldsymbol{\hat{\mfrak{O}}^{(k)}}(\hat{\psi}\pdag,\hat{\psi})\Big)\bigg)\;
\bigg(\prod_{\kappa=1,2}\delta\Big(\mfrak{v}^{(\kappa)}-
\boldsymbol{\hat{\mfrak{V}}^{(\kappa)}}(\hat{\psi}\pdag,\hat{\psi})\Big)\bigg)\;
\delta\Big(E-\boldsymbol{\hat{H}}(\hat{\psi}\pdag,\hat{\psi};B_{z})\Big)\bigg]  \\  \no &\hspace*{-0.3cm}=&\hspace*{-0.3cm}
\varrho(\mfrak{v}^{(k=0,1,2)};\mfrak{o}^{(k=1,2,3)}) =
\mbox{Tr}\bigg[\bigg(\prod_{k=1}^{3}\delta\Big(\mfrak{o}^{(k)}-
\boldsymbol{\hat{\mfrak{O}}^{(k)}}(\hat{\psi}\pdag,\hat{\psi})\Big)\bigg)\;
\bigg(\prod_{\kappa=0}^{2}\delta\Big(\mfrak{v}^{(\kappa)}-
\boldsymbol{\hat{\mfrak{V}}^{(\kappa)}}(\hat{\psi}\pdag,\hat{\psi})\Big)\bigg)\bigg] \;.
\eeq
According to Refs. \cite{precisecoh1}, we straightforwardly
transform the integral representations (\ref{s1_50}-\ref{s1_52})
of delta functions within the trace to coherent state path integrals.
The Grassmann fields \(\chi_{\vec{x},s}\), \(\chi_{\vec{x},s}^{*}\) (\ref{s1_54},\ref{s1_55})
with their anti-commuting integrations (\ref{s1_56}) replace the fermionic annihilation and creation operators
\(\hat{\psi}_{\vec{x},s}\), \(\hat{\psi}_{\vec{x},s}\pdag\) as eigenvalues of coherent states and are used
to split the total product of second quantized delta functions to separate factors. They therefore act as source
fields for the separate factors of delta functions of one-particle and two-particle operators. In consequence
we list the combined definitions of the Grassmann
fields \(\chi_{\vec{x},s}\), \(\chi_{\vec{x},s}^{*}\) (\ref{s1_54}-\ref{s1_57})
and \(\xi_{\vec{x},s}\), \(\xi_{\vec{x},s}^{*}\) (\ref{s1_58}-\ref{s1_60})
(for one-particle and two-particle components) with their
overcompleteness relations (\ref{s1_57},\ref{s1_60}) together with
the factoring of delta functions (\ref{s1_61}) within the
coherent state trace representation of the density of states \(\varrho(E,\mfrak{v}^{(\kappa=1,2)};\mfrak{o}^{(k=1,2,3)})\)
\beq\lb{s1_54}
\hat{\psi}_{\vec{x},s}\pdag\big|0\big\rangle &=&\sqrt{\mcal{N}_{x}}\,\big|\ldots\,,n_{\vec{x},s}=1,\,\ldots\big\rangle\;;
\;\;\;\;\big\langle0\big|\hat{\psi}_{\vec{x},s}=\big\langle\ldots\,,
n_{\vec{x},s}=1,\,\ldots\big|\,\sqrt{\mcal{N}_{x}}\;\;;\\  \lb{s1_55}
\big|\chi\big\rangle &=& \exp\Big\{\sum_{\vec{x},s}\chi_{\vec{x},s}\;\hat{\psi}_{\vec{x},s}\pdag\Big\}\,
\big|0\big\rangle \;\;\;;\;\;\;
\hat{\psi}_{\vec{x},s}\:\big|\chi\big\rangle =\chi_{\vec{x},s}\:\big|\chi\big\rangle\;\;\;;\;\;\;
\big\langle\chi\big|\:\hat{\psi}_{\vec{x},s}\pdag=\big\langle\chi\big|\:\chi_{\vec{x},s}^{*} \;\;\;;  \\  \lb{s1_56}
\int d\chi_{\vec{x},s}\;\chi_{\vec{x}\ppr,s\ppr} &=& \mcal{N}_{x}\;\delta_{\vec{x},\vec{x}\ppr}\;\delta_{ss\ppr}\;\;\;;\;\;\;
\int d\chi_{\vec{x},s}^{*}\;\chi_{\vec{x}\ppr,s\ppr}^{*} = \mcal{N}_{x}\;
\delta_{\vec{x},\vec{x}\ppr}\;\delta_{ss\ppr}\;\;\;; \\  \lb{s1_57}
1 &=& \int d[\chi^{*},\chi]\;\;\exp\Big\{-\sum_{\vec{x},s}\chi_{\vec{x},s}^{*}\;\chi_{\vec{x},s}\Big\}\;\;
\big|\chi\big\rangle\:\big\langle\chi\big| \;\;\;; \;\;\;
d[\chi^{*},\chi]  = \prod_{\{\vec{x},s\}}
\frac{d\chi_{\vec{x},s}^{*}\;d\chi_{\vec{x},s}}{\mcal{N}_{x}} \;\;\;; \\  \lb{s1_58}
\big|\xi\big\rangle &=& \exp\Big\{\sum_{\vec{x},s}\xi_{\vec{x},s}\;\hat{\psi}_{\vec{x},s}\pdag\Big\}\,
\big|0\big\rangle \;\;\;;\;\;\;
\hat{\psi}_{\vec{x},s}\:\big|\xi\big\rangle = \xi_{\vec{x},s}\:\big|\xi\big\rangle\;\;\;;\;\;\;
\big\langle\xi\big|\:\hat{\psi}_{\vec{x},s}\pdag=\big\langle\chi\big|\:\xi_{\vec{x},s}^{*} \;\;\;;  \\  \lb{s1_59}
\int d\xi_{\vec{x},s}\;\xi_{\vec{x}\ppr,s\ppr} &=& \mcal{N}_{x}\;\delta_{\vec{x},\vec{x}\ppr}\;\delta_{ss\ppr}\;\;\;;\;\;\;
\int d\xi_{\vec{x},s}^{*}\;\xi_{\vec{x}\ppr,s\ppr}^{*} = \mcal{N}_{x}\;
\delta_{\vec{x},\vec{x}\ppr}\;\delta_{ss\ppr}\;\;\;; \\  \lb{s1_60}
1 &=& \int d[\xi^{*},\xi]\;\;\exp\Big\{-\sum_{\vec{x},s}\xi_{\vec{x},s}^{*}\;\xi_{\vec{x},s}\Big\}\;\;
\big|\xi\big\rangle\:\big\langle\xi\big| \;\;\;;\;\;\;
d[\xi^{*},\xi]  = \prod_{\{\vec{x},s\}}\frac{d\xi_{\vec{x},s}^{*}\;d\xi_{\vec{x},s}}{\mcal{N}_{x}} \;;  \\  \lb{s1_61}
\varrho(\mfrak{v}^{(\kappa=0,1,2)};\mfrak{o}^{(k=1,2,3)}) \hspace*{-0.3cm}&=&
\hspace*{-0.3cm}\bigg(\prod_{k=1}^{3}\prod_{\kappa=0}^{2}
\int d[\chi_{\vec{x},s}^{(k)*},d\chi_{\vec{x},s}^{(k)}]\;d[\xi_{\vec{x},s}^{(\kappa)*},d\xi_{\vec{x},s}^{(\kappa)}]\;
\exp\Big\{-\sum_{\vec{x},s}\Big(\chi_{\vec{x},s}^{(k)*}\;\chi_{\vec{x},s}^{(k)}+
\xi_{\vec{x},s}^{(\kappa)*}\;\xi_{\vec{x},s}^{(\kappa)}\Big)\Big\}\bigg)\;\times \\ \no &\times&
\overbrace{\big\langle\chi^{(k=4)}\big|}^{\langle-\xi^{(\kappa=0)}|}
\delta\big(\mfrak{o}^{(k=3)}-\boldsymbol{\hat{\mfrak{O}}^{(k=3)}}(\hat{\psi}\pdag,\hat{\psi})\,\big)\big|\chi^{(k=3)}\big\rangle\;
\big\langle\chi^{(k=3)}\big|\delta\big(\mfrak{o}^{(k=2)}-\boldsymbol{\hat{\mfrak{O}}^{(k=2)}}(\hat{\psi}\pdag,\hat{\psi})\,\big)
\big|\chi^{(k=2)}\big\rangle\;\times  \\ \no &\times&
\big\langle\chi^{(k=2)}\big|
\delta\big(\mfrak{o}^{(k=1)}-\boldsymbol{\hat{\mfrak{O}}^{(k=1)}}(\hat{\psi}\pdag,\hat{\psi})\,\big)
\big|\chi^{(k=1)}\big\rangle\;\times \\ \no &\times&
\big\langle\xi^{(\kappa=3)}\big|\delta\big(\mfrak{v}^{(\kappa=2)}-\boldsymbol{\hat{\mfrak{V}}^{(\kappa=2)}}(\hat{\psi}\pdag,\hat{\psi})\,\big)
\big|\xi^{(\kappa=2)}\big\rangle\;
\big\langle\xi^{(\kappa=2)}\big|
\delta\big(\mfrak{v}^{(\kappa=1)}-\boldsymbol{\hat{\mfrak{V}}^{(\kappa=1)}}(\hat{\psi}\pdag,\hat{\psi})\,\big)\big|\xi^{(\kappa=1)}\big\rangle\;\times  \\  \no &\times&
\big\langle\xi^{(\kappa=1)}\big|\delta\big(E-\boldsymbol{\hat{H}}(\hat{\psi}\pdag,\hat{\psi};B_{z})\,\big)
\big|\xi^{(\kappa=0)}\big\rangle\;; \\ \lb{s1_62}  &&
\big|\chi^{(k=4)}\rangle=|-\xi^{(\kappa=0)}\rangle\;;\;\;\chi_{\vec{x},s}^{(k=4)}=-\xi_{\vec{x},s}^{(\kappa=0)}\;;\;\;
\big|\chi^{(k=1)}\big\rangle=\big|\xi^{(\kappa=3)}\big\rangle\;;\;\;\chi_{\vec{x},s}^{(k=1)}=\xi_{\vec{x},s}^{(\kappa=3)} \;.
\eeq
Since we regard a purely fermionic system of electrons, the Grassmann fields have to fulfill anti-periodic
boundary conditions (\ref{s1_62}) following from the overcompleteness relation for the trace operation.

\subsection{The remaining field theory of coset matrices derived in subsequent sections}\lb{s13}

Since the remaining field theory of coset matrices follows after several transformations with involved appearance,
we briefly describe the result in advance. We do not omit the precise generalized time steps following from
the coherent state representation of exponentials from the Dirac identity for delta functions in order to
testify the exactness of coherent state path integrals with exact non-hermitian parts
\(\psi_{\vec{y}\ppr}^{*}(\vartheta_{q_{\kappa}}^{(\kappa)}+\Delta\vartheta^{(\kappa)})\ldots
\psi_{\vec{y}}(\vartheta_{q_{\kappa}}^{(\kappa)})\) instead of
approximating, more appealing hermitian kinds
\(\psi_{\vec{y}\ppr}^{*}(\vartheta_{q_{\kappa}}^{(\kappa)})\ldots\psi_{\vec{y}}(\vartheta_{q_{\kappa}}^{(\kappa)})\). This exactness of
coherent state path integrals also holds for the anomalous doubling of Fermi fields with further precise
time steps within a coset decomposition for the self-energies after a HST transformation. We even keep the
precise, exact generalized time steps constrained by the normal ordering of the second quantized field operators
in the final form of the field theory with remaining coset degrees of freedom which is listed in subsequent equation
(\ref{s1_63}). Since one has to split the evolution generators (\ref{s1_50}-\ref{s1_52}) into infinitesimal,
generalized time development steps, one has to regard additional time parameters \(\tau_{p_{k}}^{(k)}\) and
\(\vartheta_{q_{\kappa}}^{(\kappa)}\) within the exponentials of corresponding one-particle and two-particle parts
\beq \lb{s1_63}
\lefteqn{\varrho(\mfrak{v}^{(\kappa=0,1,2)};\mfrak{o}^{(k=1,2,3)}) =
\prod_{k=1}^{3}\sum_{p_{k}=\pm}\;
\lim_{|\ve_{p_{k}}^{(k)}|\rightarrow0}\;\lim_{T^{(k)}\rightarrow+\infty}
\int_{0}^{T^{(k)}}\frac{dt_{p_{k}}^{(k)}}{2\pi\,\hbar}\;\times \;
\exp\Big\{-\frac{\im}{\hbar}\:t_{p_{k}}^{(k)}\:
\eta_{p_{k}}^{(k)}\:\big(\mfrak{o}^{(k)}-\im\:\ve_{p_{k}}^{(k)}\big)\Big\}\;\times   } \\ \no &\times&
\prod_{\kappa=0}^{2}\sum_{q_{\kappa}=\pm}\;
\lim_{|\mfrak{e}_{q_{\kappa}}^{(\kappa)}|\rightarrow 0} \;\lim_{\mcal{T}^{(\kappa)}\rightarrow+\infty}
\int_{0}^{\mcal{T}^{(\kappa)}}\frac{d\mfrak{t}_{q_{\kappa}}^{(\kappa)}}{2\pi\,\hbar}\;
\exp\Big\{-\im\,\zeta_{q_{\kappa}}^{(\kappa)}\:\frac{\mfrak{t}_{q_{\kappa}}^{(\kappa)}}{\hbar}\:
\big(\mfrak{v}^{(\kappa)}-\im\,\mfrak{e}_{q_{\kappa}}^{(\kappa)}\big)\Big\}\;\times
\\ \no &\times&\int d[\sigma_{D}^{(i)}(\vec{x},\mfrak{t}_{q_{\kappa}}^{(\kappa)})]\;
\exp\bigg\{-\frac{\im}{2\hbar}\int_{0}^{\mfrak{t}_{q_{\kappa}}^{(\kappa)}-\Delta\vartheta^{(\kappa)}} \hspace*{-0.6cm}
d\vartheta_{q_{\kappa}}^{(\kappa)}\;\zeta_{q_{\kappa}}^{(\kappa)}\sum_{i=i_{\kappa}}^{j_{\kappa}}\sum_{\vec{x},\vec{x}\ppr}
\sigma_{D}^{(i)}(\vec{x}\ppr,\vartheta_{q_{\kappa}}^{(\kappa)})\;\hat{V}_{|\vec{x}\ppr-\vec{x}|}^{(\kappa);\boldsymbol{-1}}
\;\sigma_{D}^{(i)}(\vec{x},\vartheta_{q_{\kappa}}^{(\kappa)})\bigg\} \\  \no &\times&
\int d[\hat{T}_{\vec{y}\ppr;\vec{y}_{1}}^{-1}\!(\vartheta_{q_{\kappa}}^{(\kappa)})\;
d\!\hat{T}_{\vec{y}_{1};\vec{y}}\!(\vartheta_{q_{\kappa}}^{(\kappa)})]\;\;
\boldsymbol{\Delta^{(\kappa)}}\Big(\hat{T}_{\vec{y}\ppr;\vec{y}}^{-1}\!(\vartheta_{q_{\kappa}}^{(\kappa)});
\hat{T}_{\vec{y}\ppr;\vec{y}}\!(\vartheta_{q_{\kappa}}^{(\kappa)})\Big)\;\times\;
\Big\{\mbox{DET}\Big(\mathring{\mscr{M}}_{\vec{y}\ppr;\vec{y}}^{ba}(\vartheta_{q_{\kappa}}^{(\kappa)\bprime}\boldsymbol{|}
\vartheta_{q_{\kappa}}^{(\kappa)})\Big)\Big\}^{1/2}\times \\ \no &\times&
\exp\Bigg\{\frac{1}{6}\trxs\ln\Bigg(\hat{1}\:\delta_{\vec{y}_{7};\vec{y}_{1}}+
\bigg(\Big(\exp\Big\{\frac{\im}{\hbar}\:\eta_{p_{3}}^{(3)}\;t_{p_{3}}^{(3)}\;\frac{1}{\mcal{N}_{x}}\,
\hat{\mfrak{O}}_{\vec{y}_{7}\ppr;\vec{y}_{6}\ppr}^{(3)}\Big\}\Big)_{\vec{y}_{7};\vec{y}_{6}}\;
\Big(\exp\Big\{\frac{\im}{\hbar}\:\eta_{p_{2}}^{(2)}\;t_{p_{2}}^{(2)}\;\frac{1}{\mcal{N}_{x}}\,
\hat{\mfrak{O}}_{\vec{y}_{6}\ppr;\vec{y}_{5}\ppr}^{(2)}\Big\}\Big)_{\vec{y}_{6};\vec{y}_{5}}\;\times \\ \no &\times&\hspace*{-0.2cm}
\Big(\exp\Big\{\frac{\im}{\hbar}\:\eta_{p_{1}}^{(1)}\;t_{p_{1}}^{(1)}\;\frac{1}{\mcal{N}_{x}}\,
\hat{\mfrak{O}}_{\vec{y}_{5}\ppr;\vec{y}_{4}\ppr}^{(1)}\Big\}\Big)_{\vec{y}_{5};\vec{y}_{4}}
\;\mathring{\mscr{M}}_{\vec{y}_{4};\vec{y}_{3}}^{-1;ba}
(\mfrak{t}_{q_{2}}^{(2)}\boldsymbol{|}-\Delta\vartheta^{(2)})\;
\;\mathring{\mscr{M}}_{\vec{y}_{3};\vec{y}_{2}}^{-1;ba}
(\mfrak{t}_{q_{1}}^{(1)}\boldsymbol{|}-\Delta\vartheta^{(1)})\;
\mathring{\mscr{M}}_{\vec{y}_{2};\vec{y}_{1}}^{-1;ba}
(\mfrak{t}_{q_{0}}^{(0)}\boldsymbol{|}-\Delta\vartheta^{(0)})\bigg)_{\vec{y}_{7};\vec{y}_{1}}^{b=a\equiv1}
\Bigg)\Bigg\}_{\mbox{.}}
\eeq
Apart from the scalar self-energy densities \(\sigma_{D}^{(i)}(\vec{x},\vartheta_{q_{\kappa}}^{(\kappa)})\),
we have the coset matrices \(\hat{T}_{\vec{y};\vec{y}\ppr}\!(\vartheta_{q_{\kappa}}^{(\kappa)})\) with anti-symmetric
generator part \(\hat{T}_{\vec{y};\vec{y}\ppr}^{ab}(\vartheta_{q_{\kappa}}^{(\kappa)})=
(\,\exp\{-\hat{Y}_{\vec{y}_{1};\vec{y}_{2}}^{a_{1}\neq b_{1}}(\vartheta_{q_{\kappa}}^{(\kappa)})\}\,)_{
\vec{y};\vec{y}\ppr}^{ab}\) in the off-diagonal blocks. The matrix
\(\mathring{\mscr{M}}_{\vec{y}\ppr;\vec{y}}^{ba}(\vartheta_{q_{\kappa}}^{(\kappa)\bprime}\boldsymbol{|}
\vartheta_{q_{\kappa}}^{(\kappa)})\) in the functional determinant contains a gradient operator of the
coset matrices and the generalized Hamiltonians with the self-energy densities. Furthermore,
we have propagator parts of all six symmetry operators in the last two lines of (\ref{s1_63})
composed of the three one-particle operators (particle number, z-components of orbital and
spin angular momentum) and the three 'interacting' two-particle parts with the total Hamilton
operator and the absolute values of the orbital and spin angular momentum. The various details,
how to attain the given structure with further definitions and precise time steps, are specified in the remainder
of this article.

\section{Transformation of the delta functions composed of one-particle operators}\lb{s2}

\subsection{Field theory for delta functions with second quantized operators} \lb{s21}

Although the case of one-particle operators \(\boldsymbol{\hat{\mfrak{O}}^{(k)}}(\hat{\psi}\pdag,\hat{\psi})\)
can be directly computed from the exponentials of bilinear, density related, second quantized Fermi operators,
we outline the calculation of the one-particle operators within Grassmann-valued matrix elements of delta functions,
separated from the total density of state relation (\ref{s1_61},\ref{s1_62}). Since one-particle, density related
operators do not allow for anomalous terms as \(\langle\hat{\psi}_{\vec{x},s}\:\hat{\psi}_{\vec{x}\ppr,s\ppr}\rangle\),
\(\langle\hat{\psi}_{\vec{x},s}\pdag\:\hat{\psi}_{\vec{x}\ppr,s\ppr}\pdag\rangle\), the calculation simplifies
considerably. We define the anti-commuting fields (\ref{s2_1},\ref{s2_2}) for the Fermi operators
in eqs. (\ref{s2_1}-\ref{s2_7}) with the integrations (\ref{s2_3},\ref{s2_4}) and overcompleteness relation (\ref{s2_5})
to be inserted at the various time development steps of exponential operators. Moreover, we abbreviate the corresponding,
Grassmann-valued, integration measures, which frequently occur in the transformation to coherent state path integrals,
by relations
(\ref{s2_6},\ref{s2_7}) \footnote{The various, dimensionless Kronecker deltas of generalized 'time' parameters
\(\tau_{p_{k}}^{(k)}\), \(\tau_{p_{k}}^{(k)\bprime}\) are symbolized by the notation
'\(\delta\big(\tau_{p_{k}}^{(k)}\boldsymbol{\big|}\tau_{p_{k}}^{(k)\bprime}\big)\)' and correspond to the analogous delta function of time \(\simeq\delta(\tau_{p_{k}}^{(k)}-\tau_{p_{k}}^{(k)\bprime})\propto\;\big(\Delta\tau^{(k)}\big)^{-1}\;
\delta\raisebox{3pt}{$_{\tau_{p_{k}}^{(k)}\boldsymbol{,}\tau_{p_{k}}^{(k)\bprime}}$}\) without the inverted
'time' interval \(\big(\Delta\tau^{(k)}\big)^{-1}\).}
\beq  \lb{s2_1}
\big|\psi\big(\tau_{p_{k}}^{(k)}\big)\big\rangle &=& \exp\Big\{\sum_{\vec{x},s}
\psi_{\vec{x},s}\big(\tau_{p_{k}}^{(k)}\big)\;\hat{\psi}_{\vec{x},s}\pdag\Big\}\,\big|0\big\rangle \;; \\ \lb{s2_2}
\hat{\psi}_{\vec{x},s}\:\big|\psi\big(\tau_{p_{k}}^{(k)}\big)\big\rangle  =
\psi_{\vec{x},s}\big(\tau_{p_{k}}^{(k)}\big)\: \big|\psi\big(\tau_{p_{k}}^{(k)}\big)\big\rangle   &;&
\big\langle\psi\big(\tau_{p_{k}}^{(k)}\big)\big|\:\hat{\psi}_{\vec{x},s}\pdag=
\big\langle\psi\big(\tau_{p_{k}}^{(k)}\big)\big|\:\psi_{\vec{x},s}^{*}\big(\tau_{p_{k}}^{(k)}\big) \;;  \\  \lb{s2_3}
\int d\psi_{\vec{x},s}(\tau_{p_{k}}^{(k)})\;\;\psi_{\vec{x}\ppr,s\ppr}(\tau_{p_{k}}^{(k)\bprime})
&=& \mcal{N}_{x}\;\delta_{\vec{x},\vec{x}\ppr}\;\delta_{ss\ppr}\;
\delta\big(\tau_{p_{k}}^{(k)}\big|\tau_{p_{k}}^{(k)\bprime}\big)  \;;  \\ \lb{s2_4}
\int d\psi_{\vec{x},s}^{*}(\tau_{p_{k}}^{(k)})\;\;\psi_{\vec{x}\ppr,s\ppr}^{*}(\tau_{p_{k}}^{(k)\bprime})
&=& \mcal{N}_{x}\;\delta_{\vec{x},\vec{x}\ppr}\;\delta_{ss\ppr}\;
\delta\big(\tau_{p_{k}}^{(k)}\big|\tau_{p_{k}}^{(k)\bprime}\big)\;; \\ \lb{s2_5}
1 = \int d[\psi^{*}(\tau_{p_{k}}^{(k)}),\psi(\tau_{p_{k}}^{(k)})]  &&\hspace*{-0.6cm}
\exp\Big\{-\sum_{\vec{x},s}\psi_{\vec{x},s}^{*}(\tau_{p_{k}}^{(k)})\;\:
\psi_{\vec{x},s}(\tau_{p_{k}}^{(k)})\Big\}\;\;
\big|\psi\big(\tau_{p_{k}}^{(k)}\big)\big\rangle\big\langle\psi\big(\tau_{p_{k}}^{(k)}\big)\big|\;;  \\  \lb{s2_6}
d[\psi^{*}(\tau_{p_{k}}^{(k)}),\psi(\tau_{p_{k}}^{(k)})] &=&\prod_{\{\vec{x},s\}}
\frac{d\psi_{\vec{x},s}^{*}(\tau_{p_{k}}^{(k)})\; \;d\psi_{\vec{x},s}(\tau_{p_{k}}^{(k)})}{\mcal{N}_{x}} \;; \\  \lb{s2_7}
d[\psi^{*}(t_{p_{k}}^{(k)}),\psi(t_{p_{k}}^{(k)})] &=&
\prod_{0\leq\{\tau_{p_{k}}^{(k)}\}\leq t_{p_{k}}^{(k)}} \;\;\prod_{\{\vec{x},s\}}
\frac{d\psi_{\vec{x},s}^{*}(\tau_{p_{k}}^{(k)})\; \;d\psi_{\vec{x},s}(\tau_{p_{k}}^{(k)})}{\mcal{N}_{x}} \;.
\eeq
After we have introduced the overcompleteness relation (\ref{s2_5}) at the various, sequentially ordered, exponential
'time' development steps of relations (\ref{s1_50},\ref{s1_61}),
we finally achieve the coherent state path integral (\ref{s2_8}) with the
fermionic source fields \(\chi_{\vec{x},s}^{(k)}\), \(\chi_{\vec{x},s}^{(k+1)*}\) which couple
to the fields \(\psi_{\vec{x},s}^{*}(\tau_{p_{k}}^{(k)}\!=\!0)\) and
\(\psi_{\vec{x},s}(\tau_{p_{k}}^{(k)}\!=\!t_{p_{k}}^{(k)})\)
within the action \(\mscr{A}^{(k)}(\chi^{(k+1)*},\chi^{(k)})\).
According to the wanted preciseness of our exact discretization,
we have additionally to incorporate a remaining Gaussian field part for zero time labels into the source action
\(\wt{\mscr{A}}^{(k)}(\chi^{(k+1)*},\chi^{(k)})\) (\ref{s2_9})
\beq \lb{s2_8}
\lefteqn{\hspace*{-2.8cm}\big\langle\chi^{(k+1)}\big|\delta(\mfrak{o}^{(k)}-
\boldsymbol{\hat{\mfrak{O}}^{(k)}}(\hat{\psi}\pdag,\hat{\psi})\,\big)\big|\chi^{(k)}\big\rangle
= \sum_{p_{k}=\pm}\;\lim_{|\ve_{p_{k}}^{(k)}|\rightarrow0}\;\lim_{T^{(k)}\rightarrow+\infty}
\int_{0}^{T^{(k)}}\frac{dt_{p_{k}}^{(k)}}{2\pi\,\hbar}\;\;d[\psi^{*}(t_{p_{k}}^{(k)}),\psi(t_{p_{k}}^{(k)})]\;\times}
\\ \no &\times& \exp\bigg\{-\frac{\im}{\hbar}\int_{0}^{t_{p_{k}}^{(k)}-\Delta\tau^{(k)}}d\!\tau_{p_{k}}^{(k)}\;
\bigg(\sum_{\vec{y}}\psi_{\vec{y}}^{*}(\tau_{p_{k}}^{(k)}\!+\!\Delta\tau^{(k)})\;
(-\im\hbar)\:\frac{\psi_{\vec{y}}(\tau_{p_{k}}^{(k)}\!+\!\Delta\tau^{(k)})-
\psi_{\vec{y}}(\tau_{p_{k}}^{(k)})}{\Delta\tau^{(k)}} + \\ \no &-& \sum_{\vec{y};\vec{y}\ppr}
\psi_{\vec{y}\ppr}^{*}(\tau_{p_{k}}^{(k)}\!+\!\Delta\tau^{(k)})\;\eta_{p_{k}}^{(k)}\;
\hat{\mfrak{O}}_{\vec{y}\ppr;\vec{y}}^{(k)}\;\psi_{\vec{y}}(\tau_{p_{k}}^{(k)})\bigg)\bigg\}
\;\times\;   \\ \no &\times& \exp\Big\{\wt{\mscr{A}}^{(k)}\big(\chi^{(k+1)*},\chi^{(k)}\big)\Big\} \;\times\;
\exp\Big\{-\frac{\im}{\hbar}\:t_{p_{k}}^{(k)}\:\eta_{p_{k}}^{(k)}\:
\big(\mfrak{o}^{(k)}-\im\:\ve_{p_{k}}^{(k)}\big)\Big\}\;;  \\ \lb{s2_9}
\exp\Big\{\wt{\mscr{A}}^{(k)}\big(\chi^{(k+1)*},\chi^{(k)}\big)\Big\} \hspace*{-0.3cm}&=&\hspace*{-0.3cm}
\exp\!\bigg\{\!\mscr{A}^{(k)}\big(\chi^{(k+1)*},\chi^{(k)}\big)-
\sum_{\vec{y}}\psi_{\vec{y}}^{*}(\tau_{p_{k}}^{(k)}\!=\!0)\;\;\psi_{\vec{y}}(\tau_{p_{k}}^{(k)}\!=\!0)\bigg\}  \;; \\ \no
\mscr{A}^{(k)}\big(\chi^{(k+1)*},\chi^{(k)}\big)\hspace*{-0.3cm}&=&\hspace*{-0.3cm}
\sum_{\vec{y}}\Big(\psi_{\vec{y}}^{*}(\tau_{p_{k}}^{(k)}\!=\!0)\;\;\chi_{\vec{y}}^{(k)}+
\chi_{\vec{y}}^{(k+1)*}\;\;\psi_{\vec{y}}(\tau_{p_{k}}^{(k)}\!=\!t_{p_{k}}^{(k)})\Big)\;; \\  \lb{s2_10}
\psi_{\vec{x},s}(\tau_{p_{k}}^{(k)}) &=&\psi_{\vec{x},s}(\tau_{p_{k}}^{(k)}\!=\!0)\,,\,
\psi_{\vec{x},s}(\tau_{p_{k}}^{(k)}\!=\!\Delta\!\tau^{(k)})\,,\, \ldots\,,\,
\psi_{\vec{x},s}(\tau_{p_{k}}^{(k)}\!=\!t_{p_{k}}^{(k)})\;; \\  \lb{s2_11}
N_{p_{k}}^{(k)}&=&t_{p_{k}}^{(k)}/\Delta\!\tau^{(k)}\;.
\eeq
Each summand with a generalized time integration in (\ref{s2_8}) consists of fermionic fields with generalized
time points between '\(\tau_{p_{k}}^{(k)}=0\)' and '\(\tau_{p_{k}}^{(k)}=t_{p_{k}}^{(k)}\)' of discrete
intervals \(\Delta\!\tau^{(k)}\) for the one-particle
matrix elements \(\hat{\mfrak{O}}_{\vec{x}\ppr,s\ppr;\vec{x},s}^{(k)}\) (\ref{s1_13}-\ref{s1_16}).
Since the matrix elements of the delta function in (\ref{s2_8}) are only composed of exponentials with
bilinear Grassmann fields, the latter can be integrated out to obtain the determinant of the characteristic one-particle
matrix \(\hat{M}_{\vec{x}\ppr,s\ppr;\vec{x},s}^{(k,p_{k})}(\tau_{p_{k}}^{(k)\bprime}\boldsymbol{|}\tau_{p_{k}}^{(k)})\)
(\ref{s2_13}) with a coupling of the source fields \(\chi_{\vec{x}\ppr,s\ppr}^{(k+1)*}\;\ldots\;\chi_{\vec{x},s}^{(k)}\)
to the corresponding propagator. Since the relevant matrix (\ref{s2_13}) with one-particle part
\(\hat{\mfrak{O}}_{\vec{x}\ppr,s\ppr;\vec{x},s}^{(k)}\) only has a non-vanishing main-diagonal and a nonzero, first lower
sub-diagonal in the 'time' indices, the determinant entirely reduces to the value one so that we are only left with the
propagator (\ref{s2_17}) for the source fields \(\chi_{\vec{x}\ppr,s\ppr}^{(k+1)*}\;\ldots\;\chi_{\vec{x},s}^{(k)}\)
\beq\lb{s2_12}
\lefteqn{\big\langle\chi^{(k+1)}\big|\delta\big(\mfrak{o}^{(k)}-
\boldsymbol{\hat{\mfrak{O}}^{(k)}}(\hat{\psi}\pdag,\hat{\psi})\,\big)\big|\chi^{(k)}\big\rangle =} \\ \no
&=& \sum_{p_{k}=\pm}\;\lim_{|\ve_{p_{k}}^{(k)}|\rightarrow0}\;\lim_{T^{(k)}\rightarrow+\infty}
\int_{0}^{T^{(k)}}\frac{dt_{p_{k}}^{(k)}}{2\pi\,\hbar}\;\times \;
\mbox{det}\Big[\hat{M}_{\vec{x}\ppr,s\ppr;\vec{x},s}^{(k,p_{k})}\Big(\tau_{p_{k}}^{(k)\boldsymbol{\prime}}\!=\!0,\,\ldots\,,
t_{p_{k}}^{(k)}\boldsymbol{\big|}\tau_{p_{k}}^{(k)}\!=\!0,\,\ldots\,,t_{p_{k}}^{(k)}\Big)\Big]\;\times \\ \no &\times&
\exp\bigg\{\sum_{\vec{x}\ppr,s\ppr;\vec{x},s}\mcal{N}_{x}\:\chi_{\vec{x}\ppr,s\ppr}^{(k+1)*}\;\;
\hat{M}_{\vec{x}\ppr,s\ppr\boldsymbol{;}\vec{x},s}^{(k,p_{k});\boldsymbol{-1}}
\big(\tau_{p_{k}}^{(k)\boldsymbol{\prime}}\!=\!t_{p_{k}}\boldsymbol{\big|}\tau_{p_{k}}^{(k)}\!=\!0\big)\;\;
\chi_{\vec{x},s}^{(k)}\bigg\}\;
\times\;\exp\Big\{-\frac{\im}{\hbar}\:t_{p_{k}}^{(k)}\:
\eta_{p_{k}}^{(k)}\:\big(\mfrak{o}^{(k)}-\im\:\ve_{p_{k}}^{(k)}\big)\Big\}\;;
\eeq
\beq\lb{s2_13}
\hat{M}_{\vec{x}\ppr,s\ppr;\vec{x},s}^{(k,p_{k})}\big(\tau_{p_{k}}^{(k)\boldsymbol{\prime}}\boldsymbol{|}\tau_{p_{k}}^{(k)}\big)
\hspace*{-0.3cm}&=&\hspace*{-0.3cm} \delta_{\vec{x}\ppr,\vec{x}}\,\delta_{s\ppr s}
\Big[\delta(\tau_{p_{k}}^{(k)\boldsymbol{\prime}}\boldsymbol{|}\tau_{p_{k}}^{(k)})-
\delta(\tau_{p_{k}}^{(k)\boldsymbol{\prime}}\boldsymbol{|}\tau_{p_{k}}^{(k)}\!+\!\Delta\!\tau^{(k)})\Big] -
\eta_{p_{k}}^{(k)}\frac{\im}{\hbar}\frac{\Delta\!\tau^{(k)}}{\mcal{N}_{x}}\,
\hat{\mfrak{O}}^{(k)}_{\vec{x}\ppr,s\ppr;\vec{x},s}\,
\delta(\tau_{p_{k}}^{(k)\boldsymbol{\prime}}\boldsymbol{|}\tau_{p_{k}}^{(k)}\!+\!\Delta\!\tau^{(k)})_{\mbox{;}}  \\ \lb{s2_14}
\tau_{p_{k}}^{(k)\boldsymbol{\prime}}\,,\,\tau_{p_{k}}^{(k)} &=&
0\,,\,\Delta\!\tau^{(k)}\,,\,\ldots\,,\,t_{p_{k}}^{(k)}\;; \\  \lb{s2_15}
t_{p_{k}}^{(k)} &=& \Delta\!\tau^{(k)}\,,\,\ldots\,,\,T^{(k)}\;;  \\  \lb{s2_16}
1&\equiv&\mbox{det}\Big[\hat{M}_{\vec{x}\ppr,s\ppr;\vec{x},s}^{(k,p_{k})}\Big(\tau_{p_{k}}^{(k)\boldsymbol{\prime}}\!=\!0,\,\ldots\,,
t_{p_{k}}^{(k)}\boldsymbol{\big|}\tau_{p_{k}}^{(k)}\!=\!0,\,\ldots\,,t_{p_{k}}^{(k)}\Big)\Big] \;;  \\ \lb{s2_17}
\lefteqn{\hspace*{-3.06cm}\hat{M}_{\vec{x}\ppr,s\ppr\boldsymbol{;}\vec{x},s}^{(k,p_{k});\boldsymbol{-1}}
\big(\tau_{p_{k}}^{(k)\boldsymbol{\prime}}\!=\!t_{p_{k}}\boldsymbol{\big|}\tau_{p_{k}}^{(k)}\!=\!0\big) =
\bigg(\delta_{\vec{x}_{2}\ppr,\vec{x}_{1}\ppr}\;\delta_{s_{2}\ppr s_{1}\ppr}+\frac{\im}{\hbar}\:
\eta_{p_{k}}^{(k)}\;\frac{t_{p_{k}}^{(k)}}{N_{p_{k}}^{(k)}}\;\frac{1}{\mcal{N}_{x}}\,
\hat{\mfrak{O}}_{\vec{x}_{2}\ppr,s_{2}\ppr;\vec{x}_{1}\ppr,s_{1}\ppr}^{(k)}\bigg)_{\vec{x}\ppr,s\ppr;\vec{x},s}^{N_{p_{k}}^{(k)}=
t_{p_{k}}^{(k)}/\Delta\tau^{(k)}}\;.}
\eeq
As we conduct the continuum limit \(N_{p_{k}}^{(k)}\rightarrow\infty\) of infinitesimal time steps \(\Delta\!\tau^{(k)}\)
(\ref{s2_11}), we accomplish the exponential (\ref{s2_18}) of one-particle matrix elements
\(\hat{\mfrak{O}}_{\vec{x}_{2}\ppr,s_{2}\ppr;\vec{x}_{1}\ppr,s_{1}\ppr}^{(k)}\) (\ref{s1_13}-\ref{s1_16}) for the propagator (\ref{s2_17})
\beq\lb{s2_18}
\lim_{N_{p_{k}}^{(k)}\rightarrow\infty}
\hat{M}_{\vec{x}\ppr,s\ppr\boldsymbol{;}\vec{x},s}^{(k,p_{k});\boldsymbol{-1}}
\big(\tau_{p_{k}}^{(k)\boldsymbol{\prime}}\!=\!t_{p_{k}}\boldsymbol{\big|}\tau_{p_{k}}^{(k)}\!=\!0\big) &=&\lim_{N_{p_{k}}^{(k)}\rightarrow\infty}
\bigg(\delta_{\vec{x}_{2}\ppr,\vec{x}_{1}\ppr}\;\delta_{s_{2}\ppr s_{1}\ppr}+\frac{\im}{\hbar}\:
\eta_{p_{k}}^{(k)}\;\frac{t_{p_{k}}^{(k)}}{N_{p_{k}}^{(k)}}\;\frac{1}{\mcal{N}_{x}}\,
\hat{\mfrak{O}}_{\vec{x}_{2}\ppr,s_{2}\ppr;\vec{x}_{1}\ppr,s_{1}\ppr}^{(k)}\bigg)_{\vec{x}\ppr,s\ppr;\vec{x},s}^{N_{p_{k}}^{(k)}=
t_{p_{k}}^{(k)}/\Delta\tau^{(k)}}  \\  \no &=&
\bigg(\exp\bigg\{\frac{\im}{\hbar}\:\eta_{p_{k}}^{(k)}\;t_{p_{k}}^{(k)}\;\frac{1}{\mcal{N}_{x}}\,
\hat{\mfrak{O}}_{\vec{x}_{2}\ppr,s_{2}\ppr;\vec{x}_{1}\ppr,s_{1}\ppr}^{(k)}\bigg\}\bigg)_{\vec{x}\ppr,s\ppr;\vec{x},s} \;.
\eeq
After substitution of the determinant and the propagator with (\ref{s2_16}) and (\ref{s2_17},\ref{s2_18}),
the delta function with source fields \(\chi_{\vec{x}\ppr,s\ppr}^{(k+1)*}\), \(\chi_{\vec{x},s}^{(k)}\)
simplifies to relation (\ref{s2_19}) with integrations over the two separate branches '\(p_{k}=\pm\)'
with metric signs '\(\eta_{p_{k}}^{(k)}\)'. The integrations over the generalized time '\(t_{p_{k}}^{(k)}\,/\,\hbar\)'
result in a Fourier transformation for the generalized frequency '\(\mfrak{o}^{(k)}\)' of the source action
with fields \(\chi_{\vec{x}\ppr,s\ppr}^{(k+1)*}\), \(\chi_{\vec{x},s}^{(k)}\) which couple to the 'time'
'\(t_{p_{k}}^{(k)}\,/\,\hbar\)' varying exponential with one-particle matrix elements (\ref{s1_13}-\ref{s1_16})
of the original second quantized, symmetry operators
\beq\lb{s2_19}
\big\langle\chi^{(k+1)}\big|\delta(\mfrak{o}^{(k)}-
\boldsymbol{\hat{\mfrak{O}}^{(k)}}(\hat{\psi}\pdag,\hat{\psi})\,\big)\big|\chi^{(k)}\big\rangle &=&\hspace*{-0.3cm}\sum_{p_{k}=\pm}\;
\lim_{|\ve_{p_{k}}^{(k)}|\rightarrow0}\;\lim_{T^{(k)}\rightarrow+\infty}
\int_{0}^{T^{(k)}}\frac{dt_{p_{k}}^{(k)}}{2\pi\,\hbar}\;
\exp\Big\{-\frac{\im}{\hbar}\:t_{p_{k}}^{(k)}\:
\eta_{p_{k}}^{(k)}\:\big(\mfrak{o}^{(k)}-\im\:\ve_{p_{k}}^{(k)}\big)\Big\}\;\times \\ \no &\times&
\exp\bigg\{\sum_{\vec{x}\ppr,s\ppr;\vec{x},s}\mcal{N}_{x}\:\chi_{\vec{x}\ppr,s\ppr}^{(k+1)*}\;\;
\bigg(\exp\bigg\{\frac{\im}{\hbar}\:\eta_{p_{k}}^{(k)}\;t_{p_{k}}^{(k)}\;\frac{1}{\mcal{N}_{x}}\,
\hat{\mfrak{O}}_{\vec{x}_{2}\ppr,s_{2}\ppr;\vec{x}_{1}\ppr,s_{1}\ppr}^{(k)}\bigg\}\bigg)_{\vec{x}\ppr,s\ppr;\vec{x},s}
\;\;\chi_{\vec{x},s}^{(k)}\bigg\}\;.
\eeq
In the case of the one-particle operators \(\boldsymbol{\hat{N}}(\hat{\psi}\pdag,\hat{\psi})\),
\(\boldsymbol{\hat{S}_{z}}(\hat{\psi}\pdag,\hat{\psi})\), the Fourier transformations can be easily performed because
the corresponding source actions only consist of a single frequency
\beq\lb{s2_20}
\lefteqn{\big\langle\chi^{(2)}\big|\delta\big(n_{0}-
\boldsymbol{\hat{N}}(\hat{\psi}\pdag,\hat{\psi})\,\big)\big|\chi^{(1)}\big\rangle = } \\ \no &=&
\sum_{p_{1}=\pm}\; \lim_{|\ve_{p_{1}}^{(1)}|\rightarrow0}\;\lim_{T^{(1)}\rightarrow+\infty}
\int_{0}^{T^{(1)}}\frac{dt_{p_{1}}^{(1)}}{2\pi\,\hbar}\;
\exp\Big\{-\frac{\im}{\hbar}\:\eta_{p_{1}}^{(1)}\: t_{p_{1}}^{(1)}\:
\big(n_{0}-\im\:\ve_{p_{1}}^{(1)}\big)\Big\}\;
\exp\bigg\{\Big(\sum_{\vec{x},s}\chi_{\vec{x},s}^{(2)*}\;\chi_{\vec{x},s}^{(1)}\Big)\;
\exp\Big\{\frac{\im}{\hbar}\:\eta_{p_{1}}^{(1)}\;t_{p_{1}}^{(1)}\Big\}\bigg\}\;=   \\ \no &=&\sum_{p_{1}=\pm}\;
\lim_{|\ve_{p_{1}}^{(1)}|\rightarrow0}\;\lim_{T^{(1)}\rightarrow+\infty}
\int_{0}^{T^{(1)}}\frac{dt_{p_{1}}^{(1)}}{2\pi\,\hbar}\;
\exp\Big\{-\frac{\im}{\hbar}\:\eta_{p_{1}}^{(1)}\: t_{p_{1}}^{(1)}\:
\big(n_{0}-\im\:\ve_{p_{1}}^{(1)}\big)\Big\}\;
\sum_{n=0}^{+\infty} \frac{\Big(\sum_{\vec{y}}\chi_{\vec{y}}^{(2)*}\;\chi_{\vec{y}}^{(1)}\Big)^{n}}{n!}\;
\exp\Big\{\frac{\im}{\hbar}\:\eta_{p_{1}}^{(1)}\;t_{p_{1}}^{(1)}\cdot n\Big\}  \\ \no &=&
\frac{1}{\Delta\tau^{(1)}}\;\frac{\Big(\sum_{\vec{y}}\chi_{\vec{y}}^{(2)*}\;
\chi_{\vec{y}}^{(1)}\Big)^{n_{0}}}{n_{0}!}\;;\;\;\;\;n_{0}\in\mathbb{N}\;;
\eeq
\beq\lb{s2_21}
\lefteqn{\big\langle\chi^{(4)}\big|\delta\big(\hbar\,s_{z}-
\boldsymbol{\hat{S}_{z}}(\hat{\psi}\pdag,\hat{\psi})\,\big)\big|\chi^{(3)}\big\rangle =
\sum_{p_{3}=\pm}\; \lim_{|\ve_{p_{3}}^{(3)}|\rightarrow0}\;\lim_{T^{(3)}\rightarrow+\infty}
\int_{0}^{T^{(3)}} \hspace*{-0.2cm} \frac{dt_{p_{3}}^{(3)}}{2\pi\,\hbar}\;
\exp\Big\{-\frac{\im}{\hbar}\:\eta_{p_{3}}^{(3)}\: t_{p_{3}}^{(3)}\:
\big(\hbar\,s_{z}-\im\:\ve_{p_{3}}^{(3)}\big)\Big\}    }   \\  \no &\times&
\exp\bigg\{\Big(\sum_{\vec{x}}\chi_{\vec{x},\uparrow}^{(4)*}\;\chi_{\vec{x},\uparrow}^{(3)}\Big)\;
\exp\Big\{\frac{\im}{2}\:\eta_{p_{3}}^{(3)}\;t_{p_{3}}^{(3)}\Big\}+
\Big(\sum_{\vec{x}}\chi_{\vec{x},\downarrow}^{(4)*}\;\chi_{\vec{x},\downarrow}^{(3)}\Big)\;
\exp\Big\{-\frac{\im}{2}\:\eta_{p_{3}}^{(3)}\;t_{p_{3}}^{(3)}\Big\}
\bigg\}\;=     \\ \no  &=&\sum_{p_{3}=\pm}\;
\lim_{|\ve_{p_{3}}^{(3)}|\rightarrow0}\;\lim_{T^{(3)}\rightarrow+\infty}\int_{0}^{T^{(3)}} \hspace*{-0.2cm}
\frac{dt_{p_{3}}^{(3)}}{2\pi\,\hbar}\;
\exp\Big\{-\frac{\im}{\hbar}\:\eta_{p_{3}}^{(3)}\: t_{p_{3}}^{(3)}\:
\big(\hbar\,s_{z}-\im\:\ve_{p_{3}}^{(3)}\big)\Big\}  \\  \no &\times&
\sum_{n=0}^{+\infty}\frac{1}{n!}\sum_{n_{1}=0}^{n}\binom{n}{n_{1}}
\Big(\sum_{\vec{x}}\chi_{\vec{x},\uparrow}^{(4)*}\;\chi_{\vec{x},\uparrow}^{(3)}\Big)^{n_{1}}\;
\Big(\sum_{\vec{x}}\chi_{\vec{x},\downarrow}^{(4)*}\;\chi_{\vec{x},\downarrow}^{(3)}\Big)^{n-n_{1}}\;
\exp\Big\{-\im\:\eta_{p_{3}}^{(3)}\;t_{p_{3}}^{(3)}\:\big({\ts\frac{n}{2}}-n_{1}\big)\Big\} \\ \no &=&
\frac{1}{\Delta\tau^{(3)}}\sum_{n_{1}=1}^{+\infty}\frac{1}{n_{1}!\;(n_{1}-1)!}\;
\Big(\sum_{\vec{x}}\chi_{\vec{x},\uparrow}^{(4)*}\;\chi_{\vec{x},\uparrow}^{(3)}\Big)^{n_{1}+s_{z}-\frac{1}{2}}\;
\Big(\sum_{\vec{x}}\chi_{\vec{x},\downarrow}^{(4)*}\;\chi_{\vec{x},\downarrow}^{(3)}\Big)^{n_{1}-s_{z}-\frac{1}{2}}
\;;\;\;\;\big(s_{z}=\pm{\ts\frac{1}{2}}\big)\;.
\eeq
whereas the orbital angular momentum component \(\boldsymbol{\hat{L}_{z}}(\hat{\psi}\pdag,\hat{\psi})\) has
an infinite number '\(m\)' ((\(-l\leq m\leq +l\)) , \(l\in[0,\infty)\)) of different frequencies with varying
coefficients following from the spin summation and radial space integration of the prevailing source fields.
However, as we define coefficients \(\mscr{C}_{m}^{(l_{0})}\) (\ref{s2_23}) with Heaviside function
\(\theta(l+{\ts\frac{1}{2}}-|m|)\), one can also describe the resulting summation (\ref{s2_22}) with Kronecker
deltas \(\delta(l_{z}\boldsymbol{|}n_{-l_{0}}\cdot (-l_{0})+\ldots+n_{m}\cdot m+
\ldots+n_{+l_{0}}\cdot (+l_{0})\,)\) which restrict the total z-orbital angular momentum to those combinations
having value '\(\hbar\:l_{z}\)'
\beq\lb{s2_22}
\lefteqn{\big\langle\chi^{(3)}\big|\delta\big(\hbar\,l_{z}-
\boldsymbol{\hat{L}_{z}}(\hat{\psi}\pdag,\hat{\psi})\,\big)\big|\chi^{(2)}\big\rangle =
\sum_{p_{2}=\pm}\; \lim_{|\ve_{p_{2}}^{(2)}|\rightarrow0}\;\lim_{T^{(2)}\rightarrow+\infty}
\int_{0}^{T^{(2)}} \hspace*{-0.2cm} \frac{dt_{p_{2}}^{(2)}}{2\pi\,\hbar}\;
\exp\Big\{-\frac{\im}{\hbar}\:\eta_{p_{2}}^{(2)}\: t_{p_{2}}^{(2)}\:
\big(\hbar\,l_{z}-\im\:\ve_{p_{2}}^{(2)}\big)\Big\} \times  }  \\  \no &\times&
\exp\bigg\{\sum_{l=0}^{+\infty}\sum_{m=-l}^{+l}
\bigg(\int_{0}^{R}\frac{d\!(|\vec{x}|\!)\;\;\;|\vec{x}|^{2}}{V^{(d)}}\;
\sum_{s}\chi_{|\vec{x}|,l,m,s}^{(3)*}\;\chi_{|\vec{x}|,l,m,s}^{(2)}\bigg)\;
\exp\Big\{\im\:\eta_{p_{2}}^{(2)}\;m\;t_{p_{2}}^{(2)}\Big\}\bigg\} =  \\ \no &=&
\sum_{n=0}^{+\infty}\frac{1}{\Delta\tau^{(2)}}
\sum_{n_{-l_{0}}+\ldots+n_{m}+\ldots+n_{+l_{0}}=n}\;
\frac{\delta\big(l_{z}\boldsymbol{|}n_{-l_{0}}\cdot (-l_{0})+\ldots+n_{m}\cdot m+
\ldots+n_{+l_{0}}\cdot (+l_{0})\,\big)}{n_{-l_{0}}!\cdot\ldots\cdot n_{m}!\cdot\ldots\cdot n_{+l_{0}}!}\;\times
\\  \no &\times&\big(\mscr{C}_{-l_{0}}^{(l_{0})}\big)^{n_{-l_{0}}}\cdot\ldots\cdot
\big(\mscr{C}_{m}^{(l_{0})}\big)^{n_{m}}\cdot\ldots\cdot
\big(\mscr{C}_{+l_{0}}^{(l_{0})}\big)^{n_{+l_{0}}} \;;
\eeq
\beq\lb{s2_23}
\mscr{C}_{m}^{(l_{0})} &=&\sum_{l=0}^{l_{0}}\bigg(\int_{0}^{R}\frac{d\!(|\vec{x}|\!)\;\;\;|\vec{x}|^{2}}{V^{(d)}}\;
\sum_{s}\chi_{|\vec{x}|,l,m,s}^{(3)*}\;\chi_{|\vec{x}|,l,m,s}^{(2)}\bigg)\;
\theta\big(l+{\ts\frac{1}{2}}-|m|\big)\;;  \\ \no &&\big(\theta(x)=1\;\;,x\geq0\big)\;;\;\;\;
\big(\theta(x)=0\;\;,x<0\big)\;; \;\;\; \big(l_{0}\in\mathbb{N}_{0}\;,\;\;
\mbox{max. angular momentum number }\rightarrow\infty \big)\;;  \\   \lb{s2_24}
\chi_{|\vec{x}|,l,m,s}^{(2)} &=&\int_{0}^{2\pi}d\beta\int_{0}^{\pi}d\alpha\;\;
\sin(\alpha)\;Y_{l,m}^{*}(\alpha,\beta)\;\;\chi_{\vec{x},s}^{(2)} \;;\hspace*{0.6cm}
\chi_{|\vec{x}|,l,m,s}^{(3)*} =\int_{0}^{2\pi}d\beta\int_{0}^{\pi}d\alpha\;\;
\sin(\alpha)\;Y_{l,m}(\alpha,\beta)\;\;\chi_{\vec{x},s}^{(3)*} \;;  \\   \lb{s2_25}
\vec{x} &=&|\vec{x}|\;\;\big\{\sin(\alpha)\;\cos(\beta)\;,\;\sin(\alpha)\;\sin(\beta)\;,\;\cos(\alpha)\big\} \;.
\eeq

\section{Transformation of the delta functions composed of two-particle operators} \lb{s3}

\subsection{HST to self-energies with anomalous doubled pairs}\lb{s31}

Various references \cite{Kleinert1} infer that coherent state path integrals depend on the
chosen, underlying kind of discrete time grid so that one can only attain unreliable results, following
from transformations of these. However, we have already pointed out in Refs. \cite{physica1,precisecoh1} how to conduct
a coset decomposition within coherent state path integrals in an {\it exact, appropriate} manner from
second quantized Hamiltonians in normal order. The fundamental point of a {\it correct} discrete time grid
of normal ordered Hamiltonians consists of two different kinds of anomalous doubled fields
\(\Psi_{\vec{x},s}^{a}(\vartheta_{q_{\kappa}}^{(\kappa)})\), \(\breve{\Psi}_{\vec{x},s}^{a}(\vartheta_{q_{\kappa}}^{(\kappa)})\)
(\ref{s3_1},\ref{s3_3}) with two different kinds (\ref{s3_2},\ref{s3_4}) of hermitian conjugation
'\(\pdag\)' and '\(^{\sharp}\)'
\beq
\lb{s3_1} \mbox{(1)}&:& \mbox{'{\it equal time}', anomalous-doubled field :} \;\;
(0\leq\vartheta_{q_{\kappa}}^{(\kappa)}\leq\mfrak{t}_{q_{\kappa}}^{(\kappa)}-\Delta\vartheta^{(\kappa)})  \\  \no
\Psi_{\vec{x},s}^{a(=1/2)}(\vartheta_{q_{\kappa}}^{(\kappa)})&=&\hspace*{-0.4cm}
\left(\bea{c}\psi_{\vec{x},s}(\vartheta_{q_{\kappa}}^{(\kappa)}) \\
\psi_{\vec{x},s}^{*}(\vartheta_{q_{\kappa}}^{(\kappa)})\eea\right)^{a}=
\Big(\underbrace{\psi_{\vec{x},s}(\vartheta_{q_{\kappa}}^{(\kappa)})}_{a=1}\;;\;
\underbrace{\psi_{\vec{x},s}^{*}(\vartheta_{q_{\kappa}}^{(\kappa)})}_{a=2}\Big)^{T}\;;
(\kappa=0,1,2)\;;\;\;(\kappa=0\simeq E)\;; \\   \lb{s3_2}
\mbox{(2)} &:& \mbox{'hermitian-conjugation' '$\pdag$' of '{\it equal time}', anomalous-doubled field :} \\  \no
\Psi_{\vec{x},s}^{\dag a(=1/2)}(\vartheta_{q_{\kappa}}^{(\kappa)})&=&
\Big(\underbrace{\psi_{\vec{x},s}^{*}(\vartheta_{q_{\kappa}}^{(\kappa)})}_{a=1}\;;\;
\underbrace{\psi_{\vec{x},s}(\vartheta_{q_{\kappa}}^{(\kappa)})}_{a=2}\Big)\;;   \\  \lb{s3_3}
\mbox{(1)}&:& \mbox{'{\it time shifted}' $\Delta\vartheta^{(\kappa)}$,
anomalous-doubled field '$\boldsymbol{\breve{\ph{\Psi}}}$' :} \;\;
(0\leq\vartheta_{q_{\kappa}}^{(\kappa)}\leq\mfrak{t}_{q_{\kappa}}^{(\kappa)}-\Delta\vartheta^{(\kappa)})   \\ \no
\breve{\Psi}_{\vec{x},s}^{a(=1/2)}(\vartheta_{q_{\kappa}}^{(\kappa)})&=&\hspace*{-0.4cm}
\left(\bea{c}\psi_{\vec{x},s}(\vartheta_{q_{\kappa}}^{(\kappa)}) \\
\psi_{\vec{x},s}^{*}(\vartheta_{q_{\kappa}}^{(\kappa)}\!+\!\Delta\vartheta^{(\kappa)})\eea\right)^{a}=
\Big(\underbrace{\psi_{\vec{x},s}(\vartheta_{q_{\kappa}}^{(\kappa)})}_{a=1}\;;\;
\underbrace{\psi_{\vec{x},s}^{*}(\vartheta_{q_{\kappa}}^{(\kappa)}\!+
\!\Delta\vartheta^{(\kappa)})}_{a=2}\Big)^{T}\;; (\kappa=0,1,2)\;;\;(\kappa=0\simeq E);  \\  \lb{s3_4}
\mbox{(2)} &:&
\mbox{'hermitian-conjugation' '$^{\sharp}$' with '{\it time shift correction}' \(\Delta\vartheta^{(\kappa)}\)} \\ \no &&
\mbox{in the resulting complex part :}    \\  \no
\breve{\Psi}_{\vec{x},s}^{a(=1/2)}(\vartheta_{q_{\kappa}}^{(\kappa)}) &\stackrel{'\sharp'}{\Longrightarrow}&
\breve{\Psi}_{\vec{x},s}^{\sharp a(=1/2)}(\vartheta_{q_{\kappa}}^{(\kappa)}) =
\Big(\underbrace{\psi_{\vec{x},s}^{*}(\vartheta_{q_{\kappa}}^{(\kappa)}\!+\!\Delta\vartheta^{(\kappa)})}_{a=1}\;;\;
\underbrace{\psi_{\vec{x},s}(\vartheta_{q_{\kappa}}^{(\kappa)})}_{a=2}\Big)\;.
\eeq
As we insert overcomplete sets of coherent states between the generalized 'time' steps \(\Delta\vartheta^{(\kappa)}\) for the
'time' development \(\vartheta_{q_{\kappa}}^{(\kappa)}\) with exponential operators, one has to take into account the sole appearance
of field combinations \(\psi_{\vec{x}\ppr,s\ppr}^{*}(\vartheta_{q_{\kappa}}^{(\kappa)}\!+\!
\Delta\vartheta^{(\kappa)})\;\ldots\;\psi_{\vec{x},s}(\vartheta_{q_{\kappa}}^{(\kappa)})\) instead of the more appealing hermitian form
\(\psi_{\vec{x}\ppr,s\ppr}^{*}(\vartheta_{q_{\kappa}}^{(\kappa)})\;\ldots\;\psi_{\vec{x},s}(\vartheta_{q_{\kappa}}^{(\kappa)})\) so that the
complex conjugated field \(\psi_{\vec{x}\ppr,s\ppr}^{*}(\vartheta_{q_{\kappa}}^{(\kappa)}\!+\!\Delta\vartheta^{(\kappa)})\) always occurs
a particular time step \(\Delta\vartheta^{(\kappa)}\) later than its corresponding field
\(\psi_{\vec{x},s}(\vartheta_{q_{\kappa}}^{(\kappa)})\). In order to preserve this property of the time shifted
'\(\Delta\vartheta^{(\kappa)}\)', complex conjugated fields
\(\psi_{\vec{x},s}^{*}(\vartheta_{q_{\kappa}}^{(\kappa)}\!+\!\Delta\vartheta^{(\kappa)})\) under a coset decomposition, we have to use
the 'time shifted' '\(\Delta\vartheta^{(\kappa)}\)', anomalous doubled field
\(\breve{\Psi}_{\vec{x},s}^{a(=1/2)}(\vartheta_{q_{\kappa}}^{(\kappa)})\) (\ref{s3_3}), additionally marked by the symbol
'\(\breve{\ph{\Psi}}\)', with its particular hermitian conjugation '\(^{\sharp}\)' (\ref{s3_4}) which includes a
time shift correction '\(\Delta\vartheta^{(\kappa)}\)' in the resulting complex part. Therefore, the anomalous
doubling for the scalar product of normal ordered fields (\ref{s3_5}) is obtained with the time shifted
'\(\Delta\vartheta^{(\kappa)}\)' anomalous doubled
field \(\breve{\Psi}_{\vec{x},s}^{a(=1/2)}(\vartheta_{q_{\kappa}}^{(\kappa)})\) (\ref{s3_3}) and the hermitian conjugation
'\(^{\sharp}\)' (\ref{s3_4}), having the time shift correction  \(\Delta\vartheta^{(\kappa)}\) in the resulting
complex part. Moreover, the diagonal metric tensor \(\hat{S}^{ab}\) (\ref{s3_6}) of the anomalous doubling
has to be introduced, due to the anti-commutativity of the Fermi fields
\beq\lb{s3_5}
\psi_{\vec{x},s}^{*}(\vartheta_{q_{\kappa}}^{(\kappa)}\!+\!\Delta\vartheta^{(\kappa)})\;\;
\psi_{\vec{x},s}(\vartheta_{q_{\kappa}}^{(\kappa)})
&=&\frac{1}{2}\;\bigg(
\psi_{\vec{x},s}^{*}(\vartheta_{q_{\kappa}}^{(\kappa)}\!+\!\Delta\vartheta^{(\kappa)})\;\;
\psi_{\vec{x},s}(\vartheta_{q_{\kappa}}^{(\kappa)})-
\psi_{\vec{x},s}(\vartheta_{q_{\kappa}}^{(\kappa)})\;\;
\psi_{\vec{x},s}^{*}(\vartheta_{q_{\kappa}}^{(\kappa)}\!+\!\Delta\vartheta^{(\kappa)})
\bigg)  \\  \no &=& \frac{1}{2}\;\Big(\breve{\Psi}_{\vec{x},s}^{\sharp a}(\vartheta_{q_{\kappa}}^{(\kappa)})\;\hat{S}\;
\breve{\Psi}_{\vec{x},s}^{a}(\vartheta_{q_{\kappa}}^{(\kappa)})\Big) \;;  \\ \lb{s3_6}
\hat{S}^{ab} &=&\delta_{ab}\;\mbox{diag}\Big(\underbrace{\hat{1}}_{a=1}\;\boldsymbol{;}\;
\underbrace{-\hat{1}}_{a=2}\Big)\;\;\;.
\eeq
According to the two different kinds (\ref{s3_1},\ref{s3_2}) and (\ref{s3_3},\ref{s3_4}) of anomalous doubling,
we have two different kinds of dyadic products (\ref{s3_7}) and (\ref{s3_8}).
As we have already described in Ref. \cite{precisecoh1}, the anomalous doubled self-energy matrix has to comply
with the symmetries (\ref{s3_9},\ref{s3_10}) of the dyadic product of the equal time anomalous doubled field and its equal time
hermitian conjugation whereas the anomalous doubled density matrices have to follow from the dyadic product (\ref{s3_11},\ref{s3_12})
of a time shifted anomalous doubled field with its time-shift corrected, hermitian conjugated field.
In correspondence to Ref. \cite{precisecoh1}, we briefly list the two kinds of dyadic products of anomalous doubled
fields, which can be regarded as the appropriate order parameter matrices of the coset decomposition,
and additionally define the transposition (\ref{s3_13},\ref{s3_16}), trace (\ref{s3_14},\ref{s3_17}) and
hermitian conjugation (\ref{s3_15},\ref{s3_18}) of these two kinds of anomalous
doubled order parameter matrices for the
self-energy (\ref{s3_7},\ref{s3_9}) and the density matrix (\ref{s3_8},\ref{s3_11}), respectively
\beq \lb{s3_7}
\hat{\Phi}_{\vec{x},s;\vec{x}\ppr,s\ppr}^{ab}(\vartheta_{q_{\kappa}}^{(\kappa)})&=&
\Psi_{\vec{x},s}^{a}(\vartheta_{q_{\kappa}}^{(\kappa)})\otimes
\Psi_{\vec{x}\ppr,s\ppr}^{\dag b}(\vartheta_{q_{\kappa}}^{(\kappa)})  \;;  \\   \lb{s3_8}
\breve{\Phi}_{\vec{x},s;\vec{x}\ppr,s\ppr}^{ab}(\vartheta_{q_{\kappa}}^{(\kappa)})&=&
\breve{\Psi}_{\vec{x},s}^{a}(\vartheta_{q_{\kappa}}^{(\kappa)})\otimes
\breve{\Psi}_{\vec{x}\ppr,s\ppr}^{\sharp b}(\vartheta_{q_{\kappa}}^{(\kappa)})   \;; \\ \lb{s3_9}
\hat{\Phi}_{\vec{y};\vec{y}\ppr}^{ab}(\vartheta_{q_{\kappa}}^{(\kappa)})&=&
\Psi_{\vec{y}}^{a}(\vartheta_{q_{\kappa}}^{(\kappa)})\;\otimes\;
\Psi_{\vec{y}\ppr}^{\dagger b}(\vartheta_{q_{\kappa}}^{(\kappa)}) = \left( \bea{c}
\psi_{\vec{y}}(\vartheta_{q_{\kappa}}^{(\kappa)}) \\ \psi_{\vec{y}}^{*}(\vartheta_{q_{\kappa}}^{(\kappa)}) \eea\right)^{a} \otimes
\Big(\psi_{\vec{y}\ppr}^{*}(\vartheta_{q_{\kappa}}^{(\kappa)})\;;\;\psi_{\vec{y}\ppr}(\vartheta_{q_{\kappa}}^{(\kappa)})\Big)^{b}=
\\ \no &=&\left( \bea{cc}
\langle\psi_{\vec{y}}(\vartheta_{q_{\kappa}}^{(\kappa)})\;\psi_{\vec{y}\ppr}^{*}(\vartheta_{q_{\kappa}}^{(\kappa)})\rangle &
\langle\psi_{\vec{y}}(\vartheta_{q_{\kappa}}^{(\kappa)})\;\psi_{\vec{y}\ppr}(\vartheta_{q_{\kappa}}^{(\kappa)})\rangle \\
\langle\psi_{\vec{y}}^{*}(\vartheta_{q_{\kappa}}^{(\kappa)})\;\psi_{\vec{y}\ppr}^{*}(\vartheta_{q_{\kappa}}^{(\kappa)})\rangle &
\langle\psi_{\vec{y}}^{*}(\vartheta_{q_{\kappa}}^{(\kappa)})\;
\psi_{\vec{y}\ppr}(\vartheta_{q_{\kappa}}^{(\kappa)})\rangle \eea\right) =
\left( \bea{cc}
\hat{\Phi}_{\vec{y};\vec{y}\ppr}^{11}(\vartheta_{q_{\kappa}}^{(\kappa)}) &
\hat{\Phi}_{\vec{y};\vec{y}\ppr}^{12}(\vartheta_{q_{\kappa}}^{(\kappa)}) \\
\hat{\Phi}_{\vec{y};\vec{y}\ppr}^{21}(\vartheta_{q_{\kappa}}^{(\kappa)}) &
\hat{\Phi}_{\vec{y};\vec{y}\ppr}^{22}(\vartheta_{q_{\kappa}}^{(\kappa)})
\eea\right)_{\mbox{;}}  \\ \lb{s3_10} &&\hspace*{-3.6cm}
\bea{rclrcl}
\hat{\Phi}_{\vec{x},s;\vec{x}\ppr,s\ppr}^{aa,\dagger}(\vartheta_{q_{\kappa}}^{(\kappa)}) &=&
\hat{\Phi}_{\vec{x},s;\vec{x}\ppr,s\ppr}^{aa}(\vartheta_{q_{\kappa}}^{(\kappa)})  & \hspace*{0.6cm}
\hat{\Phi}_{\vec{x},s;\vec{x}\ppr,s\ppr}^{22}(\vartheta_{q_{\kappa}}^{(\kappa)}) &=&
-\hat{\Phi}_{\vec{x},s;\vec{x}\ppr,s\ppr}^{11,T}(\vartheta_{q_{\kappa}}^{(\kappa)})  \\
\hat{\Phi}_{\vec{x},s;\vec{x}\ppr,s\ppr}^{21}(\vartheta_{q_{\kappa}}^{(\kappa)}) &=&
\hat{\Phi}_{\vec{x},s;\vec{x}\ppr,s\ppr}^{12,\dagger}(\vartheta_{q_{\kappa}}^{(\kappa)})  & \hspace*{0.6cm}
\hat{\Phi}_{\vec{x},s;\vec{x}\ppr,s\ppr}^{12,T}(\vartheta_{q_{\kappa}}^{(\kappa)}) &=&
-\hat{\Phi}_{\vec{x},s;\vec{x}\ppr,s\ppr}^{12}(\vartheta_{q_{\kappa}}^{(\kappa)})
\eea_{\mbox{;}}   \\  \lb{s3_11}
\lefteqn{\hspace*{-3.7cm}\breve{\Phi}_{\vec{y};\vec{y}\ppr}^{ab}(\vartheta_{q_{\kappa}}^{(\kappa)}) =
\breve{\Psi}_{\vec{y}}^{a}(\vartheta_{q_{\kappa}}^{(\kappa)})\;\otimes\;
\breve{\Psi}_{\vec{y}\ppr}^{\sharp b}(\vartheta_{q_{\kappa}}^{(\kappa)}) = \left( \bea{c}
\psi_{\vec{y}}(\vartheta_{q_{\kappa}}^{(\kappa)}) \\
\psi_{\vec{y}}^{*}(\vartheta_{q_{\kappa}}^{(\kappa)}\!+\!\Delta\vartheta^{(\kappa)}) \eea\right)^{a} \otimes
\Big(\psi_{\vec{y}\ppr}^{*}(\vartheta_{q_{\kappa}}^{(\kappa)}\!+\!
\Delta\vartheta^{(\kappa)})\;;\;\psi_{\vec{y}\ppr}(\vartheta_{q_{\kappa}}^{(\kappa)})\Big)^{b} }
\\ \no \lefteqn{\hspace*{-1.9cm}=\left(\hspace*{-0.2cm} \bea{cc}
\langle\psi_{\vec{y}}(\vartheta_{q_{\kappa}}^{(\kappa)})\;
\psi_{\vec{y}\ppr}^{*}(\vartheta_{q_{\kappa}}^{(\kappa)}\!+\!\Delta\vartheta^{(\kappa)})\rangle &
\langle\psi_{\vec{y}}(\vartheta_{q_{\kappa}}^{(\kappa)})\;\psi_{\vec{y}\ppr}(\vartheta_{q_{\kappa}}^{(\kappa)})\rangle \\
\langle\psi_{\vec{y}}^{*}(\vartheta_{q_{\kappa}}^{(\kappa)}\!+\!
\Delta\vartheta^{(\kappa)})\;\psi_{\vec{y}\ppr}^{*}(\vartheta_{q_{\kappa}}^{(\kappa)}\!+\!\Delta\vartheta^{(\kappa)})\rangle &
\langle\psi_{\vec{y}}^{*}(\vartheta_{q_{\kappa}}^{(\kappa)}\!+\!
\Delta\vartheta^{(\kappa)})\;\psi_{\vec{y}\ppr}(\vartheta_{q_{\kappa}}^{(\kappa)})\rangle \eea\hspace*{-0.2cm}\right)  =
\left(\hspace*{-0.2cm} \bea{cc}
\breve{\Phi}_{\vec{y};\vec{y}\ppr}^{11}(\vartheta_{q_{\kappa}}^{(\kappa)}) &
\breve{\Phi}_{\vec{y};\vec{y}\ppr}^{12}(\vartheta_{q_{\kappa}}^{(\kappa)}) \\
\breve{\Phi}_{\vec{y};\vec{y}\ppr}^{21}(\vartheta_{q_{\kappa}}^{(\kappa)}) &
\breve{\Phi}_{\vec{y};\vec{y}\ppr}^{22}(\vartheta_{q_{\kappa}}^{(\kappa)})
\eea\hspace*{-0.2cm}\right)_{\mbox{;}}       }        \\ \lb{s3_12} && \hspace*{-3.6cm}
\bea{rclrcl}
\breve{\Phi}_{\vec{x},s;\vec{x}\ppr,s\ppr}^{aa,\sharp}(\vartheta_{q_{\kappa}}^{(\kappa)}) &=&
\breve{\Phi}_{\vec{x},s;\vec{x}\ppr,s\ppr}^{aa}(\vartheta_{q_{\kappa}}^{(\kappa)})  & \hspace*{0.6cm}
\breve{\Phi}_{\vec{x},s;\vec{x}\ppr,s\ppr}^{22}(\vartheta_{q_{\kappa}}^{(\kappa)}) &=&
-\breve{\Phi}_{\vec{x},s;\vec{x}\ppr,s\ppr}^{11,T}(\vartheta_{q_{\kappa}}^{(\kappa)})  \\
\breve{\Phi}_{\vec{x},s;\vec{x}\ppr,s\ppr}^{21}(\vartheta_{q_{\kappa}}^{(\kappa)}) &=&
\breve{\Phi}_{\vec{x},s;\vec{x}\ppr,s\ppr}^{12,\sharp}(\vartheta_{q_{\kappa}}^{(\kappa)})  & \hspace*{0.6cm}
\breve{\Phi}_{\vec{x},s;\vec{x}\ppr,s\ppr}^{12,T}(\vartheta_{q_{\kappa}}^{(\kappa)}) &=&
-\breve{\Phi}_{\vec{x},s;\vec{x}\ppr,s\ppr}^{12}(\vartheta_{q_{\kappa}}^{(\kappa)})
\eea_{\mbox{;}}  \\ \lb{s3_13}
\Big(\hat{\Phi}_{\vec{x},s;\vec{x}\ppr,s\ppr}^{ab}(\vartheta_{q_{\kappa}}^{(\kappa)})\Big)^{T}&=&\hspace*{-0.3cm}\left( \bea{cc}
\hat{\Phi}_{\vec{x},s;\vec{x}\ppr,s\ppr}^{11}(\vartheta_{q_{\kappa}}^{(\kappa)}) &
\hat{\Phi}_{\vec{x},s;\vec{x}\ppr,s\ppr}^{12}(\vartheta_{q_{\kappa}}^{(\kappa)}) \\
\hat{\Phi}_{\vec{x},s;\vec{x}\ppr,s\ppr}^{21}(\vartheta_{q_{\kappa}}^{(\kappa)}) &
\hat{\Phi}_{\vec{x},s;\vec{x}\ppr,s\ppr}^{22}(\vartheta_{q_{\kappa}}^{(\kappa)}) \eea\right)^{T}\hspace*{-0.3cm}= \left( \bea{cc}
\big(\hat{\Phi}_{\vec{x},s;\vec{x}\ppr,s\ppr}^{11}(\vartheta_{q_{\kappa}}^{(\kappa)})\big)^{T} &
\big(\hat{\Phi}_{\vec{x},s;\vec{x}\ppr,s\ppr}^{21}(\vartheta_{q_{\kappa}}^{(\kappa)})\big)^{T} \\
\big(\hat{\Phi}_{\vec{x},s;\vec{x}\ppr,s\ppr}^{12}(\vartheta_{q_{\kappa}}^{(\kappa)})\big)^{T} &
\big(\hat{\Phi}_{\vec{x},s;\vec{x}\ppr,s\ppr}^{22}(\vartheta_{q_{\kappa}}^{(\kappa)})\big)^{T}
\eea\right)_{\mbox{;}} \\     \lb{s3_14}
\TRS\Big[\hat{\Phi}_{\vec{x},s;\vec{x}\ppr,s\ppr}^{ab}(\vartheta_{q_{\kappa}}^{(\kappa)})\Big]&=&
\trs\Big[\hat{\Phi}_{\vec{x},s;\vec{x}\ppr,s\ppr}^{11}(\vartheta_{q_{\kappa}}^{(\kappa)})\Big]+
\trs\Big[\hat{\Phi}_{\vec{x},s;\vec{x}\ppr,s\ppr}^{22}(\vartheta_{q_{\kappa}}^{(\kappa)})\Big]_{\mbox{;}} \\ \lb{s3_15}
\Big(\hat{\Phi}_{\vec{x},s;\vec{x}\ppr,s\ppr}^{ab}(\vartheta_{q_{\kappa}}^{(\kappa)})\Big)\pdag&=&\left( \bea{cc}
\hat{\Phi}_{\vec{x},s;\vec{x}\ppr,s\ppr}^{11}(\vartheta_{q_{\kappa}}^{(\kappa)}) &
\hat{\Phi}_{\vec{x},s;\vec{x}\ppr,s\ppr}^{12}(\vartheta_{q_{\kappa}}^{(\kappa)}) \\
\hat{\Phi}_{\vec{x},s;\vec{x}\ppr,s\ppr}^{21}(\vartheta_{q_{\kappa}}^{(\kappa)}) &
\hat{\Phi}_{\vec{x},s;\vec{x}\ppr,s\ppr}^{22}(\vartheta_{q_{\kappa}}^{(\kappa)}) \eea\right)\pdag= \left( \bea{cc}
\big(\hat{\Phi}_{\vec{x},s;\vec{x}\ppr,s\ppr}^{11}(\vartheta_{q_{\kappa}}^{(\kappa)})\big)\pdag &
\big(\hat{\Phi}_{\vec{x},s;\vec{x}\ppr,s\ppr}^{21}(\vartheta_{q_{\kappa}}^{(\kappa)})\big)\pdag \\
\big(\hat{\Phi}_{\vec{x},s;\vec{x}\ppr,s\ppr}^{12}(\vartheta_{q_{\kappa}}^{(\kappa)})\big)\pdag &
\big(\hat{\Phi}_{\vec{x},s;\vec{x}\ppr,s\ppr}^{22}(\vartheta_{q_{\kappa}}^{(\kappa)})\big)\pdag
\eea\right)_{\mbox{;}} \\ \lb{s3_16}  \hspace*{-0.3cm}
\Big(\breve{\Phi}_{\vec{x},s;\vec{x}\ppr,s\ppr}^{ab}(\vartheta_{q_{\kappa}}^{(\kappa)})\Big)^{T}&=&\hspace*{-0.3cm}\left( \bea{cc}
\breve{\Phi}_{\vec{x},s;\vec{x}\ppr,s\ppr}^{11}(\vartheta_{q_{\kappa}}^{(\kappa)}) &
\breve{\Phi}_{\vec{x},s;\vec{x}\ppr,s\ppr}^{12}(\vartheta_{q_{\kappa}}^{(\kappa)}) \\
\breve{\Phi}_{\vec{x},s;\vec{x}\ppr,s\ppr}^{21}(\vartheta_{q_{\kappa}}^{(\kappa)}) &
\breve{\Phi}_{\vec{x},s;\vec{x}\ppr,s\ppr}^{22}(\vartheta_{q_{\kappa}}^{(\kappa)}) \eea\right)^{T}\hspace*{-0.3cm}= \left( \bea{cc}
\big(\breve{\Phi}_{\vec{x},s;\vec{x}\ppr,s\ppr}^{11}(\vartheta_{q_{\kappa}}^{(\kappa)})\big)^{T} &
\big(\breve{\Phi}_{\vec{x},s;\vec{x}\ppr,s\ppr}^{21}(\vartheta_{q_{\kappa}}^{(\kappa)})\big)^{T} \\
\big(\breve{\Phi}_{\vec{x},s;\vec{x}\ppr,s\ppr}^{12}(\vartheta_{q_{\kappa}}^{(\kappa)})\big)^{T} &
\big(\breve{\Phi}_{\vec{x},s;\vec{x}\ppr,s\ppr}^{22}(\vartheta_{q_{\kappa}}^{(\kappa)})\big)^{T}
\eea\right)_{\mbox{;}}   \\    \lb{s3_17}
\TRS\Big[\breve{\Phi}_{\vec{x},s;\vec{x}\ppr,s\ppr}^{ab}(\vartheta_{q_{\kappa}}^{(\kappa)})\Big]&=&
\trs\Big[\breve{\Phi}_{\vec{x},s;\vec{x}\ppr,s\ppr}^{11}(\vartheta_{q_{\kappa}}^{(\kappa)})\Big]+
\trs\Big[\breve{\Phi}_{\vec{x},s;\vec{x}\ppr,s\ppr}^{22}(\vartheta_{q_{\kappa}}^{(\kappa)})\Big] \;;  \\  \lb{s3_18}
\Big(\breve{\Phi}_{\vec{x},s;\vec{x}\ppr,s\ppr}^{ab}(\vartheta_{q_{\kappa}}^{(\kappa)})\Big)^{\sharp}&=&\left( \bea{cc}
\breve{\Phi}_{\vec{x},s;\vec{x}\ppr,s\ppr}^{11}(\vartheta_{q_{\kappa}}^{(\kappa)}) &
\breve{\Phi}_{\vec{x},s;\vec{x}\ppr,s\ppr}^{12}(\vartheta_{q_{\kappa}}^{(\kappa)}) \\
\breve{\Phi}_{\vec{x},s;\vec{x}\ppr,s\ppr}^{21}(\vartheta_{q_{\kappa}}^{(\kappa)}) &
\breve{\Phi}_{\vec{x},s;\vec{x}\ppr,s\ppr}^{22}(\vartheta_{q_{\kappa}}^{(\kappa)}) \eea\right)^{\sharp}= \left( \bea{cc}
\big(\breve{\Phi}_{\vec{x},s;\vec{x}\ppr,s\ppr}^{11}(\vartheta_{q_{\kappa}}^{(\kappa)})\big)^{\sharp} &
\big(\breve{\Phi}_{\vec{x},s;\vec{x}\ppr,s\ppr}^{21}(\vartheta_{q_{\kappa}}^{(\kappa)})\big)^{\sharp} \\
\big(\breve{\Phi}_{\vec{x},s;\vec{x}\ppr,s\ppr}^{12}(\vartheta_{q_{\kappa}}^{(\kappa)})\big)^{\sharp} &
\big(\breve{\Phi}_{\vec{x},s;\vec{x}\ppr,s\ppr}^{22}(\vartheta_{q_{\kappa}}^{(\kappa)})\big)^{\sharp}
\eea\right)_{\mbox{.}}
\eeq
We can straightforwardly generalize the coherent state path integral of
the one-particle case (\ref{s2_8}-\ref{s2_11}) with
operators \(\boldsymbol{\hat{\mfrak{O}}^{(k)}}(\hat{\psi}\pdag,\hat{\psi})\) to the delta functions of the operators
\(\boldsymbol{\hat{\mfrak{V}}^{(\kappa)}}(\hat{\psi}\pdag,\hat{\psi})\). Since the operators
\(\boldsymbol{\hat{\mfrak{V}}^{(\kappa)}}(\hat{\psi}\pdag,\hat{\psi})\) finally contain the second
quantized Fermi operators only in normal order as the one-particle operators, we achieve the
coherent state path integral (\ref{s3_21}) with the additional interaction part of four anti-commuting
fields
\beq\lb{s3_19}
\lefteqn{d[\psi^{*}(\vartheta_{q_{\kappa}}^{(\kappa)}),\psi(\vartheta_{q_{\kappa}}^{(\kappa)})] = \prod_{\{\vec{x},s\}}
\frac{d\psi_{\vec{x},s}^{*}(\vartheta_{q_{\kappa}}^{(\kappa)})\; \;
d\psi_{\vec{x},s}(\vartheta_{q_{\kappa}}^{(\kappa)})}{\mcal{N}_{x}} \;;} \\ \lb{s3_20}
\lefteqn{d[\psi^{*}(\mfrak{t}_{q_{\kappa}}^{(\kappa)}),\psi(\mfrak{t}_{q_{\kappa}}^{(\kappa)})] =
\prod_{0\leq\{\vartheta_{q_{\kappa}}^{(\kappa)}\}\leq \mfrak{t}_{q_{\kappa}}^{(\kappa)}} \;\;\prod_{\{\vec{x},s\}}
\frac{d\psi_{\vec{x},s}^{*}(\vartheta_{q_{\kappa}}^{(\kappa)})\; \;
d\psi_{\vec{x},s}(\vartheta_{q_{\kappa}}^{(\kappa)})}{\mcal{N}_{x}} \;;}  \\  \lb{s3_21}
\lefteqn{\big\langle\xi^{(\kappa+1)}\big|\delta\big(\mfrak{v}^{(\kappa)}-
\boldsymbol{\hat{\mfrak{V}}^{(\kappa)}}(\hat{\psi}\pdag,\hat{\psi})\,\big)\big|\xi^{(\kappa)}\big\rangle =  }  \\ \no &=&
\sum_{q_{\kappa}=\pm}\; \lim_{|\mfrak{e}_{q_{\kappa}}^{(\kappa)}|\rightarrow 0} \;\lim_{\mcal{T}^{(\kappa)}\rightarrow+\infty}
\int_{0}^{\mcal{T}^{(\kappa)}}\frac{d\mfrak{t}_{q_{\kappa}}^{(\kappa)}}{2\pi\,\hbar}\;
\big\langle\xi^{(\kappa+1)}\big|\overleftarrow{\exp}\Big\{-\im\,\zeta_{q_{\kappa}}^{(\kappa)}\:\frac{\mfrak{t}_{q_{\kappa}}^{(\kappa)}}{\hbar}\;
\Big(\mfrak{v}^{(\kappa)}-\boldsymbol{\hat{\mfrak{V}}^{(\kappa)}}(\hat{\psi}\pdag,\hat{\psi})-\im\:
\mfrak{e}_{q_{\kappa}}^{(\kappa)}\Big)\Big\}\big|\xi^{(\kappa)}\big\rangle   \\ \no &=&\sum_{q_{\kappa}=\pm}\;
\lim_{|\mfrak{e}_{q_{\kappa}}^{(\kappa)}|\rightarrow 0} \;\lim_{\mcal{T}^{(\kappa)}\rightarrow+\infty}
\int_{0}^{\mcal{T}^{(\kappa)}}\frac{d\mfrak{t}_{q_{\kappa}}^{(\kappa)}}{2\pi\,\hbar}\;
d[\psi^{*}(\mfrak{t}_{q_{\kappa}}^{(\kappa)}),\psi(\mfrak{t}_{q_{\kappa}}^{(\kappa)})]  \;\;
\exp\Big\{-\im\,\zeta_{q_{\kappa}}^{(\kappa)}\:\frac{\mfrak{t}_{q_{\kappa}}^{(\kappa)}}{\hbar}\:
\big(\mfrak{v}^{(\kappa)}-\im\,\mfrak{e}_{q_{\kappa}}^{(\kappa)}\big)\Big\}\;\times  \\ \no &\times&
\exp\bigg\{-\frac{\im}{\hbar}\int_{0}^{\mfrak{t}_{q_{\kappa}}^{(\kappa)}-\Delta\vartheta^{(\kappa)}}
\hspace*{-0.6cm} d\vartheta_{q_{\kappa}}^{(\kappa)}
\bigg(\sum_{\vec{y}}\psi_{\vec{y}}^{*}(\vartheta_{q_{\kappa}}^{(\kappa)}\!+\!\Delta\vartheta^{(\kappa)})\:(-\im\hbar)
\frac{\psi_{\vec{y}}(\vartheta_{q_{\kappa}}^{(\kappa)}\!+\!\Delta\vartheta^{(\kappa)})-
\psi_{\vec{y}}(\vartheta_{q_{\kappa}}^{(\kappa)})}{\Delta\vartheta^{(\kappa)}} +  \\ \no &-&\sum_{\vec{y};\vec{y}\ppr}
\psi_{\vec{y}\ppr}^{*}(\vartheta_{q_{\kappa}}^{(\kappa)}\!+\!\Delta\vartheta^{(\kappa)})\:\zeta_{q_{\kappa}}^{(\kappa)}\:
\hat{v}_{\vec{y}\ppr;\vec{y}}^{(\kappa)}\:\psi_{\vec{y}}(\vartheta_{q_{\kappa}}^{(\kappa)})\bigg)\bigg\}\;\times\;
\exp\Big\{\wt{\mfrak{A}}^{(\kappa)}(\xi^{(\kappa+1)*},\xi^{(\kappa)})\Big\}\,\times  \\ \no &\times&
\exp\bigg\{\frac{\im}{\hbar}\int_{0}^{\mfrak{t}_{q_{\kappa}}^{(\kappa)}-\Delta\vartheta^{(\kappa)}} \hspace*{-0.6cm}
d\vartheta_{q_{\kappa}}^{(\kappa)}\;\:\zeta_{q_{\kappa}}^{(\kappa)}\hspace*{-0.2cm}
\sum_{\vec{y}_{1/2};\vec{y}_{1/2}\ppr}\hspace*{-0.2cm}
\hat{\mscr{V}}_{\vec{y}_{2},\vec{y}_{2}\ppr;\vec{y}_{1}\ppr,\vec{y}_{1}}^{(\kappa)}\;
\psi_{\vec{y}_{2}}^{*}(\vartheta_{q_{\kappa}}^{(\kappa)}\!+\!\Delta\vartheta^{(\kappa)})\;\;
\psi_{\vec{y}_{1}}(\vartheta_{q_{\kappa}}^{(\kappa)})\;\;\;
\psi_{\vec{y}_{2}\ppr}^{*}(\vartheta_{q_{\kappa}}^{(\kappa)}\!+\!\Delta\vartheta^{(\kappa)})\;\;
\psi_{\vec{y}_{1}\ppr}(\vartheta_{q_{\kappa}}^{(\kappa)})\bigg\}\;.
\eeq
Apart from the two-particle part, the coherent state path integral (\ref{s3_21}) with
\(\boldsymbol{\hat{\mfrak{V}}^{(\kappa)}}(\hat{\psi}\pdag,\hat{\psi})\) encompasses the analogous terms
as e.\ g.\ the source action \(\wt{\mfrak{A}}^{(\kappa)}\big(\xi^{(\kappa+1)*},\xi^{(\kappa)}\big)\) (\ref{s3_22})
with additional Gaussian term
for the 'zero' time fields in order to achieve the precise discretization so that each field from the overcomplete sets
is exactly contained in (\ref{s3_21}) within the normal ordered operators
\(\boldsymbol{\hat{\mfrak{V}}^{(\kappa)}}(\hat{\psi}\pdag,\hat{\psi})\) of the total Hamiltonian and the
absolute values of orbital and spin angular momentum
\beq\lb{s3_22}\hspace*{-0.6cm}
\exp\Big\{\wt{\mfrak{A}}^{(\kappa)}\big(\xi^{(\kappa+1)*},\xi^{(\kappa)}\big)\Big\} &\hspace*{-0.5cm}=&\hspace*{-0.5cm}
\exp\bigg\{\mfrak{A}^{(\kappa)}\big(\xi^{(\kappa+1)*},\xi^{(\kappa)}\big)-\sum_{\vec{y}}
\psi_{\vec{y}}^{*}(\vartheta_{q_{\kappa}}^{(\kappa)}\!=\!0)\;\;
\psi_{\vec{y}}(\vartheta_{q_{\kappa}}^{(\kappa)}\!=\!0)\bigg\}  \;\;; \\ \no
\mfrak{A}^{(\kappa)}\big(\xi^{(\kappa+1)*},\xi^{(\kappa)}\big) &=&
\sum_{\vec{y}}\Big(\psi_{\vec{y}}^{*}(\vartheta_{q_{\kappa}}^{(\kappa)}\!=\!0)\;\;\xi_{\vec{y}}^{(\kappa)}+
\xi_{\vec{y}}^{(\kappa+1)*}\;\;\psi_{\vec{y}}(\vartheta_{q_{\kappa}}^{(\kappa)}\!=\!\mfrak{t}_{q_{\kappa}}^{(\kappa)})\Big)\;;  \\  \lb{s3_23}
\psi_{\vec{x},s}(\vartheta_{q_{\kappa}}^{(\kappa)}) &\in&\psi_{\vec{x},s}(\vartheta_{q_{\kappa}}^{(\kappa)}\!=\!0)\,,\,
\psi_{\vec{x},s}(\vartheta_{q_{\kappa}}^{(\kappa)}\!=\!\Delta\vartheta^{(\kappa)})\,,\, \ldots\,,\,
\psi_{\vec{x},s}(\vartheta_{q_{\kappa}}^{(\kappa)}\!=\!\mfrak{t}_{q_{\kappa}}^{(\kappa)})\;; \\  \lb{s3_24}
\mfrak{N}_{q_{\kappa}}^{(\kappa)}&=&\mfrak{t}_{q_{\kappa}}^{(\kappa)}/\Delta\vartheta^{(\kappa)}\;.
\eeq
However, the two-particle interaction term of four Fermi fields allows for nontrivial anomalous pairings of fields,
as e.\ g.\ in the order parameters (\ref{s3_8},\ref{s3_11}) from the dyadic products,
so that we have also to take the anomalous doubling
of the one-particle parts of (\ref{s3_21}) which is combined into the anomalous doubled,
one-particle Hamiltonian
\(\breve{\mfrak{H}}_{\vec{x}\ppr,s\ppr;\vec{x},s}^{(\kappa)ba}(\vartheta_{q_{\kappa}}^{(\kappa)\bprime}\boldsymbol{|}
\vartheta_{q_{\kappa}}^{(\kappa)})\) (\ref{s3_25}) with one-particle matrix elements
\(\hat{v}_{\vec{x}\ppr,s\ppr;\vec{x},s}^{(\kappa)}\) (\ref{s1_36},\ref{s1_38},\ref{s1_41}) and Kronecker deltas
\(\delta(\vartheta_{q_{\kappa}}^{(\kappa)\bprime}\boldsymbol{|}\vartheta_{q_{\kappa}}^{(\kappa)})\) of generalized time variables
'\(\vartheta_{q_{\kappa}}^{(\kappa)\bprime}\)', '\(\vartheta_{q_{\kappa}}^{(\kappa)}\)'
\beq \lb{s3_25}\hspace*{-0.8cm}
\breve{\mfrak{H}}_{\vec{x}\ppr,s\ppr;\vec{x},s}^{(\kappa)ba}(\vartheta_{q_{\kappa}}^{(\kappa)\bprime}
\boldsymbol{|}\vartheta_{q_{\kappa}}^{(\kappa)})
\hspace*{-0.3cm}&=&\hspace*{-0.3cm}
\mbox{diag}\Big(\breve{\mfrak{H}}_{\vec{x}\ppr,s\ppr;
\vec{x},s}^{(\kappa)11}(\vartheta_{q_{\kappa}}^{(\kappa)\bprime}\boldsymbol{|}\vartheta_{q_{\kappa}}^{(\kappa)})
\;\boldsymbol{;}\;
\breve{\mfrak{H}}_{\vec{x}\ppr,s\ppr;\vec{x},s}^{(\kappa)22}(\vartheta_{q_{\kappa}}^{(\kappa)\bprime}
\boldsymbol{|}\vartheta_{q_{\kappa}}^{(\kappa)})\Big)
\;; \\  \lb{s3_26}
\hspace*{-0.8cm}\breve{\mfrak{H}}_{\vec{x}\ppr,s\ppr;
\vec{x},s}^{(\kappa)11}(\vartheta_{q_{\kappa}}^{(\kappa)\bprime}\boldsymbol{|}\vartheta_{q_{\kappa}}^{(\kappa)})
\hspace*{-0.3cm}&=&\hspace*{-0.3cm}
\frac{\delta_{\vec{x}\ppr,\vec{x}}\,\delta_{s\ppr s}}{\big(\Delta\vartheta^{(\kappa)}\big)^{2}}\:
\big[\delta(\vartheta_{q_{\kappa}}^{(\kappa)\bprime}\boldsymbol{|}\vartheta_{q_{\kappa}}^{(\kappa)}\!-\!\Delta\vartheta^{(\kappa)})-
\delta(\vartheta_{q_{\kappa}}^{(\kappa)\bprime}\boldsymbol{|}\vartheta_{q_{\kappa}}^{(\kappa)})\big]-\frac{\im}{\hbar}\:
\frac{\zeta_{q_{\kappa}}^{(\kappa)}}{\Delta\vartheta^{(\kappa)}\; \mcal{N}_{x}}\:\hat{v}_{\vec{x}\ppr,s\ppr;\vec{x},s}^{(\kappa)}\;
\delta(\vartheta_{q_{\kappa}}^{(\kappa)\bprime}\boldsymbol{|}\vartheta_{q_{\kappa}}^{(\kappa)}) \;;   \\  \lb{s3_27}
\hspace*{-0.8cm}\breve{\mfrak{H}}_{\vec{x}\ppr,s\ppr;
\vec{x},s}^{(\kappa)22}(\vartheta_{q_{\kappa}}^{(\kappa)\bprime}\boldsymbol{|}\vartheta_{q_{\kappa}}^{(\kappa)})
\hspace*{-0.3cm}&=&\hspace*{-0.3cm}
\frac{\delta_{\vec{x}\ppr,\vec{x}}\,\delta_{s\ppr s}}{\big(\Delta\vartheta^{(\kappa)}\big)^{2}}\:
\big[\delta(\vartheta_{q_{\kappa}}^{(\kappa)\bprime}\!-\!\Delta\vartheta^{(\kappa)} \boldsymbol{|}\vartheta_{q_{\kappa}}^{(\kappa)})-
\delta(\vartheta_{q_{\kappa}}^{(\kappa)\bprime}\boldsymbol{|}\vartheta_{q_{\kappa}}^{(\kappa)})\big]-\frac{\im}{\hbar}\:
\frac{\zeta_{q_{\kappa}}^{(\kappa)}}{\Delta\vartheta^{(\kappa)}\; \mcal{N}_{x}}
\:\hat{v}_{\vec{x}\ppr,s\ppr;\vec{x},s}^{(\kappa)\boldsymbol{T}}\;
\delta(\vartheta_{q_{\kappa}}^{(\kappa)\bprime}\boldsymbol{|}\vartheta_{q_{\kappa}}^{(\kappa)})   \\  \no \hspace*{-0.3cm}&=&\hspace*{-0.3cm}
\Big(\breve{\mfrak{H}}_{\vec{x}\ppr,s\ppr;
\vec{x},s}^{(\kappa)11}(\vartheta_{q_{\kappa}}^{(\kappa)\bprime}\boldsymbol{|}\vartheta_{q_{\kappa}}^{(\kappa)})\Big)^{T}\;.
\eeq
In advance we emphasize that one has to take into account an important, further correction which concerns
the precise discretization of anomalous doubled operators as the doubled Hamiltonian
\(\breve{\mfrak{H}}_{\vec{x}\ppr,s\ppr;\vec{x},s}^{(\kappa)ba}(\vartheta_{q_{\kappa}}^{(\kappa)\bprime}
\boldsymbol{|}\vartheta_{q_{\kappa}}^{(\kappa)})\) and the anomalous doubled self-energies. In order to preserve the exact
discretization for a normal ordered operator, we have to define the extended, anomalous doubled fields
\(\mathring{\Psi}_{\vec{y}}^{a}(\vartheta_{q_{\kappa}}^{(\kappa)})\) (\ref{s3_28},\ref{s3_29})
with a further change of the time shift corrections in
the hermitian conjugation '\(^{\sharp}\)'. We label the discrete steps of generalized time variables
\(\vartheta_{q_{\kappa}}^{(\kappa)}\) by \(\Delta\vartheta^{(\kappa)}\cdot \mfrak{n}_{q_{\kappa}}^{(\kappa)}\) with
\(\mfrak{n}_{q_{\kappa}}^{(\kappa)}\in\mathbb{N}_{0}=[0,1,2,\ldots,\mfrak{N}_{q_{\kappa}}^{(\kappa)}]\) and the corresponding end point
\(\mfrak{t}_{q_{\kappa}}^{(\kappa)}=\Delta\vartheta^{(\kappa)}\cdot\mfrak{N}_{q_{\kappa}}^{(\kappa)}\) in order to present the precise form of
anomalous doubled fields \(\mathring{\Psi}_{\vec{y}}^{a}(\vartheta_{q_{\kappa}}^{(\kappa)})\) for an exact discretization
and an exact coset decomposition (Note that we have introduced a different symbol '\(\mathring{\ph{\Psi}}\)'
above \(\mathring{\Psi}_{\vec{y}}^{a}(\vartheta_{q_{\kappa}}^{(\kappa)})\) to be distinguished from '\(\breve{\ph{\Psi}}\)' in
\(\breve{\Psi}_{\vec{y}}^{a}(\vartheta_{q_{\kappa}}^{(\kappa)})\) and
\(\breve{\Psi}_{\vec{y}}^{\sharp a}(\vartheta_{q_{\kappa}}^{(\kappa)})\) of eqs. (\ref{s3_3},\ref{s3_4}))
\beq \lb{s3_28}
\mathring{\Psi}_{\vec{y}}^{a}(\vartheta_{q_{\kappa}}^{(\kappa)}) &=&
\left(\bea{c}\psi_{\vec{y}}^{*}(0\cdot\Delta\vartheta^{(\kappa)}) \\ \breve{\Psi}_{\vec{y}}^{a}(0\leq\vartheta_{q_{\kappa}}^{(\kappa)}\leq
\mfrak{t}_{q_{\kappa}}^{(\kappa)}-\Delta\vartheta^{(\kappa)}) \\ \psi_{\vec{y}}(\mfrak{t}_{q_{\kappa}}^{(\kappa)}=\mfrak{N}_{q_{\kappa}}^{(\kappa)}\cdot\Delta\vartheta^{(\kappa)})
\eea\right) =
\left(\bea{c}\mathring{\Psi}_{\vec{y}}^{(a\equiv1)}(\vartheta_{q_{\kappa}}^{(\kappa)}=-\Delta\vartheta^{(\kappa)})
\\ \mathring{\Psi}_{\vec{y}}^{(a=1/2)}(0\leq\vartheta_{q_{\kappa}}^{(\kappa)}\leq
\mfrak{t}_{q_{\kappa}}^{(\kappa)}-\Delta\vartheta^{(\kappa)}) \\
\mathring{\Psi}_{\vec{y}}^{(a\equiv1)}(\mfrak{t}_{q_{\kappa}}^{(\kappa)})
\eea\right) =      \\ \no  &=&
\Big(\psi_{\vec{y}}^{*}(0\cdot\Delta\vartheta^{(\kappa)}) \,\boldsymbol{\big|}\,
\breve{\Psi}_{\vec{y}}^{a}(0\leq\vartheta_{q_{\kappa}}^{(\kappa)}\leq
\mfrak{t}_{q_{\kappa}}^{(\kappa)}-\Delta\vartheta^{(\kappa)}) \,\boldsymbol{\big|}\,
\psi_{\vec{y}}(\mfrak{t}_{q_{\kappa}}^{(\kappa)}=\mfrak{N}_{q_{\kappa}}^{(\kappa)}\cdot\Delta\vartheta^{(\kappa)}) \Big)^{T} \\ \no &=&
\bigg(\psi_{\vec{y}}^{*}\big(0_{q_{\kappa}}^{(\kappa)}\big) \,\boldsymbol{\Big|}\,
\underbrace{\psi_{\vec{y}}\big(0_{q_{\kappa}}^{(\kappa)}\big)}_{a=1}\,;\,
\underbrace{\psi_{\vec{y}}^{*}\big(1_{q_{\kappa}}^{(\kappa)}\;\Delta\vartheta^{(\kappa)}\big)}_{a=2}\,\boldsymbol{\Big|}\,
\underbrace{\psi_{\vec{y}}\big(1_{q_{\kappa}}^{(\kappa)}\;\Delta\vartheta^{(\kappa)}\big)}_{a=1}\,;\,
\underbrace{\psi_{\vec{y}}^{*}\big(2_{q_{\kappa}}^{(\kappa)}\;\Delta\vartheta^{(\kappa)}\big)}_{a=2}\,\boldsymbol{\Big|}\,\ldots
\\ \no &\ldots&\,\boldsymbol{\Big|}\,
\underbrace{\psi_{\vec{y}}\big(\mfrak{n}_{q_{\kappa}}^{(\kappa)}\cdot\Delta\vartheta^{(\kappa)}\big)}_{a=1}\,;\,
\underbrace{\psi_{\vec{y}}^{*}\big(\,(\mfrak{n}_{q_{\kappa}}^{(\kappa)}+1)\;\Delta\vartheta^{(\kappa)}\big)}_{a=2}\,\boldsymbol{\Big|}\,
\ldots \,\boldsymbol{\Big|}\,
\underbrace{\psi_{\vec{y}}\big(\mfrak{t}_{q_{\kappa}}^{(\kappa)}-\Delta\vartheta^{(\kappa)}\big)}_{a=1}\,;\,
\underbrace{\psi_{\vec{y}}^{*}\big(\mfrak{t}_{q_{\kappa}}^{(\kappa)}\big)}_{a=2}\,\boldsymbol{\Big|}\,
\psi_{\vec{y}}\big(\mfrak{t}_{q_{\kappa}}^{(\kappa)}\big)\bigg)^{T}\;;   \\   \lb{s3_29}
\mathring{\Psi}_{\vec{y}}^{\sharp a}(\vartheta_{q_{\kappa}}^{(\kappa)}) &=&
\left(\bea{c} \psi_{\vec{y}}^{*}(0\cdot\Delta\vartheta^{(\kappa)}) \\
\breve{\Psi}_{\vec{y}}^{\sharp a}(0\leq\vartheta_{q_{\kappa}}^{(\kappa)}\leq
\mfrak{t}_{q_{\kappa}}^{(\kappa)}-\Delta\vartheta^{(\kappa)})  \\
\psi_{\vec{y}}(\mfrak{t}_{q_{\kappa}}^{(\kappa)}=\mfrak{N}_{q_{\kappa}}^{(\kappa)}\cdot\Delta\vartheta^{(\kappa)}) \eea\right)^{T} =
\left(\bea{c}\mathring{\Psi}_{\vec{y}}^{\sharp(a\equiv1)}(\vartheta_{q_{\kappa}}^{(\kappa)}=-\Delta\vartheta^{(\kappa)})
\\ \mathring{\Psi}_{\vec{y}}^{\sharp(a=1/2)}(0\leq\vartheta_{q_{\kappa}}^{(\kappa)}\leq
\mfrak{t}_{q_{\kappa}}^{(\kappa)}-\Delta\vartheta^{(\kappa)}) \\
\mathring{\Psi}_{\vec{y}}^{\sharp(a\equiv1)}(\mfrak{t}_{q_{\kappa}}^{(\kappa)})
\eea\right) =   \\ \no &=&
\Big(\psi_{\vec{y}}^{*}(0\cdot\Delta\vartheta^{(\kappa)}) \,\boldsymbol{\big|}\,
\breve{\Psi}_{\vec{y}}^{\sharp a}(0\leq\vartheta_{q_{\kappa}}^{(\kappa)}\leq
\mfrak{t}_{q_{\kappa}}^{(\kappa)}-\Delta\vartheta^{(\kappa)}) \,\boldsymbol{\big|}\,
\psi_{\vec{y}}(\mfrak{t}_{q_{\kappa}}^{(\kappa)}=\mfrak{N}_{q_{\kappa}}^{(\kappa)}\cdot\Delta\vartheta^{(\kappa)}) \Big) \\ \no &=&
\bigg(\psi_{\vec{y}}^{*}\big(0_{q_{\kappa}}^{(\kappa)}\big) \,\boldsymbol{\Big|}\,
\underbrace{\psi_{\vec{y}}^{*}\big(1_{q_{\kappa}}^{(\kappa)}\;\Delta\vartheta^{(\kappa)}\big)}_{a=1}\,;\,
\underbrace{\psi_{\vec{y}}\big(0_{q_{\kappa}}^{(\kappa)}\big)}_{a=2}
\,\boldsymbol{\Big|}\,
\underbrace{\psi_{\vec{y}}^{*}\big(2_{q_{\kappa}}^{(\kappa)}\;\Delta\vartheta^{(\kappa)}\big)}_{a=1}\,;\,
\underbrace{\psi_{\vec{y}}\big(1_{q_{\kappa}}^{(\kappa)}\;\Delta\vartheta^{(\kappa)}\big)}_{a=2}
\,\boldsymbol{\Big|}\,\ldots
\\ \no &\ldots&\,\boldsymbol{\Big|}\,
\underbrace{\psi_{\vec{y}}^{*}\big(\,(\mfrak{n}_{q_{\kappa}}^{(\kappa)}+1)\;\Delta\vartheta^{(\kappa)}\big)}_{a=1}\,;\,
\underbrace{\psi_{\vec{y}}\big(\mfrak{n}_{q_{\kappa}}^{(\kappa)}\cdot\Delta\vartheta^{(\kappa)}\big)}_{a=2}
\,\boldsymbol{\Big|}\,
\ldots \,\boldsymbol{\Big|}\,
\underbrace{\psi_{\vec{y}}^{*}\big(\mfrak{t}_{q_{\kappa}}^{(\kappa)}\big)}_{a=1}\,;\,
\underbrace{\psi_{\vec{y}}\big(\mfrak{t}_{q_{\kappa}}^{(\kappa)}-\Delta\vartheta^{(\kappa)}\big)}_{a=2}\,\boldsymbol{\Big|}\,
\psi_{\vec{y}}\big(\mfrak{t}_{q_{\kappa}}^{(\kappa)}\big)\bigg)\;;
\eeq
Moreover,
we briefly list the detailed form of the corrected, anomalous doubled one-particle Hamiltonian
\(\mathring{\mfrak{H}}_{\vec{y}\ppr;\vec{y}}^{(\kappa)ba}(\vartheta_{q_{\kappa}}^{(\kappa)\bprime}
\boldsymbol{|}\vartheta_{q_{\kappa}}^{(\kappa)})\), corresponding to the fields
\(\mathring{\Psi}_{\vec{y}}^{a}(\vartheta_{q_{\kappa}}^{(\kappa)})\),
\(\mathring{\Psi}_{\vec{y}}^{\sharp a}(\vartheta_{q_{\kappa}}^{(\kappa)})\) in eqs. (\ref{s3_28},\ref{s3_29}). For a better appearance
we omit multiplicative factors of \((\Delta\vartheta^{(\kappa)})^{2}\) in the list (\ref{s3_30}) for the one-particle
Hamiltonian \(\mathring{\mfrak{H}}_{\vec{y}\ppr;\vec{y}}^{(\kappa)ba}(\vartheta_{q_{\kappa}}^{(\kappa)\bprime}
\boldsymbol{|}\vartheta_{q_{\kappa}}^{(\kappa)})\) with matrix entries of \(\breve{\mfrak{H}}_{\vec{y}\ppr;\vec{y}}^{(\kappa)ba}(\vartheta_{q_{\kappa}}^{(\kappa)\bprime}\boldsymbol{|}\vartheta_{q_{\kappa}}^{(\kappa)})\) (\ref{s3_26}-\ref{s3_27}). Moreover, note the abbreviation of the
discrete indices \(\dot{\mfrak{N}}_{q_{\kappa}}^{(\kappa)}=\mfrak{N}_{q_{\kappa}}^{(\kappa)}-1\), \(\ddot{\mfrak{N}}_{q_{\kappa}}^{(\kappa)}=
\mfrak{N}_{q_{\kappa}}^{(\kappa)}-2\) with one dot \(\dot{\mfrak{N}}_{q_{\kappa}}^{(\kappa)}\) and two dots
\(\ddot{\mfrak{N}}_{q_{\kappa}}^{(\kappa)}\) indicating the subtraction of the integer numbers one and two
from the discrete number \(\mfrak{N}_{q_{\kappa}}^{(\kappa)}\) for the endpoint \(\mfrak{t}_{q_{\kappa}}^{(\kappa)}\)
\beq\lb{s3_30}
\lefteqn{\mathring{\mfrak{H}}_{\vec{x}\ppr,s\ppr;\vec{x},s}^{(\kappa)ba}(\vartheta_{q_{\kappa}}^{(\kappa)\bprime}
\boldsymbol{|}\vartheta_{q_{\kappa}}^{(\kappa)}) = }  \\ \no \hspace*{-0.4cm}&\hspace*{-0.4cm}=&\hspace*{-0.4cm}\left[
\bea{ccc}
\bea{|c|c|c|}\hline
{\scrscr 0} & {\scrscr 1} & {\scrscr0} \\ \hline {\scrscr 0} &
{\scrscr\breve{\mfrak{H}}_{\vec{y}\ppr;\vec{y}}^{(\kappa)11}(0_{q_{\kappa}}^{(\kappa)\bprime}\boldsymbol{|}0_{q_{\kappa}}^{(\kappa)})} &
{\scrscr 0} \\ \hline {\scrscr-1} & \scrscr{0} &
{\scrscr\breve{\mfrak{H}}_{\vec{y}\ppr;\vec{y}}^{(\kappa)22}(0_{q_{\kappa}}^{(\kappa)\bprime}\boldsymbol{|}0_{q_{\kappa}}^{(\kappa)})  } \\ \hline
\eea  & \bea{|c|c|}\hline
{\scrscr 0} & {\scrscr0} \\ \hline
{\scrscr\breve{\mfrak{H}}_{\vec{y}\ppr;\vec{y}}^{(\kappa)11}(0_{q_{\kappa}}^{(\kappa)\bprime}\boldsymbol{|}1_{q_{\kappa}}^{(\kappa)})} &
{\scrscr 0} \\ \hline  \scrscr{0} &
{\scrscr\breve{\mfrak{H}}_{\vec{y}\ppr;\vec{y}}^{(\kappa)22}(0_{q_{\kappa}}^{(\kappa)\bprime}\boldsymbol{|}1_{q_{\kappa}}^{(\kappa)})  } \\ \hline
\eea  \\  \bea{|c|c|c|} \hline {\scrscr \ph{-}0} &
{\scrscr\breve{\mfrak{H}}_{\vec{y}\ppr;\vec{y}}^{(\kappa)11}(1_{q_{\kappa}}^{(\kappa)\bprime}\boldsymbol{|}0_{q_{\kappa}}^{(\kappa)})} &
{\scrscr 0} \\ \hline {\scrscr\ph{-}0} & \scrscr{0} &
{\scrscr\breve{\mfrak{H}}_{\vec{y}\ppr;\vec{y}}^{(\kappa)22}(1_{q_{\kappa}}^{(\kappa)\bprime}\boldsymbol{|}0_{q_{\kappa}}^{(\kappa)})  } \\ \hline
\eea   &  \ddots \\   \\ \ddots \hspace*{0.3cm}
{\scrscr \big(\mfrak{n}_{q_{\kappa}}^{(\kappa)\bprime}\,,\;\mfrak{n}_{q_{\kappa}}^{(\kappa)}
\in [0,1,2,\ldots,\mfrak{N}_{q_{\kappa}}^{(\kappa)}-1]\big)}\hspace*{-1.3cm} &
\bea{|c|c|}\hline
{\scrscr\breve{\mfrak{H}}_{\vec{y}\ppr;\vec{y}}^{(\kappa)11}(\mfrak{n}_{q_{\kappa}}^{(k)\bprime}
\boldsymbol{|} \mfrak{n}_{q_{\kappa}}^{(\kappa)})} & {\scrscr0} \\ \hline {\scrscr0} &
{\scrscr\breve{\mfrak{H}}_{\vec{y}\ppr;\vec{y}}^{(\kappa)22}(\mfrak{n}_{q_{\kappa}}^{(k)\bprime}
\boldsymbol{|}\mfrak{n}_{q_{\kappa}}^{(\kappa)})}   \\ \hline
\eea & \hspace*{-1.6cm}
{\scrscr \big(\mfrak{n}_{q_{\kappa}}^{(\kappa)\bprime}=\mfrak{n}_{q_{\kappa}}^{(\kappa)}\,\&\,
\mfrak{n}_{q_{\kappa}}^{(\kappa)\bprime}=\mfrak{n}_{q_{\kappa}}^{(\kappa)}\pm1\big)}\hspace*{0.3cm}
 \ddots \\        \\ &\ddots&
\bea{|c|c|c|}\hline
{\scrscr\breve{\mfrak{H}}_{\vec{y}\ppr;\vec{y}}^{(\kappa)11}(\ddot{\mfrak{N}}_{q_{\kappa}}^{(\kappa)\bprime}
\boldsymbol{|} \dot{\mfrak{N}}_{q_{\kappa}}^{(\kappa)})} & {\scrscr0} & {\scrscr0} \\ \hline {\scrscr0} &
{\scrscr\breve{\mfrak{H}}_{\vec{y}\ppr;\vec{y}}^{(\kappa)22}(\ddot{\mfrak{N}}_{q_{\kappa}}^{(\kappa)\bprime}
\boldsymbol{|}\dot{\mfrak{N}}_{q_{\kappa}}^{(\kappa)})}  & {\scrscr0} \\
\eea   \\  & \bea{|c|c|}\hline
{\scrscr\breve{\mfrak{H}}_{\vec{y}\ppr;\vec{y}}^{(\kappa)11}(\dot{\mfrak{N}}_{q_{\kappa}}^{(\kappa)\bprime}
\boldsymbol{|} \ddot{\mfrak{N}}_{q_{\kappa}}^{(\kappa)})} & {\scrscr0}  \\ \hline {\scrscr0} &
{\scrscr\breve{\mfrak{H}}_{\vec{y}\ppr;\vec{y}}^{(\kappa)22}(\dot{\mfrak{N}}_{q_{\kappa}}^{(\kappa)\bprime}
\boldsymbol{|}\ddot{\mfrak{N}}_{q_{\kappa}}^{(\kappa)})}   \\ \hline {\scrscr0} & {\scrscr0}  \\ \hline
\eea  &
\bea{|c|c|c|}\hline
{\scrscr\breve{\mfrak{H}}_{\vec{y}\ppr;\vec{y}}^{(\kappa)11}(\dot{\mfrak{N}}_{q_{\kappa}}^{(\kappa)\bprime}
\boldsymbol{|} \dot{\mfrak{N}}_{q_{\kappa}}^{(\kappa)})} & {\scrscr0} & {\scrscr1} \\ \hline {\scrscr0} &
{\scrscr\breve{\mfrak{H}}_{\vec{y}\ppr;\vec{y}}^{(\kappa)22}(\dot{\mfrak{N}}_{q_{\kappa}}^{(\kappa)\bprime}
\boldsymbol{|}\dot{\mfrak{N}}_{q_{\kappa}}^{(\kappa)})}  & {\scrscr0}
\\ \hline {\scrscr0} & {\scrscr-1} & {\scrscr0} \\ \hline
\eea
\eea\right]_{\mbox{;}}  \\ \lb{s3_31}
 && \breve{\mfrak{H}}_{\vec{x}\ppr,s\ppr;\vec{x},s}^{(\kappa)ba}(\vartheta_{q_{\kappa}}^{(\kappa)\bprime}
\boldsymbol{|}\vartheta_{q_{\kappa}}^{(\kappa)}) \;\mbox{ to be replaced by }\; \breve{\mfrak{H}}_{\vec{x}\ppr,s\ppr;\vec{x},s}^{(\kappa)ba}(\vartheta_{q_{\kappa}}^{(\kappa)\bprime}
\boldsymbol{|}\vartheta_{q_{\kappa}}^{(\kappa)})\:\cdot\:\big(\Delta\vartheta^{(\kappa)}\big)^{2}\;\mbox{ in above list (\ref{s3_30})} \;.
\eeq
As we multiply \(\mathring{\mfrak{H}}_{\vec{x}\ppr,s\ppr;\vec{x},s}^{(\kappa)ba}(\vartheta_{q_{\kappa}}^{(\kappa)\bprime}
\boldsymbol{|}\vartheta_{q_{\kappa}}^{(\kappa)})\) (\ref{s3_30}) by the adapted Pauli matrix \((\mathring{\tau}_{1})^{ba}\) (\ref{s3_32})
of 'Nambu' doubled space and by the modified metric tensor \((\mathring{S})^{ba}\) (\ref{s3_5},\ref{s3_6},\ref{s3_32}),
also adapted to the discrete form of the fields \(\mathring{\Psi}_{\vec{y}}^{a}(\vartheta_{q_{\kappa}}^{(\kappa)})\) (\ref{s3_28}),
\(\mathring{\Psi}_{\vec{y}}^{\sharp a}(\vartheta_{q_{\kappa}}^{(\kappa)})\) (\ref{s3_29})
\beq \lb{s3_32}  \hspace*{-0.3cm}
(\mathring{\tau_{1}})^{ba}&\hspace*{-0.3cm}:=&\hspace*{-0.3cm}
\left[\bea{ccccc}
\bea{|c|c|c|}\hline
{\scrscr 1} & {\scrscr 0} & {\scrscr0} \\ \hline
{\scrscr 0} & {\scrscr0} & {\scrscr 1} \\ \hline
{\scrscr0} & \scrscr{1} & {\scrscr0  } \\ \hline
\eea  &  \\ \vspace*{-0.3cm}   &  {\scrscr\ddots} & \\
{\scr\delta_{\vec{y}\ppr;\vec{y}}\;\times} &&
\underbrace{\bea{|c|c|}\hline {\scrscr0} & {\scrscr1} \\ \hline
{\scrscr1} & {\scrscr0} \\ \hline\eea}_{(\hat{\tau}_{1})^{ba}} & &
{\scr\times\;\delta(\vartheta_{q_{\kappa}}^{(\kappa)\bprime}\boldsymbol{|}\vartheta_{q_{\kappa}}^{(\kappa)})}
\\ \vspace*{-0.3cm} &&& {\scrscr\ddots} & \\  &&&&
\bea{|c|c|c|}\hline
{\scrscr0} & {\scrscr1} & {\scrscr0} \\ \hline {\scrscr1} &
{\scrscr0}  & {\scrscr0}
\\ \hline {\scrscr0} & {\scrscr0} & {\scrscr1} \\ \hline
\eea
\eea\right]_{\mbox{;}}\;
(\mathring{S})^{ba}:=
\left[\bea{ccccc}
\bea{|c|c|c|}\hline
{\scrscr 1} & {\scrscr 0} & {\scrscr0} \\ \hline
{\scrscr 0} & {\scrscr1} & {\scrscr 0} \\ \hline
{\scrscr0} & \scrscr{0} & {\scrscr-1} \\ \hline
\eea  &  \\  \vspace*{-0.3cm}  &  {\scrscr\ddots} & \\ {\scr\delta_{\vec{y}\ppr;\vec{y}}\;\times} &&
\underbrace{\bea{|c|c|}\hline {\scrscr1} & {\scrscr0} \\ \hline
{\scrscr0} & {\scrscr-1} \\ \hline\eea}_{(\hat{S})^{ba}} & &
{\scr\times\;\delta(\vartheta_{q_{\kappa}}^{(\kappa)\bprime}\boldsymbol{|}\vartheta_{q_{\kappa}}^{(\kappa)})}
 \\ \vspace*{-0.3cm} &&& {\scrscr\ddots} & \\  &&&&
\bea{|c|c|c|}\hline
{\scrscr1} & {\scrscr0} & {\scrscr0} \\ \hline {\scrscr0} &
{\scrscr-1}  & {\scrscr0}
\\ \hline {\scrscr0} & {\scrscr0} & {\scrscr1} \\ \hline
\eea
\eea\right]_{\mbox{.}}
\eeq
we accomplish an anti-symmetric matrix \((\,(\mathring{\tau}_{1})^{bb_{2}}(\mathring{S})^{b_{2}b_{1}}
\mathring{\mfrak{H}}_{\vec{x}\ppr,s\ppr;\vec{x},s}^{(\kappa)b_{1}a}(\vartheta_{q_{\kappa}}^{(\kappa)\bprime}
\boldsymbol{|}\vartheta_{q_{\kappa}}^{(\kappa)})\,)^{ba}\)
so that bilinear, anomalous doubled fields (\ref{s3_28}, \ref{s3_29}) can be removed by integration
in later steps of the following HST. This explains the particular, involved  appearance of the one-particle
Hamiltonian matrix \(\mathring{\mfrak{H}}_{\vec{x}\ppr,s\ppr;\vec{x},s}^{(\kappa)ba}(\vartheta_{q_{\kappa}}^{(\kappa)\bprime}
\boldsymbol{|}\vartheta_{q_{\kappa}}^{(\kappa)})\) (\ref{s3_30}). Note that the matrices \((\mathring{\tau_{1}})^{ba}\),
\((\mathring{S})^{ba}\) have their respective form with Pauli matrix \((\hat{\tau}_{1})^{ba}\) and 'Nambu' metric
\(\hat{S}^{ba}\) along the main diagonal, but differ by the additional unit matrix \(\hat{1}\) in the uppermost
left corner and lowest right corner just outside the range of fields
\(\breve{\Psi}_{\vec{y}}^{a}(\vartheta_{q_{\kappa}}^{(\kappa)})\), \(\breve{\Psi}_{\vec{y}}^{\sharp a}(\vartheta_{q_{\kappa}}^{(\kappa)})\)
(\ref{s3_1},\ref{s3_3}) with further adaption for the fields
\(\mathring{\Psi}_{\vec{y}}^{a}(\vartheta_{q_{\kappa}}^{(\kappa)})\),
\(\mathring{\Psi}_{\vec{y}}^{\sharp a}(\vartheta_{q_{\kappa}}^{(\kappa)})\) having extra entries
\(\psi_{\vec{y}}^{*}(0_{q_{\kappa}}^{(\kappa)})\), \(\psi_{\vec{y}}(\mfrak{t}_{q_{\kappa}}^{(\kappa)})\) at the beginning and end of their
arrays. We define these various, involved appearing matrices and arrays in order to testify by their usage
{\it the possibility of exact discrete time grids for a coset decomposition} enforced from
the original normal ordering of operators !

Similarly to relations (\ref{s3_25}-\ref{s3_27}),
we perform the anomalous doubling in the quartic interaction of Fermi fields so that two factors of
'\(\frac{1}{2}\)' have to be incorporated (cf. Refs. \cite{pop1,pop2})
\beq\lb{s3_33}
\lefteqn{\exp\bigg\{\frac{\im}{\hbar}\int_{0}^{\mfrak{t}_{q_{\kappa}}^{(\kappa)}-\Delta\vartheta^{(\kappa)}} \hspace*{-0.6cm}
d\vartheta_{q_{\kappa}}^{(\kappa)}\:\zeta_{q_{\kappa}}^{(\kappa)}
\sum_{\vec{y}_{1/2};\vec{y}_{1/2}\ppr}
\hat{\mscr{V}}_{\vec{y}_{2},\vec{y}_{2}\ppr;\vec{y}_{1}\ppr,\vec{y}_{1}}^{(\kappa)}\;
\psi_{\vec{y}_{2}}^{*}(\vartheta_{q_{\kappa}}^{(\kappa)}\!+\!\Delta\vartheta^{(\kappa)})\;\;
\psi_{\vec{y}_{1}}(\vartheta_{q_{\kappa}}^{(\kappa)})\;\;\;
\psi_{\vec{y}_{2}\ppr}^{*}(\vartheta_{q_{\kappa}}^{(\kappa)}\!+\!\Delta\vartheta^{(\kappa)})\;\;
\psi_{\vec{y}_{1}\ppr}(\vartheta_{q_{\kappa}}^{(\kappa)})\bigg\} \;= } \\ \no \hspace*{-0.7cm}&\hspace*{-0.4cm}=&\hspace*{-0.3cm}
\exp\bigg\{\hspace*{-0.1cm}\frac{\im}{\hbar}\int_{0}^{\mfrak{t}_{q_{\kappa}}^{(\kappa)}-\Delta\vartheta^{(\kappa)}} \hspace*{-0.9cm}
d\vartheta_{q_{\kappa}}^{(\kappa)}\:\zeta_{q_{\kappa}}^{(\kappa)}\sum_{i=i_{\kappa}}^{j_{\kappa}}
\sum_{\vec{y}_{1/2};\vec{y}_{1/2}\ppr} \hspace*{-0.2cm}
V_{|\vec{x}_{1}\ppr-\vec{x}_{1}|}^{(\kappa)} \;
\frac{1}{2}\breve{\Psi}_{\vec{y}_{2}}^{\sharp a}(\vartheta_{q_{\kappa}}^{(\kappa)})\;
\hat{\mscr{S}}_{\vec{y}_{2};\vec{y}_{1}}^{(\kappa;i);aa}\;\hat{S}^{aa}\;
\breve{\Psi}_{\vec{y}_{1}}^{a}(\vartheta_{q_{\kappa}}^{(\kappa)})\;
\frac{1}{2}\breve{\Psi}_{\vec{y}_{2}\ppr}^{\sharp b}(\vartheta_{q_{\kappa}}^{(\kappa)})\;
\hat{\mscr{S}}_{\vec{y}_{2}\ppr;\vec{y}_{1}\ppr}^{(\kappa;i);bb}\;\hat{S}^{bb}\;
\breve{\Psi}_{\vec{y}_{1}\ppr}^{b}(\vartheta_{q_{\kappa}}^{(\kappa)})\bigg\}_{.}
\eeq
In order to obtain the anomalous doubling in (\ref{s3_33}), we have to use the 'Nambu' doubled fields
\(\breve{\Psi}_{\vec{x},s}^{a}(\vartheta_{q_{\kappa}}^{(\kappa)})\),
\(\breve{\Psi}_{\vec{x},s}^{\sharp a}(\vartheta_{q_{\kappa}}^{(\kappa)})\) (\ref{s3_3},\ref{s3_4}) with the time shift
correction \(\Delta\vartheta^{(\kappa)}\) in the complex conjugated parts and have to consider the metric
\(\hat{S}^{ab}\) (\ref{s3_34}) of the anomalous doubling which has further to be factorized into
\(\hat{I}\cdot\hat{I}\) (\ref{s3_35},\ref{s3_36}) for the application of the Weyl unitary trick in the
coset decompositions \cite{Weyl1,BoYouHou1}. As we have already pointed out, the two-particle matrix elements
\(\hat{\mscr{V}}_{\vec{y}_{2},\vec{y}_{2}\ppr;\vec{y}_{1}\ppr,\vec{y}_{1}}^{(\kappa)}\) separate into two
independent factors for each orbital and spin angular momentum component \(i=1,2,3\) whereas we have a 'true'
interaction \(V_{|\vec{x}_{1}\ppr-\vec{x}_{1}|}^{(\kappa=0)}\) of Coulomb repulsion (\ref{s3_37}) between primed
\(\vec{x}_{1}\ppr\) and unprimed \(\vec{x}_{1}\) parts of two-particle matrix elements. This different
behaviour between the Hamiltonian part \(\boldsymbol{\hat{\mfrak{V}}^{(\kappa=0\simeq E)}}(\hat{\psi}\pdag,\hat{\psi})=
\boldsymbol{\hat{H}}(\hat{\psi}\pdag,\hat{\psi};B_{z})\) and the absolute values
\(\boldsymbol{\hat{\mfrak{V}}^{(\kappa=1)}}(\hat{\psi}\pdag,\hat{\psi})=\boldsymbol{\vec{L}}(\hat{\psi}\pdag,\hat{\psi})\boldsymbol{\cdot}\boldsymbol{\vec{L}}(\hat{\psi}\pdag,\hat{\psi})\),
\(\boldsymbol{\hat{\mfrak{V}}^{(\kappa=2)}}(\hat{\psi}\pdag,\hat{\psi})=\boldsymbol{\vec{S}}(\hat{\psi}\pdag,\hat{\psi})\boldsymbol{\cdot}\boldsymbol{\vec{S}}(\hat{\psi}\pdag,\hat{\psi})\)
of orbital and spin angular momentum is regarded by the prevailing, adapted matrix elements
\(\hat{\mscr{S}}_{\vec{x}_{2},s_{2};\vec{x}_{1},s_{1}}^{(\kappa;i);a=b}\) (\ref{s3_37}), (\ref{s3_38},\ref{s3_39}),
which are also doubled in anomalous kind (\ref{s3_40}) for the Fermi fields and which accompany the metric tensor
\(\hat{S}^{ab}\) for the anti-commuting property
\beq\lb{s3_34}
\hat{S}^{ab} &=& \delta_{ab}\;\;\mbox{diag}\Big(\hat{S}^{11}=
\hat{1}\;\boldsymbol{;}\;\hat{S}^{22}=-\hat{1}\Big)\;;  \\  \lb{s3_35}
\hat{S} &=& \hat{I}\cdot \hat{I}\;;  \\  \lb{s3_36}
\hat{I}^{ab} &=& \delta_{ab}\;\;\mbox{diag}\Big(\hat{I}^{11}=
\hat{1}\;\boldsymbol{;}\;\hat{I}^{22}=\hat{\im}\Big)\;;
\eeq\vspace*{-0.3cm}
\beq\lb{s3_37}
&\hspace*{-3.0cm}\bea{rclrcl}
\kappa&=&0 \;;& i&=&i_{0}=j_{0}\equiv0 \;; \\
V_{|\vec{x}_{1}\ppr-\vec{x}_{1}|}^{(\kappa=0)} &=&\frac{e^{2}}{4\pi\,\ve_{0}}\:
\frac{1}{|\vec{x}_{1}\ppr-\vec{x}_{1}|+k_{e}} \;;&
\delta_{\vec{x}_{2},\vec{x}_{1}}\;\mcal{N}_{x} &=&
\sum_{\vec{x}_{1}\ppr}\hat{V}^{(0);\boldsymbol{-1}}_{|\vec{x}_{2}-\vec{x}_{1}\ppr|}\;
\big(V_{|\vec{x}_{1}\ppr-\vec{x}_{1}|}^{(0)}-\im\;\mfrak{e}_{q_{0}}^{(0)}\;\delta_{\vec{x}_{1}\ppr,\vec{x}_{1}}\big)
\;; \\
\hat{\mscr{S}}_{\vec{x}_{2},s_{2};\vec{x}_{1},s_{1}}^{(\kappa=0;i=0);11} &=&
\delta_{\vec{x}_{2},\vec{x}_{1}}\,\mcal{N}_{x}\;\delta_{s_{2}s_{1}} \;; &
\hat{\mscr{S}}_{\vec{x}_{2}\ppr,s_{2}\ppr;\vec{x}_{1}\ppr,s_{1}\ppr}^{(\kappa=0;i=0);22}
&=&\delta_{\vec{x}_{2}\ppr,\vec{x}_{1}\ppr}\,\mcal{N}_{x}\;\delta_{s_{2}\ppr s_{1}\ppr} \;; \\
\eea& \\  \lb{s3_38}
&\bea{rclrcl}
\kappa&=&1 \;;& i_{1}&=&1 \;;\;\;\; j_{1}=3\;;\;\;\;i=1,2,3 \;; \\
V_{|\vec{x}_{1}\ppr-\vec{x}_{1}|}^{(\kappa=1)} &=&\hbar^{2} \;; &
\delta_{\vec{x}_{2},\vec{x}_{1}}\;\mcal{N}_{x}&=&
\sum_{\vec{x}_{1}\ppr}\hat{V}^{(1);\boldsymbol{-1}}_{|\vec{x}_{2}-\vec{x}_{1}\ppr|}\;
\big(V_{|\vec{x}_{1}\ppr-\vec{x}_{1}|}^{(1)}-\im\;\mfrak{e}_{q_{1}}^{(1)}\;
\delta_{\vec{x}_{1}\ppr,\vec{x}_{1}} \big) \;;     \\
\hat{\mscr{S}}_{\vec{x}_{2},s_{2};\vec{x}_{1},s_{1}}^{(\kappa=1;i);11} &=&{\ds\frac{1}{\hbar}}
\big(\vec{x}_{2}\times\vec{p}_{2}\big)_{i}\delta_{\vec{x}_{2},\vec{x}_{1}}\,\mcal{N}_{x}\;\delta_{s_{2}s_{1}} \;; &
\hat{\mscr{S}}_{\vec{x}_{2}\ppr,s_{2}\ppr;\vec{x}_{1}\ppr,s_{1}\ppr}^{(\kappa=1;i);22}
&=&{\ds\frac{1}{\hbar}}
\big(\vec{x}_{2}\ppr\times\vec{p}_{2}\ppr\big)_{i}^{T}\delta_{\vec{x}_{2}\ppr,\vec{x}_{1}\ppr}\,\mcal{N}_{x}\;
\delta_{s_{2}\ppr s_{1}\ppr} \;;
\eea& \\   \lb{s3_39}
&\hspace*{-0.8cm}\bea{rclrcl}
\kappa&=&2 \;;& i_{2}&=&1 \;;\;\;\; j_{2}=3\;;\;\;\;i=1,2,3 \;; \\
V_{|\vec{x}_{1}\ppr-\vec{x}_{1}|}^{(\kappa=2)} &=&\hbar^{2} \;; &
\delta_{\vec{x}_{2},\vec{x}_{1}}\;\mcal{N}_{x}&=&
\sum_{\vec{x}_{1}\ppr}\hat{V}^{(2);\boldsymbol{-1}}_{|\vec{x}_{2}-\vec{x}_{1}\ppr|}\;
\big(V_{|\vec{x}_{1}\ppr-\vec{x}_{1}|}^{(2)}-\im\;\mfrak{e}_{q_{2}}^{(2)}\;
\delta_{\vec{x}_{1}\ppr,\vec{x}_{1}} \big) \;;     \\
\hat{\mscr{S}}_{\vec{x}_{2},s_{2};\vec{x}_{1},s_{1}}^{(\kappa=2;i);11} &=&
\frac{1}{2}\:\big(\hat{\sigma}_{i}\big)_{s_{2}s_{1}}
\;\delta_{\vec{x}_{2},\vec{x}_{1}}\,\mcal{N}_{x} \;; &
\hat{\mscr{S}}_{\vec{x}_{2}\ppr,s_{2}\ppr;\vec{x}_{1}\ppr,s_{1}\ppr}^{(\kappa=2;i);22}
&=&\frac{1}{2}\:\big(\hat{\sigma}_{i}\big)_{s_{2}\ppr s_{1}\ppr}^{T}
\;\delta_{\vec{x}_{2}\ppr,\vec{x}_{1}\ppr}\,\mcal{N}_{x} \;;
\eea& \\  \lb{s3_40}
&\hat{\mscr{S}}_{\vec{x}_{2},s_{2};\vec{x}_{1},s_{1}}^{(\kappa;i);22} = \Big(
\hat{\mscr{S}}_{\vec{x}_{2},s_{2};\vec{x}_{1},s_{1}}^{(\kappa;i);11}\Big)^{T}\;.&
\eeq
The above relations (\ref{s3_34}-\ref{s3_40}) with (\ref{s3_25}-\ref{s3_33}) allow to transform the
two-particle delta functions (\ref{s3_21}) to their anomalous doubled form. This is first determined
for the Hamiltonian delta function (\ref{s3_41}) because one can straightforwardly follow subsequent steps
in correspondence to Refs. \cite{physica6,pop1,pop2,precisecoh1}, due to the 'true' interaction term
(\ref{s3_37}) without factoring into primed and unprimed labels. According to the considerations for an exact
discrete time grid with eqs. (\ref{s3_28}-\ref{s3_32}),
we point out the occurrence of the two anomalous doubled fields
\(\mathring{\Psi}_{\vec{x},s}^{a}(\vartheta_{q_{E}}^{(E)})\),
\(\breve{\Psi}_{\vec{x},s}^{a}(\vartheta_{q_{E}}^{(E)})\) with matrix
\(\mathring{\mfrak{H}}_{\vec{x}\ppr,s\ppr;\vec{x},s}^{(E)ba}(\vartheta_{q_{E}}^{(E)\bprime}\boldsymbol{|}
\vartheta_{q_{E}}^{(E)})\) (\ref{s3_30},\ref{s3_31})
having entries with \(\breve{\mfrak{H}}_{\vec{x}\ppr,s\ppr;
\vec{x},s}^{(E)ba}(\vartheta_{q_{E}}^{(E)\bprime}\boldsymbol{|}
\vartheta_{q_{E}}^{(E)})\cdot(\Delta\vartheta^{(0)})^{2}\) (\ref{s3_25}-\ref{s3_27})
\beq\lb{s3_41}
\lefteqn{\big\langle\xi^{(1)}\big|\delta\big(E-
\boldsymbol{\hat{H}}(\hat{\psi}\pdag,\hat{\psi};B_{z})\,\big)\big|\xi^{(0)}\big\rangle =  }  \\ \no &=& \sum_{q_{E}=\pm}\;
\lim_{|\mfrak{e}_{q_{E}}^{(E)}|\rightarrow 0} \;\lim_{\mcal{T}^{(E)}\rightarrow+\infty}
\int_{0}^{\mcal{T}^{(E)}}\frac{d\mfrak{t}_{q_{E}}^{(E)}}{2\pi\,\hbar}\;
\big\langle\xi^{(1)}\big|\overleftarrow{\exp}\Big\{-\im\,\zeta_{q_{E}}^{(E)}\:\frac{\mfrak{t}_{q_{E}}^{(E)}}{\hbar}\;
\Big(E-\boldsymbol{\hat{H}^{(k)}}(\hat{\psi}\pdag,\hat{\psi};B_{z})-\im\:
\mfrak{e}_{q_{E}}^{(E)}\Big)\Big\}\big|\xi^{(0)}\big\rangle   \\ \no &=&\sum_{q_{E}=\pm}\;
\lim_{|\mfrak{e}_{q_{E}}^{(E)}|\rightarrow 0} \;\lim_{\mcal{T}^{(E)}\rightarrow+\infty}
\int_{0}^{\mcal{T}^{(E)}}\frac{d\mfrak{t}_{q_{E}}^{(E)}}{2\pi\,\hbar}\;
d[\psi^{*}(\mfrak{t}_{q_{E}}^{(E)}),\psi(\mfrak{t}_{q_{E}}^{(E)})]  \;\;
\exp\Big\{-\im\,\zeta_{q_{E}}^{(E)}\:\frac{\mfrak{t}_{q_{E}}^{(E)}}{\hbar}\:
\big(E-\im\,\mfrak{e}_{q_{E}}^{(E)}\big)\Big\}\;\times  \\ \no &\times&
\exp\bigg\{-\frac{1}{2}\int_{-\Delta\vartheta^{(E)}}^{\mfrak{t}_{q_{E}}^{(E)}} \hspace*{-0.3cm}
 d\vartheta_{q_{E}}^{(E)}\;d\vartheta_{q_{E}}^{(E)\bprime}
\sum_{\vec{x},s;\vec{x}\ppr,s\ppr}\mathring{\Psi}_{\vec{x}\ppr,s\ppr}^{\sharp b}(\vartheta_{q_{E}}^{(E)\bprime})\;
\mathring{S}\;\mathring{\mfrak{H}}_{\vec{x}\ppr,s\ppr;
\vec{x},s}^{(E)ba}(\vartheta_{q_{E}}^{(E)\bprime}\boldsymbol{|}\vartheta_{q_{E}}^{(E)})\;
\mathring{\Psi}_{\vec{x},s}^{a}(\vartheta_{q_{E}}^{(E)})\bigg\}\;
\exp\Big\{\mfrak{A}^{(0)}(\xi^{(1)*},\xi^{(0)})\Big\}   \\ \no &\times&
\exp\bigg\{\frac{\im}{\hbar}\int_{0}^{\mfrak{t}_{q_{E}}^{(E)}-\Delta\vartheta^{(E)}} \hspace*{-0.6cm}
d\vartheta_{q_{E}}^{(E)}\;\zeta_{q_{E}}^{(E)}
\sum_{\vec{x},s;\vec{x}\ppr,s\ppr}
V_{|\vec{x}-\vec{x}\ppr|}^{(\kappa=0\simeq E)}\;
\frac{1}{2}\,\breve{\Psi}_{\vec{x},s}^{\sharp a}(\vartheta_{q_{E}}^{(E)})\;\hat{S}\;
\breve{\Psi}_{\vec{x},s}^{a}(\vartheta_{q_{E}}^{(E)})\;\;\;
\frac{1}{2}\,\breve{\Psi}_{\vec{x}\ppr,s\ppr}^{\sharp b}(\vartheta_{q_{E}}^{(E)})\;\hat{S}\;
\breve{\Psi}_{\vec{x}\ppr,s\ppr}^{b}(\vartheta_{q_{E}}^{(E)})\bigg\}\;.
\eeq
According to Refs. \cite{physica6,pop1,pop2,precisecoh1}, we introduce the self-energies
\(\wt{\Sigma}_{\vec{y};\vec{y}\ppr}^{ab}(\vartheta_{q_{E}}^{(E)})\),
\(\delta\wt{\Sigma}_{\vec{y};\vec{y}\ppr}^{ab}(\vartheta_{q_{E}}^{(E)})\),
\(\sigma_{D}^{(0)}\!(\vec{x},\vartheta_{q_{E}}^{(E)})\)  which are further split into density related terms with
\(\delta\hat{\Sigma}_{D;\vec{y}_{1};\vec{y}_{2}}^{aa}(\vartheta_{q_{E}}^{(E)})\),
\(\sigma_{D}^{(0)}\!(\vec{x},\vartheta_{q_{E}}^{(E)})\) and anomalous paired terms with coset matrices
\(\hat{T}_{\vec{y};\vec{y}\ppr}^{ab}(\vartheta_{q_{E}}^{(E)})=
(\,\exp\{-\hat{Y}_{\vec{y}_{1};\vec{y}_{2}}^{a_{1}\neq b_{1}}(\vartheta_{q_{E}}^{(E)})\}\,)_{
\vec{y};\vec{y}\ppr}^{ab}\) , \hspace*{0.3cm}
(\(\wt{\delta}_{\vec{y};\vec{y}\ppr}=\mcal{N}_{x}\;\delta_{\vec{y};\vec{y}\ppr}\))
\beq \lb{s3_42}
\wt{\Sigma}_{\vec{y};\vec{y}\ppr}^{ab}(\vartheta_{q_{E}}^{(E)}) &=&
\left(\bea{cc}
\sigma_{D}^{(0)}(\vec{x},\vartheta_{q_{E}}^{(E)})\;\delta_{\vec{y};\vec{y}\ppr}+
\delta\hat{\Sigma}_{\vec{y};\vec{y}\ppr}^{11}(\vartheta_{q_{E}}^{(E)})  &
\im\;\delta\hat{\Sigma}_{\vec{y};\vec{y}\ppr}^{12}(\vartheta_{q_{E}}^{(E)})  \\
\im\;\delta\hat{\Sigma}_{\vec{y};\vec{y}\ppr}^{21}(\vartheta_{q_{E}}^{(E)}) &
\sigma_{D}^{(0)}(\vec{x},\vartheta_{q_{E}}^{(E)})\;\delta_{\vec{y};\vec{y}\ppr}+
\delta\hat{\Sigma}_{\vec{y};\vec{y}\ppr}^{22}(\vartheta_{q_{E}}^{(E)}) \eea\right)^{ab}  \\ \no &=&
\sigma_{D}^{(0)}(\vec{x},\vartheta_{q_{E}}^{(E)})\;\delta_{ab}\;\delta_{\vec{y};\vec{y}\ppr}+
\delta\wt{\Sigma}_{\vec{y};\vec{y}\ppr}^{ab}(\vartheta_{q_{E}}^{(E)}) \\  \no &=&
\Big(\hat{T}(\vartheta_{q_{E}}^{(E)})\Big)_{\vec{y};\vec{y}_{1}}^{aa_{1}}
\bigg(\sigma_{D}^{(0)}(\vec{x}_{1},\vartheta_{q_{E}}^{(E)})\;\delta_{a_{1}a_{2}}\;
\wt{\delta}_{\vec{y}_{1};\vec{y}_{2}}+
\delta\hat{\Sigma}_{D;\vec{y}_{1};\vec{y}_{2}}^{a_{1}=a_{2}}(\vartheta_{q_{E}}^{(E)})\bigg)
\Big(\hat{T}^{-1}(\vartheta_{q_{E}}^{(E)})\Big)_{\vec{y}_{2};\vec{y}\ppr}^{a_{2}b}  =
\Big(\hat{T}(\vartheta_{q_{E}}^{(E)})\Big)_{\vec{y};\vec{y}_{1}}^{aa_{1}} \:\times \\ \no &\times&
\bigg(\sigma_{D}^{(0)}(\vec{x}_{1},\vartheta_{q_{E}}^{(E)})\;\delta_{a_{1}a_{2}}\;
\wt{\delta}_{\vec{y}_{1};\vec{y}_{2}} +
\hat{Q}_{\vec{y}_{1};\vec{y}_{2}\ppr}^{-1;a_{1}a_{1}}(\vartheta_{q_{E}}^{(E)})\;\;
\delta\hat{\Lambda}_{s_{2}\ppr}^{a_{1}=a_{2}}(\vec{x}_{2}\ppr,\vartheta_{q_{E}}^{(E)})\;\;
\hat{Q}_{\vec{y}_{2}\ppr;\vec{y}_{2}}^{a_{2}a_{2}}(\vartheta_{q_{E}}^{(E)})\bigg)_{
\vec{y}_{1};\vec{y}_{2}}^{a_{1}a_{2}}\!
\Big(\hat{T}^{-1}(\vartheta_{q_{E}}^{(E)})\Big)_{\vec{y}_{2};\vec{y}\ppr}^{a_{2}b}    \\ \no &=&
\Big(\hat{T}_{0}(\vartheta_{q_{E}}^{(E)})\Big)_{\vec{y};\vec{y}_{1}}^{aa_{1}}
\bigg(\sigma_{D}^{(0)}(\vec{x}_{1},\vartheta_{q_{E}}^{(E)})\;\delta_{a_{1}a_{2}}\;
\wt{\delta}_{\vec{y}_{1};\vec{y}_{2}}+
\delta\hat{\Lambda}_{s_{1}=s_{2}}^{a_{1}=a_{2}}(\vec{x}_{1},\vartheta_{q_{E}}^{(E)})\;
\wt{\delta}_{\vec{y}_{1};\vec{y}_{2}}\bigg)
\Big(\hat{T}_{0}^{-1}(\vartheta_{q_{E}}^{(E)})\Big)_{\vec{y}_{2};\vec{y}\ppr}^{a_{2}b} \;;  \\ \lb{s3_43}
\delta\hat{\Sigma}_{\vec{y};\vec{y}\ppr}^{11}(\vartheta_{q_{E}}^{(E)}) &=&\hspace*{-0.3cm}
\delta\hat{\Sigma}_{\vec{y};\vec{y}\ppr}^{11,\dagger}(\vartheta_{q_{E}}^{(E)}) \;;\;
\delta\hat{\Sigma}_{\vec{y};\vec{y}\ppr}^{22}(\vartheta_{q_{E}}^{(E)}) =
\delta\hat{\Sigma}_{\vec{y};\vec{y}\ppr}^{22,\dagger}(\vartheta_{q_{E}}^{(E)}) \;;  \hspace*{0.2cm}
\delta\hat{\Sigma}_{\vec{y};\vec{y}\ppr}^{22}(\vartheta_{q_{E}}^{(E)}) =-
\delta\hat{\Sigma}_{\vec{y};\vec{y}\ppr}^{11,T}(\vartheta_{q_{E}}^{(E)}) \,;   \\  \lb{s3_44}
\delta\hat{\Sigma}_{\vec{y};\vec{y}\ppr}^{12}(\vartheta_{q_{E}}^{(E)}) &=&\hspace*{-0.3cm}
-\delta\hat{\Sigma}_{\vec{y};\vec{y}\ppr}^{12,T}(\vartheta_{q_{E}}^{(E)})\;;\hspace*{0.1cm}
\delta\hat{\Sigma}_{\vec{y};\vec{y}}^{12}(\vartheta_{q_{E}}^{(E)})\equiv 0 \;;  \hspace*{0.2cm}
\delta\hat{\Sigma}_{\vec{y};\vec{y}\ppr}^{21}(\vartheta_{q_{E}}^{(E)}) =
\delta\hat{\Sigma}_{\vec{y};\vec{y}\ppr}^{12,\dagger}(\vartheta_{q_{E}}^{(E)}) \;;  \\ \lb{s3_45}
\delta\hat{\Sigma}_{\vec{y};\vec{y}\ppr}^{21}(\vartheta_{q_{E}}^{(E)}) &=&\hspace*{-0.3cm}
-\delta\hat{\Sigma}_{\vec{y};\vec{y}\ppr}^{21,T}(\vartheta_{q_{E}}^{(E)}) \;; \hspace*{0.1cm}
\delta\hat{\Sigma}_{\vec{y};\vec{y}}^{21}(\vartheta_{q_{E}}^{(E)})\equiv 0 \;.
\eeq
The self-energies (\ref{s3_42}-\ref{s3_45}) allow to conduct the HST transformation of quartic Fermi fields
in (\ref{s3_41}) so that bilinear, anomalous doubled Grassmann fields only remain in the one-particle part
with matrix elements \(\mathring{S}\;
\mathring{\mfrak{H}}_{\vec{x}\ppr,s\ppr;
\vec{x},s}^{(E)ba}(\vartheta_{q_{E}}^{(E)\bprime}\boldsymbol{|}\vartheta_{q_{E}}^{(E)})\)
(\ref{s3_25}-\ref{s3_27}) and in the density matrix
\(\breve{R}_{\vec{x},s;\vec{x}\ppr,s\ppr}^{ba}(\vartheta_{q_{E}}^{(E)})\) which has exactly the similar form as the
anomalous doubled order parameter \(\breve{\Phi}_{\vec{x},s;\vec{x}\ppr,s\ppr}^{ba}(\vartheta_{q_{E}}^{(E)})\)
(\ref{s3_8},\ref{s3_11}). We therefore accomplish relation (\ref{s3_46}) for the delta function with the total
energy term where various Gaussian integrals of combined self-energies (\ref{s3_42}-\ref{s3_45}) replace the
quartic interaction part with remaining bilinear integrations of Grassmann fields
\beq \lb{s3_46}
\lefteqn{\big\langle\xi^{(1)}\big|\delta\big(E-
\boldsymbol{\hat{H}}(\hat{\psi}\pdag,\hat{\psi};B_{z})\,\big)\big|\xi^{(0)}\big\rangle = }   \\ \no &=&
\sum_{q_{E}=\pm}\; \lim_{|\mfrak{e}_{q_{E}}^{(E)}|\rightarrow 0} \;\lim_{\mcal{T}^{(E)}\rightarrow+\infty}
\int_{0}^{\mcal{T}^{(E)}}\frac{d\mfrak{t}_{q_{E}}^{(E)}}{2\pi\,\hbar}\;
\big\langle\xi^{(1)}\big|\overleftarrow{\exp}\Big\{-\im\,\zeta_{q_{E}}^{(E)}\:\frac{\mfrak{t}_{q_{E}}^{(E)}}{\hbar}\;
\Big(E-\boldsymbol{\hat{H}^{(k)}}(\hat{\psi}\pdag,\hat{\psi};B_{z})-\im\:
\mfrak{e}_{q_{E}}^{(E)}\Big)\Big\}\big|\xi^{(0)}\big\rangle      \\ \no &=&\sum_{q_{E}=\pm}\;
\lim_{|\mfrak{e}_{q_{E}}^{(E)}|\rightarrow 0} \;\lim_{\mcal{T}^{(E)}\rightarrow+\infty}
\int_{0}^{\mcal{T}^{(E)}}\frac{d\mfrak{t}_{q_{E}}^{(E)}}{2\pi\,\hbar}\;
d[\psi^{*}(\mfrak{t}_{q_{E}}^{(E)}),\psi(\mfrak{t}_{q_{E}}^{(E)})]  \;\;
\exp\Big\{-\im\,\zeta_{q_{E}}^{(E)}\:\frac{\mfrak{t}_{q_{E}}^{(E)}}{\hbar}\:
\big(E-\im\,\mfrak{e}_{q_{E}}^{(E)}\big)\Big\}\;\times  \\ \no &\times&
\exp\bigg\{-\frac{1}{2}\int_{-\Delta\vartheta^{(E)}}^{\mfrak{t}_{q_{E}}^{(E)}}\!\!
d\vartheta_{q_{E}}^{(E)}\;d\vartheta_{q_{E}}^{(E)\bprime}
\sum_{\vec{x},s;\vec{x}\ppr,s\ppr}
\mathring{\Psi}_{\vec{x}\ppr,s\ppr}^{\sharp b}(\vartheta_{q_{E}}^{(E)\bprime})\;\mathring{S}\;
\mathring{\mfrak{H}}_{\vec{x}\ppr,s\ppr;\vec{x},s}^{(E)ba}(\vartheta_{q_{E}}^{(E)\bprime}
\boldsymbol{|}\vartheta_{q_{E}}^{(E)})\;
\mathring{\Psi}_{\vec{x},s}^{a}(\vartheta_{q_{E}}^{(E)})\bigg\}\;
\exp\Big\{\mfrak{A}^{(0)}(\xi^{(1)*},\xi^{(0)})\Big\}\; \times    \\ \no &\times&
\int\;\;d[\sigma_{D}^{(E)}(\vec{x},\mfrak{t}_{q_{E}}^{(E)})]\;\;
\exp\bigg\{-\frac{\im}{2\hbar}\int_{0}^{\mfrak{t}_{q_{E}}^{(E)}-\Delta\vartheta^{(E)}} \hspace*{-0.6cm}
d\vartheta_{q_{E}}^{(E)}\;\zeta_{q_{E}}^{(E)}
\sum_{\vec{x},\vec{x}\ppr}
\sigma_{D}^{(0)}(\vec{x}\ppr,\vartheta_{q_{E}}^{(E)})\;\hat{V}_{|\vec{x}\ppr-\vec{x}|}^{(E);\boldsymbol{-1}}\;
\sigma_{D}^{(0)}(\vec{x},\vartheta_{q_{E}}^{(E)})\bigg\}\;\times
\\ \no &\times&
\int d[\delta\wt{\Sigma}_{\vec{x},s;\vec{x}\ppr,s\ppr}^{ab}(\mfrak{t}_{q_{E}}^{(E)})]\;\;
\exp\Bigg\{\frac{\im}{4\hbar}
\int_{0}^{\mfrak{t}_{q_{E}}^{(E)}-\Delta\vartheta^{(E)}} \hspace*{-0.6cm}
d\vartheta_{q_{E}}^{(E)}\;\zeta_{q_{E}}^{(E)}\sum_{\vec{x},\vec{x}\ppr}
\TRS\bigg[\frac{\delta\wt{\Sigma}_{\vec{x},s;\vec{x}\ppr,s\ppr}^{ab}(\vartheta_{q_{E}}^{(E)})\;
\delta\wt{\Sigma}_{\vec{x}\ppr,s\ppr;\vec{x},s}^{ba}(\vartheta_{q_{E}}^{(E)})}{V_{|\vec{x}\ppr-\vec{x}|}^{(E)}-
\im\:\mfrak{e}_{q_{E}}^{(E)}\:\big(\delta_{a=b}-\delta_{a\neq b}\big)}
\bigg]\Bigg\}\times  \\ \no &\times&
\exp\Bigg\{-\frac{\im}{2\hbar}\int_{0}^{\mfrak{t}_{q_{E}}^{(E)}-\Delta\vartheta^{(E)}} \hspace*{-0.9cm}
d\vartheta_{q_{E}}^{(E)}\;\zeta_{q_{E}}^{(E)}
\sum_{\vec{x},\vec{x}\ppr} \TRS\Bigg[\left(\breve{R}_{\vec{y};\vec{y}\ppr}^{ba}(\vartheta_{q_{E}}^{(E)})
\right)\;\hat{I}\;
\left(  \delta_{\vec{y}\ppr;\vec{y}}\:\delta_{ab}\:\mcal{N}_{x}\:\sigma_{D}^{(0)}(\vec{x},\vartheta_{q_{E}}^{(E)})+
\delta\wt{\Sigma}_{\vec{y}\ppr;\vec{y}}^{ab}(\vartheta_{q_{E}}^{(E)})\:\right)\;\hat{I}\Bigg]\Bigg\}_{\mbox{;}}
\\  \lb{s3_47} &&
\hat{V}_{|\vec{x}\ppr-\vec{x}|}^{(E)ab}=V_{|\vec{x}\ppr-\vec{x}|}^{(E)}-\im\:\mfrak{e}_{q_{E}}^{(E)}
\:\big(\delta_{a=b}-\delta_{a\neq b}\big)\;\;\;;\;\;\;\mfrak{e}_{q_{E}}^{(E)}=\zeta_{q_{E}}^{(E)}\;
\big|\mfrak{e}_{q_{E}}^{(E)}\big|\;;  \\ \lb{s3_48} &&
\delta_{\vec{x}_{2},\vec{x}_{1}}\;\mcal{N}_{x} =
\sum_{\vec{x}_{1}\ppr}\hat{V}^{(0);\boldsymbol{-1}}_{|\vec{x}_{2}-\vec{x}_{1}\ppr|}\;
\big(V_{|\vec{x}_{1}\ppr-\vec{x}_{1}|}^{(0)}+\im\;\mfrak{e}_{q_{0}}^{(0)}\;
\delta_{\vec{x}_{1}\ppr,\vec{x}_{1}} \big)\;.
\eeq
We proceed to transform the delta functions of the absolute values of orbital and spin angular momentum
((\(\kappa=1,2\)) instead of (\(\kappa=0\simeq E\))) in analogous manner and define the similar, three component self-energies
\(\wt{\Sigma}_{\vec{y};\vec{y}\ppr}^{(i)ab}(\vartheta_{q_{\kappa}}^{(\kappa)})\),
\(\delta\wt{\Sigma}_{\vec{y};\vec{y}\ppr}^{(i)ab}(\vartheta_{q_{\kappa}}^{(\kappa)})\),
\(\sigma_{D}^{(i)}(\vartheta_{q_{\kappa}}^{(\kappa)})\) with \(i=1,2,3\); however, the appropriate coset decomposition
into block diagonal density terms \(\delta\hat{\Sigma}_{D;\vec{y}_{1};\vec{y}_{2}}^{aa}(\vartheta_{q_{\kappa}}^{(\kappa)})\),
as hinge fields in a SSB, and into anomalous terms with coset matrices
\(\hat{T}_{\vec{y};\vec{y}\ppr}^{ab}(\vartheta_{q_{\kappa}}^{(\kappa)})=
(\,\exp\{-\hat{Y}_{\vec{y}_{1};\vec{y}_{2}}^{a_{1}\neq b_{1}}(\vartheta_{q_{\kappa}}^{(\kappa)})\}\,)_{
\vec{y};\vec{y}\ppr}^{ab}\) (\(\kappa=1,2\)) has to be performed for the sum of the three component self-energies
\(\delta\wt{\Sigma}_{\vec{y};\vec{y}\ppr}^{(i)ab}(\vartheta_{q_{\kappa}}^{(\kappa)})\),
\(\sigma_{D}^{(i)}(\vartheta_{q_{\kappa}}^{(\kappa)})\) with \(i=1,2,3\)
in order to allow for a possible factoring of the coset matrices within the functional determinant and within
the corresponding propagator
\beq \lb{s3_49}
\wt{\Sigma}_{\vec{y}_{1}\ppr;\vec{y}_{1}}^{(i)ab}(\vartheta_{q_{\kappa}}^{(\kappa)})  &=&
\delta_{\vec{y}_{1}\ppr;\vec{y}_{2}}\;\delta_{\vec{y}_{2};\vec{y}_{1}}\;\delta_{ab}\;\mcal{N}_{x}^{2}\;
\sigma_{D}^{(i)}(\vartheta_{q_{\kappa}}^{(\kappa)})+
\delta\wt{\Sigma}_{\vec{y}_{1}\ppr;\vec{y}_{2}}^{(i)ab}(\vartheta_{q_{\kappa}}^{(\kappa)})\;
\hat{\mscr{S}}_{\vec{y}_{2};\vec{y}_{1}}^{(\kappa;i);bb}  \\  \lb{s3_50}
\lefteqn{\hspace*{-2.8cm}\sum_{i=i_{\kappa}}^{j_{\kappa}}
\hat{\mscr{S}}_{\vec{y};\vec{y}_{1}\ppr}^{(\kappa;i);aa}\;
\delta\wt{\Sigma}_{\vec{y}_{1}\ppr;\vec{y}_{2}}^{(i)ab}(\vartheta_{q_{\kappa}}^{(\kappa)})\;
\hat{\mscr{S}}_{\vec{y}_{2};\vec{y}\ppr}^{(\kappa;i);bb} =
\Big(\hat{T}(\vartheta_{q_{\kappa}}^{(\kappa)})\Big)_{\vec{y};\vec{y}_{1}\ppr}^{aa_{1}}\;
\delta\hat{\Sigma}_{D;\vec{y}_{1}\ppr;\vec{y}_{2}}^{a_{1}=a_{2}}(\vartheta_{q_{\kappa}}^{(\kappa)})\;
\Big(\hat{T}^{-1}(\vartheta_{q_{\kappa}}^{(\kappa)})\Big)_{\vec{y}_{2};\vec{y}\ppr}^{a_{2}b}     }   \\  \no &=&
\Big(\hat{T}(\vartheta_{q_{\kappa}}^{(\kappa)})\Big)_{\vec{y};\vec{y}_{1}}^{aa_{1}}
\bigg(\hat{Q}_{\vec{y}_{1};\vec{y}_{2}\ppr}^{-1;a_{1}a_{1}}(\vartheta_{q_{\kappa}}^{(\kappa)})\;\;
\delta\hat{\Lambda}_{\vec{y}_{2}\ppr}^{a_{1}=a_{2}}(\vartheta_{q_{\kappa}}^{(\kappa)})\;\;
\hat{Q}_{\vec{y}_{2}\ppr;\vec{y}_{2}}^{a_{2}a_{2}}(\vartheta_{q_{\kappa}}^{(\kappa)})\bigg)_{
\vec{y}_{1};\vec{y}_{2}}^{a_{1}a_{2}}\;
\Big(\hat{T}^{-1}(\vartheta_{q_{\kappa}}^{(\kappa)})\Big)_{\vec{y}_{2};\vec{y}\ppr}^{a_{2}b}  \\ \no &=&
\Big(\hat{T}_{0}(\vartheta_{q_{\kappa}}^{(\kappa)})\Big)_{\vec{y};\vec{y}_{1}}^{aa_{1}}\;
\delta\hat{\Lambda}_{\vec{y}_{1}=\vec{y}_{2}}^{a_{1}=a_{2}}(\vartheta_{q_{\kappa}}^{(\kappa)})\;
\Big(\hat{T}_{0}^{-1}(\vartheta_{q_{\kappa}}^{(\kappa)})\Big)_{\vec{y}_{2};\vec{y}\ppr}^{a_{2}b} \;;  \\ \lb{s3_51}
\delta\hat{\Sigma}_{\vec{y};\vec{y}\ppr}^{11}(\vartheta_{q_{\kappa}}^{(\kappa)}) &=&\hspace*{-0.3cm}
\delta\hat{\Sigma}_{\vec{y};\vec{y}\ppr}^{11,\dagger}(\vartheta_{q_{\kappa}}^{(\kappa)}) \;;\;
\delta\hat{\Sigma}_{\vec{y};\vec{y}\ppr}^{22}(\vartheta_{q_{\kappa}}^{(\kappa)}) =
\delta\hat{\Sigma}_{\vec{y};\vec{y}\ppr}^{22,\dagger}(\vartheta_{q_{\kappa}}^{(\kappa)}) \;;  \hspace*{0.2cm}
\delta\hat{\Sigma}_{\vec{y};\vec{y}\ppr}^{22}(\vartheta_{q_{\kappa}}^{(\kappa)}) =-
\delta\hat{\Sigma}_{\vec{y};\vec{y}\ppr}^{11,T}(\vartheta_{q_{\kappa}}^{(\kappa)}) \,;   \\  \lb{s3_52}
\delta\hat{\Sigma}_{\vec{y};\vec{y}\ppr}^{12}(\vartheta_{q_{\kappa}}^{(\kappa)}) &=&\hspace*{-0.3cm}
-\delta\hat{\Sigma}_{\vec{y};\vec{y}\ppr}^{12,T}(\vartheta_{q_{\kappa}}^{(\kappa)})\;;\hspace*{0.1cm}
\delta\hat{\Sigma}_{\vec{x},s;\vec{x},s}^{12}(\vartheta_{q_{\kappa}}^{(\kappa)})\equiv 0 \;;  \hspace*{0.2cm}
\delta\hat{\Sigma}_{\vec{y};\vec{y}\ppr}^{21}(\vartheta_{q_{\kappa}}^{(\kappa)}) =
\delta\hat{\Sigma}_{\vec{y};\vec{y}\ppr}^{12,\dagger}(\vartheta_{q_{\kappa}}^{(\kappa)}) \;;  \\ \lb{s3_53}
\delta\hat{\Sigma}_{\vec{y};\vec{y}\ppr}^{21}(\vartheta_{q_{\kappa}}^{(\kappa)}) &=&\hspace*{-0.3cm}
-\delta\hat{\Sigma}_{\vec{y};\vec{y}\ppr}^{21,T}(\vartheta_{q_{\kappa}}^{(\kappa)}) \;; \hspace*{0.1cm}
\delta\hat{\Sigma}_{\vec{x},s;\vec{x},s}^{21}(\vartheta_{q_{\kappa}}^{(\kappa)})\equiv 0 \;.
\eeq
According to the factorization (\ref{s3_33}) of the two-particle matrix elements (\ref{s3_38}-\ref{s3_40})
for the absolute values of the
orbital and spin angular momentum, the self-energy density \(\sigma_{D}^{(i)}(\vartheta_{q_{\kappa}}^{(\kappa)})\)
does not depend on the spatial vector \(\vec{x}\), but has to take three independent components \(i=1,2,3\)
which, however, have to be combined with
\(\delta\wt{\Sigma}_{\vec{y}_{1}\ppr;\vec{y}_{2}}^{(i)ab}(\vartheta_{q_{\kappa}}^{(\kappa)})\;
\hat{\mscr{S}}_{\vec{y}_{2};\vec{y}_{1}}^{(\kappa;i);bb}\) for an appropriate coset decomposition with factorization
of coset matrices of anomalous terms for the functional determinant and propagator
\beq \lb{s3_54}
\lefteqn{\big\langle\xi^{(\kappa+1)}\big|\delta\big(\mfrak{v}^{(\kappa)}-
\boldsymbol{\hat{\mfrak{V}}^{(\kappa)}}(\hat{\psi}\pdag,\hat{\psi})\,\big)\big|\xi^{(\kappa)}\big\rangle =  }  \\ \no &=&
\sum_{q_{\kappa}=\pm} \;\lim_{|\mfrak{e}_{q_{\kappa}}^{(\kappa)}|\rightarrow 0} \;\lim_{\mcal{T}^{(\kappa)}\rightarrow+\infty}
\int_{0}^{\mcal{T}^{(\kappa)}}\frac{d\mfrak{t}_{q_{\kappa}}^{(\kappa)}}{2\pi\,\hbar}\;
\big\langle\xi^{(\kappa+1)}\big|\overleftarrow{\exp}\Big\{-\im\,\zeta_{q_{\kappa}}^{(\kappa)}\:\frac{\mfrak{t}_{q_{\kappa}}^{(\kappa)}}{\hbar}\;
\Big(\mfrak{v}^{(\kappa)}-\boldsymbol{\hat{\mfrak{V}}^{(\kappa)}}(\hat{\psi}\pdag,\hat{\psi})-\im\:
\mfrak{e}_{q_{\kappa}}^{(\kappa)}\Big)\Big\}\big|\xi^{(\kappa)}\big\rangle   \\ \no &=&\sum_{q_{\kappa}=\pm}\;
\lim_{|\mfrak{e}_{q_{\kappa}}^{(\kappa)}|\rightarrow 0} \;\lim_{\mcal{T}^{(\kappa)}\rightarrow+\infty}
\int_{0}^{\mcal{T}^{(\kappa)}}\frac{d\mfrak{t}_{q_{\kappa}}^{(\kappa)}}{2\pi\,\hbar}\;
d[\psi^{*}(\mfrak{t}_{q_{\kappa}}^{(\kappa)}),\psi(\mfrak{t}_{q_{\kappa}}^{(\kappa)})]  \;\;
\exp\Big\{-\im\,\zeta_{q_{\kappa}}^{(\kappa)}\:\frac{\mfrak{t}_{q_{\kappa}}^{(\kappa)}}{\hbar}\:
\big(\mfrak{v}^{(\kappa)}-\im\,\mfrak{e}_{q_{\kappa}}^{(\kappa)}\big)\Big\}\;\times  \\ \no &\times&
\exp\bigg\{-\frac{1}{2}\int_{-\Delta\vartheta^{(\kappa)}}^{\mfrak{t}_{q_{\kappa}}^{(\kappa)}}
d\vartheta_{q_{\kappa}}^{(\kappa)}\;d\vartheta_{q_{\kappa}}^{(\kappa)\bprime}
\sum_{\vec{y};\vec{y}\ppr}\mathring{\Psi}_{\vec{y}\ppr}^{\sharp b}(\vartheta_{q_{\kappa}}^{(\kappa)\bprime})\;\mathring{S}\;
\mathring{\mfrak{H}}_{\vec{y}\ppr;\vec{y}}^{(\kappa)ba}(\vartheta_{q_{\kappa}}^{(\kappa)\bprime}
\boldsymbol{|}\vartheta_{q_{\kappa}}^{(\kappa)})\;
\mathring{\Psi}_{\vec{y}}^{a}(\vartheta_{q_{\kappa}}^{(\kappa)})\bigg\}\;
\exp\Big\{\mfrak{A}^{(\kappa)}(\xi^{(\kappa+1)*},\xi^{(\kappa)})\Big\}\;\times  \\ \no &\times&
\int d[\sigma_{D}^{(i)}(\mfrak{t}_{q_{\kappa}}^{(\kappa)})]\;
d[\delta\wt{\Sigma}_{\vec{y};\vec{y}\ppr}^{(i)ab}(\mfrak{t}_{q_{\kappa}}^{(\kappa)})]\;
\exp\Bigg\{-\frac{\im}{2\hbar}\int_{0}^{\mfrak{t}_{q_{\kappa}}^{(\kappa)}-\Delta\vartheta^{(\kappa)}} \hspace*{-0.6cm}
d\vartheta_{q_{\kappa}}^{(\kappa)}\;\zeta_{q_{\kappa}}^{(\kappa)}\sum_{i=i_{\kappa}}^{j_{\kappa}}\Bigg(\sum_{\vec{x},\vec{x}\ppr}
\sigma_{D}^{(i)}(\vec{x}\ppr,\vartheta_{q_{\kappa}}^{(\kappa)})\;\hat{V}_{|\vec{x}\ppr-\vec{x}|}^{(\kappa);\boldsymbol{-1}}
\;\sigma_{D}^{(i)}(\vec{x},\vartheta_{q_{\kappa}}^{(\kappa)})+  \\ \no &-&
\frac{1}{2}\sum_{\vec{x}_{1/2};\vec{x}\ppr_{1/2}}
\TRS\bigg[\frac{\delta\wt{\Sigma}_{\vec{y}_{1}\ppr;\vec{y}_{2}}^{(i)ab}(\vartheta_{q_{\kappa}}^{(\kappa)})\;
\hat{\mscr{S}}_{\vec{y}_{2};\vec{y}_{1}}^{(\kappa;i);bb}\;
\delta\wt{\Sigma}_{\vec{y}_{1};\vec{y}_{2}\ppr}^{(i)ba}(\vartheta_{q_{\kappa}}^{(\kappa)})\;
\hat{\mscr{S}}_{\vec{y}_{2}\ppr;\vec{y}_{1}\ppr}^{(\kappa;i);aa}}{V_{|\vec{x}_{1}\ppr-\vec{x}_{1}|}^{(\kappa)}-
\im\:\mfrak{e}_{q_{\kappa}}^{(\kappa)}
\:\big(\delta_{a=b}-\delta_{a\neq b}\big)}\bigg] +
\sum_{\vec{x}_{1/2};\vec{x}_{1/2}\ppr} \times \\ \no &\times&
\TRS\bigg[\breve{R}_{\vec{y}_{1};\vec{y}_{2}\ppr}^{ba}(\vartheta_{q_{\kappa}}^{(\kappa)})
\;\hat{I}\;\hat{\mscr{S}}_{\vec{y}_{2}\ppr;\vec{y}_{1}\ppr}^{(\kappa;i);aa}\;
\left(  \delta_{\vec{y}_{1}\ppr;\vec{y}_{2}}\;\delta_{\vec{y}_{2};\vec{y}_{1}}\;
\delta_{ab}\;\mcal{N}_{x}^{2}\;\sigma_{D}^{(i)}(\vec{x}_{2},\vartheta_{q_{\kappa}}^{(\kappa)})+
\delta\wt{\Sigma}_{\vec{y}_{1}\ppr;\vec{y}_{2}}^{(i)ab}(\vartheta_{q_{\kappa}}^{(\kappa)})\;
\hat{\mscr{S}}_{\vec{y}_{2};\vec{y}_{1}}^{(\kappa;i);bb}\right)\;\hat{I}\bigg]
\Bigg)  \Bigg\}\;;  \\  \lb{s3_55} &&
\hat{V}_{|\vec{x}\ppr-\vec{x}|}^{(\kappa)ab}=
\underbrace{V_{|\vec{x}\ppr-\vec{x}|}^{(\kappa)}}_{=\hbar^{2}\:,\;(\kappa=1,2)}-\im\:\mfrak{e}_{q_{\kappa}}^{(\kappa)}
\:\big(\delta_{a=b}-\delta_{a\neq b}\big)\;\;\;;\;\;\;\mfrak{e}_{q_{\kappa}}^{(\kappa)}=\zeta_{q_{\kappa}}^{(\kappa)}\;
\big|\mfrak{e}_{q_{\kappa}}^{(\kappa)}\big|\;\;;   \\  \lb{s3_56}  &&
\sigma_{D}^{(i)}(\vec{x},\vartheta_{q_{\kappa}}^{(\kappa)}) := \sigma_{D}^{(i)}(\vartheta_{q_{\kappa}}^{(\kappa)})\,;
\hspace*{0.1cm}\mbox{definition with independence on spatial coordinates and angular momentum}; \\ \lb{s3_57} &&
\delta_{\vec{x}_{2},\vec{x}_{1}}\;\mcal{N}_{x} =
\sum_{\vec{x}_{1}\ppr}\hat{V}^{(\kappa);\boldsymbol{-1}}_{|\vec{x}_{2}-\vec{x}_{1}\ppr|}\;
\big(V_{|\vec{x}_{1}\ppr-\vec{x}_{1}|}^{(\kappa)}+\im\;\mfrak{e}_{q_{\kappa}}^{(\kappa)}\;
\delta_{\vec{x}_{1}\ppr,\vec{x}_{1}} \big) \;.
\eeq
The above relation (\ref{s3_54}) is the result of HST transformations for the two-particle delta
functions which have been combined into similar notations for further reduction to anomalous pairs
in following section \ref{s32}.

\subsection{Coset decomposition of self-energies with anomalous pairs as remaining field degrees of freedom} \lb{s32}

The delta functions of the two-particle operators
\(\boldsymbol{\hat{\mfrak{V}}^{(\kappa=0,1,2)}}(\hat{\psi}\pdag,\hat{\psi})\) have been changed by HST's to
bilinear, anomalous doubled Fermi fields and self-energy matrices which have been separated into density
related and 'Nambu' related parts for anomalous pairs in a coset decomposition. As we take into account
the appropriate Jacobian for the coset decomposition from the invariant integration measure with its metric
tensor and project onto remaining coset degrees of freedom, we finally succeed in relation (\ref{s3_85}) for the delta
functions with two-particle terms after removal of bilinear Fermi fields by their integration. Note that we have
adapted notations of the Hamiltonian case (\(\kappa=0\simeq E\)) to the orbital and spin angular momentum case (\(\kappa=1,2\)).
Since the angular momentum cases (\(\kappa=1,2\)) involve the sum over three independent self-energy components
\(i=1,2,3\), we have to include delta functions (\ref{s3_58},\ref{s3_59})
which reduce these sums to a single self-energy so that a coset
decomposition and projection within the Fermi determinant can be obtained for a factorization of the complete
self-energy parts (compare Refs. \cite{pop1,precisecoh1} for calculating the Jacobian)
\beq \lb{s3_58}
\lefteqn{\boldsymbol{\Delta^{(\kappa=E)}}\Big(\hat{T}_{\vec{y}\ppr;\vec{y}}^{-1}\!(\vartheta_{q_{E}}^{(E)});
\hat{T}_{\vec{y}\ppr;\vec{y}}\!(\vartheta_{q_{E}}^{(E)})\Big) =
\int d[\delta\wt{\Sigma}_{\vec{y};\vec{y}\ppr}^{ab}(\mfrak{t}_{q_{E}}^{(E)})]
\int d[\delta\hat{\Sigma}_{D;\vec{y}\ppr;\vec{y}}^{aa}(\mfrak{t}_{q_{E}}^{(E)})]\;\;
\mscr{P}\Big(\delta\hat{\lambda}_{\vec{y}}(\mfrak{t}_{q_{E}}^{(E)})\Big)\;\times } \\ \no &\times&
\exp\Bigg\{\frac{\im}{4\hbar}
\int_{0}^{\mfrak{t}_{q_{E}}^{(E)}-\Delta\vartheta^{(E)}} \hspace*{-0.6cm}
d\vartheta_{q_{E}}^{(E)}\;\zeta_{q_{E}}^{(E)}\sum_{\vec{x},\vec{x}\ppr}
\TRS\bigg[\frac{\delta\wt{\Sigma}_{\vec{x},s;\vec{x}\ppr,s\ppr}^{ab}(\vartheta_{q_{E}}^{(E)})\;
\delta\wt{\Sigma}_{\vec{x}\ppr,s\ppr;\vec{x},s}^{ba}(\vartheta_{q_{E}}^{(E)})}{V_{|\vec{x}\ppr-\vec{x}|}^{(E)}-
\im\:\mfrak{e}_{q_{E}}^{(E)}\:\big(\delta_{a=b}-\delta_{a\neq b}\big)}
\bigg]\Bigg\}\times  \\ \no &\times&
\delta\bigg(
\Big(\hat{T}(\vartheta_{q_{E}}^{(E)})\Big)_{\vec{y};\vec{y}_{1}}^{aa_{1}}
\delta\hat{\Sigma}_{D;\vec{y}_{1};\vec{y}_{2}}^{a_{1}=a_{2}}(\vartheta_{q_{E}}^{(E)})
\Big(\hat{T}^{-1}(\vartheta_{q_{E}}^{(E)})\Big)_{\vec{y}_{2};\vec{y}\ppr}^{a_{2}b} -
\delta\wt{\Sigma}_{\vec{y};\vec{y}\ppr}^{ab}(\vartheta_{q_{E}}^{(E)}) \bigg)   \;;
\\ \lb{s3_59}
\lefteqn{\boldsymbol{\Delta^{(\kappa=1,2)}}\Big(\hat{T}_{\vec{y}\ppr;\vec{y}}^{-1}\!(\vartheta_{q_{\kappa}}^{(\kappa)});
\hat{T}_{\vec{y}\ppr;\vec{y}}\!(\vartheta_{q_{\kappa}}^{(\kappa)})\Big) =
\int d[\delta\wt{\Sigma}_{\vec{y};\vec{y}\ppr}^{(i)ab}(\mfrak{t}_{q_{\kappa}}^{(\kappa)})]
\int d[\delta\hat{\Sigma}_{D;\vec{y}\ppr;\vec{y}}^{aa}(\mfrak{t}_{q_{\kappa}}^{(\kappa)})]\;\;
\mscr{P}\Big(\delta\hat{\lambda}_{\vec{y}}(\mfrak{t}_{q_{\kappa}}^{(\kappa)})\Big)\;\times } \\ \no &\times&
\exp\Bigg\{\frac{\im}{4\hbar}
\int_{0}^{\mfrak{t}_{q_{\kappa}}^{(\kappa)}-\Delta\vartheta^{(\kappa)}} \hspace*{-0.6cm}
d\vartheta_{q_{\kappa}}^{(\kappa)}\;\zeta_{q_{\kappa}}^{(\kappa)}\sum_{i=i_{\kappa}}^{j_{\kappa}}\sum_{\vec{x}_{1/2};\vec{x}\ppr_{1/2}}
\TRS\bigg[\frac{\delta\wt{\Sigma}_{\vec{y}_{1}\ppr;\vec{y}_{2}}^{(i)ab}(\vartheta_{q_{\kappa}}^{(\kappa)})\;
\hat{\mscr{S}}_{\vec{y}_{2};\vec{y}_{1}}^{(\kappa;i);bb}\;
\delta\wt{\Sigma}_{\vec{y}_{1};\vec{y}_{2}\ppr}^{(i)ba}(\vartheta_{q_{\kappa}}^{(\kappa)})\;
\hat{\mscr{S}}_{\vec{y}_{2}\ppr;\vec{y}_{1}\ppr}^{(\kappa;i);aa}}{V_{|\vec{x}_{1}\ppr-\vec{x}_{1}|}^{(\kappa)}-
\im\:\mfrak{e}_{q_{\kappa}}^{(\kappa)}
\:\big(\delta_{a=b}-\delta_{a\neq b}\big)}\bigg]\Bigg\} \;\times \\ \no &\times&
\delta\bigg(\Big(\hat{T}(\vartheta_{q_{\kappa}}^{(\kappa)})\Big)_{\vec{y};\vec{y}_{1}\ppr}^{aa_{1}}\;
\delta\hat{\Sigma}_{D;\vec{y}_{1}\ppr;\vec{y}_{2}}^{a_{1}=a_{2}}(\vartheta_{q_{\kappa}}^{(\kappa)})\;
\Big(\hat{T}^{-1}(\vartheta_{q_{\kappa}}^{(\kappa)})\Big)_{\vec{y}_{2};\vec{y}\ppr}^{a_{2}b} -
\sum_{i=i_{\kappa}}^{j_{\kappa}}
\hat{\mscr{S}}_{\vec{y};\vec{y}_{1}\ppr}^{(\kappa;i);aa}\;
\delta\wt{\Sigma}_{\vec{y}_{1}\ppr;\vec{y}_{2}}^{(i)ab}(\vartheta_{q_{\kappa}}^{(\kappa)})\;
\hat{\mscr{S}}_{\vec{y}_{2};\vec{y}\ppr}^{(\kappa;i);bb} \bigg) \;.
\eeq
We can straightforwardly follow Refs. \cite{physica6,pop1,pop2,precisecoh1}
for the coset decomposition of the self-energies after the HST transformations
above, but have to specify various matrices for adaption to the anomalous doubled fields
\(\breve{\Psi}_{\vec{y}}^{a}(\vartheta_{q_{\kappa}}^{(\kappa)})\), \(\breve{\Psi}_{\vec{y}}^{\sharp a}(\vartheta_{q_{\kappa}}^{(\kappa)})\),
\(\mathring{\Psi}_{\vec{y}}^{a}(\vartheta_{q_{\kappa}}^{(\kappa)})\),
\(\mathring{\Psi}_{\vec{y}}^{\sharp a}(\vartheta_{q_{\kappa}}^{(\kappa)})\)
and corresponding discrete time grids. Despite of involved appearance this proves that coherent state path integrals
definitely allow for exact kinds of discrete time grids taking into account the various limit processes of
many-particle theory for the second quantized operators \(\hat{\psi}_{\vec{y}}\), \(\hat{\psi}_{\vec{y}}\pdag\)
where the hermitian conjugated fields should always follow a particular, infinitesimal time step later than their
correspondents without complex conjugation, due to the normal ordering.

In the following we reduce the path integral (\ref{s3_54}) to the anomalous doubled one-particle Hamiltonian
with self-energy density \(\sigma_{D}^{(i)}(\vec{x_{1}},\vartheta_{q_{\kappa}}^{(\kappa)\bprime})\) (\ref{s3_60}-\ref{s3_66})
and gradient term (\ref{s3_67}) of the coset matrices
\(\mathring{T}_{\vec{y}\ppr;\vec{y}_{1}}^{-1;bb_{1}}(\vartheta_{q_{\kappa}}^{(\kappa)\bprime})\),
\(\mathring{T}_{\vec{y}_{2};\vec{y}\ppr}^{a_{1}a}(\vartheta_{q_{\kappa}}^{(\kappa)})\) which are similarly constructed as the matrices \((\mathring{\tau}_{1})^{ba}\), \((\mathring{S})^{ba}\) with matrix entries
\(\hat{T}_{\vec{y}\ppr;\vec{y}_{1}}^{-1;bb_{1}}(\vartheta_{q_{\kappa}}^{(\kappa)\bprime})\),
\(\hat{T}_{\vec{y}_{2};\vec{y}}^{a_{1}a}(\vartheta_{q_{\kappa}}^{(\kappa)})\)
(\(0\leq\vartheta_{q_{\kappa}}^{(\kappa)}\,,\,\vartheta_{q_{\kappa}}^{(\kappa)\bprime}\leq\mfrak{t}_{q_{\kappa}}^{(\kappa)}-\Delta\vartheta^{(\kappa)}\))
along the block diagonal instead of \(\hat{\tau}_{1}^{ba}\), \(\hat{S}^{ba}\), but with additional unit matrix
\(\hat{1}\) in the uppermost left corner and lowest right corner with adaption to the range of generalized
time labels for the fields (\ref{s3_28},\ref{s3_29})
\beq \lb{s3_60}
\Delta\vartheta^{(\kappa)}\;\mathring{\mscr{H}}_{\vec{y}\ppr;\vec{y}}^{(\kappa)ba}\big(\vartheta_{q_{\kappa}}^{(\kappa)\bprime}\boldsymbol{|}
\vartheta_{q_{\kappa}}^{(\kappa)}\boldsymbol{|}\hat{\mscr{S}}^{(\kappa;i)}\cdot\sigma_{D}^{(i)}\big) &=&
\mathring{\mfrak{H}}_{\vec{y}\ppr;\vec{y}}^{(\kappa)ba}(\vartheta_{q_{\kappa}}^{(\kappa)\bprime}
\boldsymbol{|}\vartheta_{q_{\kappa}}^{(\kappa)})+  \\ \no &-& \frac{\im}{\hbar}\Delta\vartheta^{(\kappa)}\:\zeta_{q_{\kappa}}^{(\kappa)}
\sum_{\vec{y}_{1}}\hat{\mscr{S}}_{\vec{y}\ppr;\vec{y}_{1}}^{(\kappa;i)}\;\sigma_{D}^{(i)}(\vec{x}_{1},\vartheta_{q_{\kappa}}^{(\kappa)\bprime})\;\delta_{\vec{y}_{1};\vec{y}}\;\delta(\vartheta_{q_{\kappa}}^{(\kappa)\bprime}\boldsymbol{|}\vartheta_{q_{\kappa}}^{(\kappa)})
\;;  \\ \lb{s3_61}
\sigma_{D}^{(i)}(\vec{x},\vartheta_{q_{\kappa}}^{(\kappa)}) &;&0\leq\vartheta_{q_{\kappa}}^{(\kappa)}\leq\mfrak{t}_{q_{\kappa}}^{(\kappa)}-
\Delta\vartheta^{(\kappa)}   \\ \lb{s3_62}
i=i_{\kappa},\ldots,j_{\kappa} &;&(\kappa=0;i_{0}=j_{0}=0;i=0) \;; \\ \lb{s3_63}
i=i_{\kappa},\ldots,j_{\kappa} &;&(\kappa=1;i_{1}=1;j_{1}=3;i=1,2,3) \;;  \\ \lb{s3_64}
i=i_{\kappa},\ldots,j_{\kappa} &;&(\kappa=2;i_{2}=1;j_{2}=3;i=1,2,3) \;; \\
\sigma_{D}^{(i)}(\vec{x},\vartheta_{q_{\kappa}}^{(\kappa)}) &:&
\mbox{diagonal in \lb{s3_65} spin space and 'Nambu' space} \;; \\  \lb{s3_66}
\sigma_{D}^{(i)}(\vec{x},\vartheta_{q_{\kappa=1,2}}^{(\kappa=1,2)}) &=&
\sigma_{D}^{(i)}(\vartheta_{q_{\kappa=1,2}}^{(\kappa=1,2)})\;\mbox{independence on spatial coordinates,} \\ \no &&
\mbox{due to factorization of quartic interaction in two density parts}\;;   \\    \lb{s3_67}
\Delta\vartheta^{(\kappa)}\;\delta\mathring{\mscr{H}}_{\vec{y}\ppr;\vec{y}}^{(\kappa)ba}\!\!\left(
\mathring{T}^{-1}(\vartheta_{q_{\kappa}}^{(\kappa)\bprime})\boldsymbol{|}
\mathring{T}(\vartheta_{q_{\kappa}}^{(\kappa)})\boldsymbol{|}\hat{\mscr{S}}^{(\kappa;i)}\cdot\sigma_{D}^{(i)}
\right) &=& \\ \no
\lefteqn{\hspace*{-6.6cm}=
\Delta\vartheta^{(\kappa)}\sum_{\vec{y}_{1};\vec{y}_{2}}
\mathring{T}_{\vec{y}\ppr;\vec{y}_{1}}^{-1;bb_{1}}\!(\vartheta_{q_{\kappa}}^{(\kappa)\bprime})\;\;
\mathring{\mscr{H}}_{\vec{y}_{1};\vec{y}_{2}}^{(\kappa)b_{1}a_{1}}\!\!
\left(\vartheta_{q_{\kappa}}^{(\kappa)\bprime}\boldsymbol{|}
\vartheta_{q_{\kappa}}^{(\kappa)}\boldsymbol{|}\hat{\mscr{S}}^{(\kappa;i)}\cdot\sigma_{D}^{(i)}\right) \;\;
\mathring{T}_{\vec{y}_{2};\vec{y}}^{a_{1}a}\!(\vartheta_{q_{\kappa}}^{(\kappa)}) - \Delta\vartheta^{(\kappa)}\;
\mathring{\mscr{H}}_{\vec{y}\ppr;\vec{y}}^{(\kappa)ba}\!\!\left(\vartheta_{q_{\kappa}}^{(\kappa)\bprime}\boldsymbol{|}
\vartheta_{q_{\kappa}}^{(\kappa)}\boldsymbol{|}\hat{\mscr{S}}^{(\kappa;i)}\cdot\sigma_{D}^{(i)}\right) \;.}
\eeq
According to our various definitions (\ref{s3_42}-\ref{s3_57}) and (\ref{s3_58}-\ref{s3_67}),
we obtain the path integral (\ref{s3_68}) for the two-particle
delta functions which has the particular form (\ref{s3_69},\ref{s3_70}) with Grassmann fields \(\varrho_{i}\)
and anti-symmetric matrix \(\mscr{M}_{ij}\). (The integration intervals \(d\vartheta_{q_{\kappa}}^{(\kappa)}\),
\(d\vartheta_{q_{\kappa}}^{(\kappa)\bprime}\) in the sixth line of (\ref{s3_68}) are absorbed into the matrix
\(\mathring{\mscr{N}}_{\vec{y}\ppr;\vec{y}}^{ba}(\vartheta_{q_{\kappa}}^{(\kappa)\bprime}\boldsymbol{|}
\vartheta_{q_{\kappa}}^{(\kappa)})\) (\ref{s3_71}).)
\beq \lb{s3_68}
\lefteqn{\big\langle\xi^{(\kappa+1)}\big|\delta\big(\mfrak{v}^{(\kappa)}-
\boldsymbol{\hat{\mfrak{V}}^{(\kappa)}}(\hat{\psi}\pdag,\hat{\psi})\,\big)\big|\xi^{(\kappa)}\big\rangle =  }  \\ \no &=&
\sum_{q_{\kappa}=\pm} \;\lim_{|\mfrak{e}_{q_{\kappa}}^{(\kappa)}|\rightarrow 0} \;\lim_{\mcal{T}^{(\kappa)}\rightarrow+\infty}
\int_{0}^{\mcal{T}^{(\kappa)}}\frac{d\mfrak{t}_{q_{\kappa}}^{(\kappa)}}{2\pi\,\hbar}\;
\big\langle\xi^{(\kappa+1)}\big|\overleftarrow{\exp}\Big\{-\im\,\zeta_{q_{\kappa}}^{(\kappa)}\:\frac{\mfrak{t}_{q_{\kappa}}^{(\kappa)}}{\hbar}\;
\Big(\mfrak{v}^{(\kappa)}-\boldsymbol{\hat{\mfrak{V}}^{(\kappa)}}(\hat{\psi}\pdag,\hat{\psi})-\im\:
\mfrak{e}_{q_{\kappa}}^{(\kappa)}\Big)\Big\}\big|\xi^{(\kappa)}\big\rangle   \\ \no &=&\sum_{q_{\kappa}=\pm}\;
\lim_{|\mfrak{e}_{q_{\kappa}}^{(\kappa)}|\rightarrow 0} \;\lim_{\mcal{T}^{(\kappa)}\rightarrow+\infty}
\int_{0}^{\mcal{T}^{(\kappa)}}\frac{d\mfrak{t}_{q_{\kappa}}^{(\kappa)}}{2\pi\,\hbar}\;
\exp\Big\{-\im\,\zeta_{q_{\kappa}}^{(\kappa)}\:\frac{\mfrak{t}_{q_{\kappa}}^{(\kappa)}}{\hbar}\:
\big(\mfrak{v}^{(\kappa)}-\im\,\mfrak{e}_{q_{\kappa}}^{(\kappa)}\big)\Big\}\;\times
\\ \no &\times&\int d[\sigma_{D}^{(i)}(\vec{x},\mfrak{t}_{q_{\kappa}}^{(\kappa)})]\;
\exp\bigg\{-\frac{\im}{2\hbar}\int_{0}^{\mfrak{t}_{q_{\kappa}}^{(\kappa)}-\Delta\vartheta^{(\kappa)}} \hspace*{-0.6cm}
d\vartheta_{q_{\kappa}}^{(\kappa)}\;\zeta_{q_{\kappa}}^{(\kappa)}\sum_{i=i_{\kappa}}^{j_{\kappa}}\sum_{\vec{x},\vec{x}\ppr}
\sigma_{D}^{(i)}(\vec{x}\ppr,\vartheta_{q_{\kappa}}^{(\kappa)})\;\hat{V}_{|\vec{x}\ppr-\vec{x}|}^{(\kappa);\boldsymbol{-1}}
\;\sigma_{D}^{(i)}(\vec{x},\vartheta_{q_{\kappa}}^{(\kappa)})\bigg\} \\  \no &&
\int d[\hat{T}_{\vec{y}\ppr;\vec{y}_{1}}^{-1}\!(\vartheta_{q_{\kappa}}^{(\kappa)})\;
d\!\hat{T}_{\vec{y}_{1};\vec{y}}\!(\vartheta_{q_{\kappa}}^{(\kappa)})]\;\;
\boldsymbol{\Delta^{(\kappa)}}\!\Big(\hat{T}_{\vec{y}\ppr;\vec{y}}^{-1}\!(\vartheta_{q_{\kappa}}^{(\kappa)});
\hat{T}_{\vec{y}\ppr;\vec{y}}\!(\vartheta_{q_{\kappa}}^{(\kappa)})\Big)\;\int
d[\psi^{*}(\mfrak{t}_{q_{\kappa}}^{(\kappa)}),\psi(\mfrak{t}_{q_{\kappa}}^{(\kappa)})]  \\ \no &\times&
\exp\bigg\{-\frac{1}{2}
\int_{0}^{\mfrak{t}_{q_{\kappa}}^{(\kappa)}} \;\int_{0}^{\mfrak{t}_{q_{\kappa}}^{(\kappa)}}
\sum_{\vec{y};\vec{y}\ppr}\mathring{\Psi}_{\vec{y}\ppr}^{\sharp b}(\vartheta_{q_{\kappa}}^{(\kappa)\bprime})\;\;
\mathring{\mscr{N}}_{\vec{y}\ppr;\vec{y}}^{(k)ba}(\vartheta_{q_{\kappa}}^{(\kappa)\bprime}
\boldsymbol{|}\vartheta_{q_{\kappa}}^{(\kappa)})\;\;
\mathring{\Psi}_{\vec{y}}^{a}(\vartheta_{q_{\kappa}}^{(\kappa)})\bigg\}\;\times   \\ \no &\times&
\exp\bigg\{\sum_{\vec{y}}\Big(\psi_{\vec{y}}^{*}(\vartheta_{q_{\kappa}}^{(\kappa)}\!=\!0)\;\;
\xi_{\vec{y}}^{(\kappa)}+\xi_{\vec{y}}^{(\kappa+1)*}\;\;\psi_{\vec{y}}(\vartheta_{q_{\kappa}}\!=\!\mfrak{t}_{q_{\kappa}}^{(\kappa)})\Big)\bigg\}\;;
\eeq
\beq\lb{s3_69}
\int d\varrho\;\exp\big\{-{\ts\frac{1}{2}}\;\varrho_{i}^{T}\;\mscr{M}_{ij}\;\varrho_{j}\big\} &=&
\big\{\mbox{DET}\big(\mscr{M}_{ij}\big)\big\}^{1/2}  \;; \\   \lb{s3_70}
\mscr{M}_{ij}^{T} =-\mscr{M}_{ij} &;&(\,\mscr{M}_{ij}^{T}=+\mscr{M}_{ij}\;\;\mbox{does not contribute !})\;.
\eeq
In order to verify the anti-symmetry of matrix \(\mathring{\mscr{N}}_{\vec{y}\ppr;\vec{y}}^{ba}(\vartheta_{q_{\kappa}}^{(\kappa)\bprime}\boldsymbol{|}\vartheta_{q_{\kappa}}^{(\kappa)})\) (\ref{s3_71}) analogous to (\ref{s3_69},\ref{s3_70})
\beq \lb{s3_71}
\mathring{\mscr{N}}_{\vec{y}\ppr;\vec{y}}^{ba}(\vartheta_{q_{\kappa}}^{(\kappa)\bprime}\boldsymbol{|}
\vartheta_{q_{\kappa}}^{(\kappa)}) &=&\mathring{S}\;\Delta\vartheta^{(\kappa)}\;
\mathring{\mscr{H}}_{\vec{y}\ppr;\vec{y}}^{(\kappa)ba}\big(\vartheta_{q_{\kappa}}^{(\kappa)\bprime}\boldsymbol{|}
\vartheta_{q_{\kappa}}^{(\kappa)}\boldsymbol{|}\hat{\mscr{S}}^{(\kappa;i)}\cdot\sigma_{D}^{(i)}\big) +  \\ \no &+&
\hat{I}\sum_{\vec{y}_{1};\vec{y}_{2}}\hat{T}_{\vec{y}\ppr;\vec{y}_{2}}^{bb_{2}}\!(\vartheta_{q_{\kappa}}^{(\kappa)\bprime})\;
\delta\hat{\Sigma}_{D;\vec{y}_{2};\vec{y}_{1}}^{b_{2}=a_{1}}\!(\vartheta_{q_{\kappa}}^{(\kappa)\bprime})\;
\delta(\vartheta_{q_{\kappa}}^{(\kappa)\bprime}\boldsymbol{|}\vartheta_{q_{\kappa}}^{(\kappa)})\;
\hat{T}_{\vec{y}_{1};\vec{y}}^{-1;a_{1}a}\!(\vartheta_{q_{\kappa}}^{(\kappa)})\;\hat{I}\;\;,
\eeq
we perform transformations (\ref{s3_72}-\ref{s3_75}) so that the integral relation (\ref{s3_69},\ref{s3_70})
can be applied
\beq  \lb{s3_72}
(\mathring{\tau}_{1})^{bb\ppr}\;(\mathring{\tau}_{1})^{b\ppr a} &=&\mathring{1}^{ba} \;; \\ \lb{s3_73}
\mathring{\Psi}_{\vec{y}\ppr}^{\sharp b}\!(\vartheta_{q_{\kappa}}^{(\kappa)\bprime})\;\;
(\mathring{\tau}_{1})^{bb\ppr} &=&
\big(\mathring{\Psi}_{\vec{y}\ppr}^{b\ppr}\!(\vartheta_{q_{\kappa}}^{(\kappa)\bprime})\big)^{T} \;; \\  \lb{s3_74}
\Big((\mathring{\tau}_{1})^{b\ppr b_{1}}\;\mathring{S}\;
\mathring{\mscr{H}}_{\vec{y}\ppr;\vec{y}}^{(\kappa)b_{1}a}\big(\vartheta_{q_{\kappa}}^{(\kappa)\bprime}\boldsymbol{|}
\vartheta_{q_{\kappa}}^{(\kappa)}\boldsymbol{|}\hat{\mscr{S}}^{(\kappa;i)}\cdot\sigma_{D}^{(i)}\big)\Big)^{T} &=&-
(\mathring{\tau}_{1})^{b\ppr b_{1}}\;\mathring{S}\;
\mathring{\mscr{H}}_{\vec{y}\ppr;\vec{y}}^{(\kappa)b_{1}a}\big(\vartheta_{q_{\kappa}}^{(\kappa)\bprime}\boldsymbol{|}
\vartheta_{q_{\kappa}}^{(\kappa)}\boldsymbol{|}\hat{\mscr{S}}^{(\kappa;i)}\cdot\sigma_{D}^{(i)}\big) \;;
\eeq
\beq\lb{s3_75}
\lefteqn{\Big((\mathring{\tau}_{1})^{b\ppr b_{1}}\;
\hat{I}\sum_{\vec{y}_{1};\vec{y}_{2}}\hat{T}_{\vec{y}\ppr;\vec{y}_{2}}^{b_{1}b_{2}}\!(\vartheta_{q_{\kappa}}^{(\kappa)\bprime})\;
\delta\hat{\Sigma}_{D;\vec{y}_{2};\vec{y}_{1}}^{b_{2}=a_{1}}\!(\vartheta_{q_{\kappa}}^{(\kappa)\bprime})\;
\delta(\vartheta_{q_{\kappa}}^{(\kappa)\bprime}\boldsymbol{|}\vartheta_{q_{\kappa}}^{(\kappa)})\;
\hat{T}_{\vec{y}_{1};\vec{y}}^{-1;a_{1}a}\!(\vartheta_{q_{\kappa}}^{(\kappa)})\;\hat{I}\Big)^{T} = }  \\ \no &=&
\big(-\delta_{b\ppr=a}+\delta_{b\ppr\neq a}\big)\;
(\mathring{\tau}_{1})^{b\ppr b_{1}}\;
\hat{I}\sum_{\vec{y}_{1};\vec{y}_{2}}
\hat{T}_{\vec{y}\ppr;\vec{y}_{2}}^{b_{1}b_{2}}\!(\vartheta_{q_{\kappa}}^{(\kappa)\bprime})\;
\delta\hat{\Sigma}_{D;\vec{y}_{2};\vec{y}_{1}}^{b_{2}=a_{1}}\!(\vartheta_{q_{\kappa}}^{(\kappa)\bprime})\;
\delta(\vartheta_{q_{\kappa}}^{(\kappa)\bprime}\boldsymbol{|}\vartheta_{q_{\kappa}}^{(\kappa)})\;
\hat{T}_{\vec{y}_{1};\vec{y}}^{-1;a_{1}a}\!(\vartheta_{q_{\kappa}}^{(\kappa)})\;\hat{I} \;.
\eeq
The relevant kind of Grassmann integration of bilinear fields (\ref{s3_76}) has to be computed in analogy
to (\ref{s3_69},\ref{s3_70}), but with additional coupling to the anti-commuting fields \(\xi_{\vec{y}}^{(\kappa)}\),
\(\xi_{\vec{y}}^{(\kappa+1)*}\) . Similarly to eq. (\ref{s3_68}), we absorb
the integration intervals \(d\vartheta_{q_{\kappa}}^{(\kappa)}\),
\(d\vartheta_{q_{\kappa}}^{(\kappa)\bprime}\) into the matrix
\(\mathring{\mscr{N}}_{\vec{y}\ppr;\vec{y}}^{ba}(\vartheta_{q_{\kappa}}^{(\kappa)\bprime}\boldsymbol{|}
\vartheta_{q_{\kappa}}^{(\kappa)})\) (\ref{s3_71}). (This abbreviation is also used in further transformation steps till the
end of this section \ref{s3}, especially concerning the matrix
\(\mathring{\mscr{M}}_{\vec{y}\ppr;\vec{y}}^{ba}
(\vartheta_{q_{\kappa}}^{(\kappa)\bprime}\boldsymbol{|}\vartheta_{q_{\kappa}}^{(\kappa)})\) (\ref{s3_84}).)
\beq\lb{s3_76}
\mfrak{I}^{(\kappa)}[\mathring{\mscr{N}}_{\vec{y}\ppr;\vec{y}}^{(\kappa)ba}] &=&
\int d[\psi^{*}(\mfrak{t}_{q_{\kappa}}^{(\kappa)}),\psi(\mfrak{t}_{q_{\kappa}}^{(\kappa)})]  \;\times \\ \no &\times&
\exp\bigg\{-\frac{1}{2}
\int_{-\Delta\vartheta^{(\kappa)}}^{\mfrak{t}_{q_{\kappa}}^{(\kappa)}}
\int_{-\Delta\vartheta^{(\kappa)}}^{\mfrak{t}_{q_{\kappa}}^{(\kappa)}}
\sum_{\vec{y};\vec{y}\ppr}\mathring{\Psi}_{\vec{y}\ppr}^{T,b}(\vartheta_{q_{\kappa}}^{(\kappa)\bprime})\;
\Big((\mathring{\tau}_{1})^{bb_{1}}\;
\mathring{\mscr{N}}_{\vec{y}\ppr;\vec{y}}^{(k)b_{1}a}(\vartheta_{q_{\kappa}}^{(\kappa)\bprime}\boldsymbol{|}
\vartheta_{q_{\kappa}}^{(\kappa)})\Big)\;
\mathring{\Psi}_{\vec{y}}^{a}(\vartheta_{q_{\kappa}}^{(\kappa)})\bigg\}\;\times  \\  \no   &\times&
\exp\bigg\{\sum_{\vec{y}}\Big(\psi_{\vec{y}}^{*}(\vartheta_{q_{\kappa}}^{(\kappa)}\!=\!0)\;\;
\xi_{\vec{y}}^{(\kappa)}+\xi_{\vec{y}}^{(\kappa+1)*}\;\;\psi_{\vec{y}}(\vartheta_{q_{\kappa}}\!=
\!\mfrak{t}_{q_{\kappa}}^{(\kappa)})\Big)\bigg\}   \;.
\eeq
Therefore, we define the 'source' fields \(\mathring{\Xi}_{\vec{y}}^{a}(\vartheta_{q_{\kappa}}^{(\kappa)})\)
(\ref{s3_77}) with adaption to the time grid of fields \(\mathring{\Psi}_{\vec{y}}^{a}(\vartheta_{q_{\kappa}}^{(\kappa)})\),
\(\mathring{\Psi}_{\vec{y}}^{\sharp a}(\vartheta_{q_{\kappa}}^{(\kappa)})\) (\ref{s3_28},\ref{s3_29})
\beq\lb{s3_77}
\mathring{\Xi}_{\vec{y}}^{a}(\vartheta_{q_{\kappa}}^{(\kappa)})&=&
\Big(\underbrace{\xi_{\vec{y}}^{(\kappa)}}_{-\Delta\vartheta^{(\kappa)}}
\underbrace{\boldsymbol{\big|}0;0\boldsymbol{\big|}0;0\boldsymbol{\big|}\ldots\boldsymbol{\big|}0;0\boldsymbol{\big|}}_{
\breve{\Xi}_{\vec{y}}^{a}(\vartheta_{q_{\kappa}}^{(\kappa)})\equiv 0}\underbrace{-\xi_{\vec{y}}^{(\kappa+1)*}}_{
\mfrak{t}_{q_{\kappa}}^{(\kappa)}}\Big)^{T}\;,
\eeq
so that the bilinear Grassmann fields can be removed by integration in (\ref{s3_68},\ref{s3_71},\ref{s3_76})
corresponding to the general relation (\ref{s3_69},\ref{s3_70})
\beq \lb{s3_78}
\lefteqn{\mfrak{I}^{(\kappa)}[\mathring{\mscr{N}}_{\vec{y}\ppr;\vec{y}}^{(\kappa)ba}] =
\int d[\psi^{*}(\mfrak{t}_{q_{\kappa}}^{(\kappa)}),\psi(\mfrak{t}_{q_{\kappa}}^{(\kappa)})]  \;\times}  \\ \no &\times&
\exp\bigg\{-\frac{1}{2}
\int_{-\Delta\vartheta^{(\kappa)}}^{\mfrak{t}_{q_{\kappa}}^{(\kappa)}} \hspace*{-0.1cm}
\int_{-\Delta\vartheta^{(\kappa)}}^{\mfrak{t}_{q_{\kappa}}^{(\kappa)}}
\sum_{\vec{y};\vec{y}\ppr}\bigg(\mathring{\Psi}_{\vec{y}\ppr}^{T,b\ppr}\!\!\!(\vartheta_{q_{\kappa}}^{(\kappa)\bprime})-
\int_{-\Delta\vartheta^{(\kappa)}}^{\mfrak{t}_{q_{\kappa}}^{(\kappa)}}
\sum_{\vec{y}_{2}}\mathring{\Xi}_{\vec{y}_{2}}^{T,b_{2}}\!(\vartheta_{q_{\kappa}}^{(\kappa)\bprime\bprime\bprime})\;
\big((\mathring{\tau}_{1})\mathring{\mscr{N}}\big)_{\vec{y}_{2};\vec{y}\ppr}^{T;-1;b_{2}b\ppr}\!\!\!
(\vartheta_{q_{\kappa}}^{(\kappa)\bprime\bprime\bprime}\boldsymbol{|}\vartheta_{q_{\kappa}}^{(\kappa)\bprime})\bigg)\times \\ \no &\times&
\bigg(\big(\mathring{\tau}_{1}\big)^{b\ppr b_{1}}\;
\mathring{\mscr{N}}_{\vec{y}\ppr;\vec{y}}^{b_{1}a}\!(\vartheta_{q_{\kappa}}^{(\kappa)\bprime}\boldsymbol{|}
\vartheta_{q_{\kappa}}^{(\kappa)})\bigg)\times
\bigg(\mathring{\Psi}_{\vec{y}}^{a}\!(\vartheta_{q_{\kappa}}^{(\kappa)})-
\int_{-\Delta\vartheta^{(\kappa)}}^{\mfrak{t}_{q_{\kappa}}^{(\kappa)}} \sum_{\vec{y}_{1}}
\big((\mathring{\tau}_{1})\mathring{\mscr{N}}\big)_{\vec{y};\vec{y}_{1}}^{-1;aa_{1}}\!\!
(\vartheta_{q_{\kappa}}^{(\kappa)}\boldsymbol{|}\vartheta_{q_{\kappa}}^{(\kappa)\bprime\bprime})\;
\mathring{\Xi}_{\vec{y}_{1}}^{a_{1}}\!(\vartheta_{q_{\kappa}}^{(\kappa)\bprime\bprime}) \bigg)\bigg\}\times \\ \no &\times&
\exp\bigg\{-\frac{1}{2}\int_{-\Delta\vartheta^{(\kappa)}}^{\mfrak{t}_{q_{\kappa}}^{(\kappa)}} \hspace*{-0.1cm}
\int_{-\Delta\vartheta^{(\kappa)}}^{\mfrak{t}_{q_{\kappa}}^{(\kappa)}} \sum_{\vec{y};\vec{y}\ppr}
\mathring{\Xi}_{\vec{y}\ppr}^{T,b}\!(\vartheta_{q_{\kappa}}^{(\kappa)\bprime})\;
\big((\mathring{\tau}_{1})\;\mathring{\mscr{N}}\big)_{\vec{y}\ppr;\vec{y}}^{-1;ba}\!\!
(\vartheta_{q_{\kappa}}^{(\kappa)\bprime}\boldsymbol{|}\vartheta_{q_{\kappa}}^{(\kappa)})\;
\mathring{\Xi}_{\vec{y}}^{a}\!(\vartheta_{q_{\kappa}}^{(\kappa)}) \bigg\} =  \\ \no &=&
\bigg\{\mbox{DET}\Big(\big(\mathring{\tau}_{1}\big)^{b\ppr b_{1}}\hspace*{-0.1cm}
\mathring{\mscr{N}}_{\vec{y}\ppr;\vec{y}}^{b_{1}a}(\vartheta_{q_{\kappa}}^{(\kappa)\bprime}\boldsymbol{|}
\vartheta_{q_{\kappa}}^{(\kappa)})\Big)\bigg\}^{1/2}\hspace*{-0.3cm}
\exp\bigg\{-\frac{1}{2}\int_{-\Delta\vartheta^{(\kappa)}}^{\mfrak{t}_{q_{\kappa}}^{(\kappa)}} \hspace*{-0.1cm}
\int_{-\Delta\vartheta^{(\kappa)}}^{\mfrak{t}_{q_{\kappa}}^{(\kappa)}} \sum_{\vec{y};\vec{y}\ppr}
\mathring{\Xi}_{\vec{y}\ppr}^{T,b}\!(\vartheta_{q_{\kappa}}^{(\kappa)\bprime})\;
\mathring{\mscr{N}}_{\vec{y}\ppr;\vec{y}}^{-1;ba\ppr}\!\!
(\vartheta_{q_{\kappa}}^{(\kappa)\bprime}\boldsymbol{|}\vartheta_{q_{\kappa}}^{(\kappa)})\;(\mathring{\tau}_{1})^{a\ppr a}\;
\mathring{\Xi}_{\vec{y}}^{a}\!(\vartheta_{q_{\kappa}}^{(\kappa)}) \bigg\}  \\ \no  &=&
\bigg\{\mbox{DET}\Big(\mathring{\mscr{N}}_{\vec{y}\ppr;\vec{y}}^{ba}(\vartheta_{q_{\kappa}}^{(\kappa)\bprime}\boldsymbol{|}
\vartheta_{q_{\kappa}}^{(\kappa)})\Big)\bigg\}^{1/2}\hspace*{-0.3cm}
\exp\bigg\{-\frac{1}{2}\int_{-\Delta\vartheta^{(\kappa)}}^{\mfrak{t}_{q_{\kappa}}^{(\kappa)}} \hspace*{-0.1cm}
\int_{-\Delta\vartheta^{(\kappa)}}^{\mfrak{t}_{q_{\kappa}}^{(\kappa)}} \sum_{\vec{y};\vec{y}\ppr}
\mathring{\Xi}_{\vec{y}\ppr}^{T,b}\!(\vartheta_{q_{\kappa}}^{(\kappa)\bprime})\;
\mathring{\mscr{N}}_{\vec{y}\ppr;\vec{y}}^{-1;ba}\!
(\vartheta_{q_{\kappa}}^{(\kappa)\bprime}\boldsymbol{|}\vartheta_{q_{\kappa}}^{(\kappa)})\;
\mathring{\Xi}_{\vec{y}}^{a}\!(\vartheta_{q_{\kappa}}^{(\kappa)}) \bigg\} \;.
\eeq
It remains to project the matrix \(\mathring{\mscr{N}}_{\vec{y}\ppr;\vec{y}}^{ba}(\vartheta_{q_{\kappa}}^{(\kappa)\bprime}
\boldsymbol{|}\vartheta_{q_{\kappa}}^{(\kappa)})\) (\ref{s3_71}) onto the coset degrees of freedom for the anomalous
doubled pairs of fields. Therefore, we factor the coset matrices
\(\mathring{T}_{\vec{y}\ppr;\vec{y}_{2}}^{bb_{2}}(\vartheta_{q_{\kappa}}^{(\kappa)})\),
\(\mathring{T}_{\vec{y}_{1};\vec{y}}^{-1;a_{1}a}(\vartheta_{q_{\kappa}}^{(\kappa)\bprime})\) outside of the matrix
\(\mathring{\mscr{N}}_{\vec{y}\ppr;\vec{y}}^{ba}(\vartheta_{q_{\kappa}}^{(\kappa)\bprime}\boldsymbol{|}
\vartheta_{q_{\kappa}}^{(\kappa)})\) (\ref{s3_71}) in order to achieve following relation (\ref{s3_79}) for (\ref{s3_71})
\beq \lb{s3_79}
\lefteqn{
\mathring{\mscr{N}}_{\vec{y}\ppr;\vec{y}}^{ba}\!(\vartheta_{q_{\kappa}}^{(\kappa)\bprime}\boldsymbol{|}\vartheta_{q_{\kappa}}^{(\kappa)})=
\mathring{I}\;\mathring{T}_{\vec{y}\ppr;\vec{y}_{2}}^{bb_{2}}\!(\vartheta_{q_{\kappa}}^{(\kappa)\bprime})\;\mathring{I}^{-1}
\times }  \\ \no &\times&
\bigg(\Delta\vartheta^{(\kappa)}\;\mathring{I}\;
\mathring{\mscr{H}}_{\vec{y}_{2};\vec{y}_{1}}^{(\kappa)b_{2}a_{1}}\!(\vartheta_{q_{\kappa}}^{(\kappa)\bprime}\boldsymbol{|}
\vartheta_{q_{\kappa}}^{(\kappa)}\boldsymbol{|}\hat{\mscr{S}}^{(\kappa;i)}\cdot\sigma_{D}^{(i)})\;\mathring{I} +
\Delta\vartheta^{(\kappa)}\;\mathring{I}\;\delta\mathring{\mscr{H}}_{\vec{y}_{2};\vec{y}_{1}}^{(\kappa)b_{2}a_{1}}\!(
\mathring{T}^{-1}(\vartheta_{q_{\kappa}}^{(\kappa)\bprime})\boldsymbol{|}
\mathring{T}(\vartheta_{q_{\kappa}}^{(\kappa)})\boldsymbol{|}\hat{\mscr{S}}^{(\kappa;i)}\cdot\sigma_{D}^{(i)}) \;\mathring{I} +
\\   \no  &+& \underbrace{\mathring{I}\;
\delta\hat{\Sigma}_{D;\vec{y}_{2};\vec{y}_{1}}^{b_{2}=a_{1}}\!(\vartheta_{q_{\kappa}}^{(\kappa)\bprime})\;\mathring{I}\;
\delta(\vartheta_{q_{\kappa}}^{(\kappa)\bprime}\boldsymbol{|}\vartheta_{q_{\kappa}}^{(\kappa)})}_{\rightarrow 0}\bigg) \times
\mathring{I}^{-1}\;\mathring{T}_{\vec{y}_{1};\vec{y}}^{-1;a_{1}a}\!(\vartheta_{q_{\kappa}}^{(\kappa)})\;
\mathring{I}\;.
\eeq
According to (\ref{s3_69},\ref{s3_70}), the anti-symmetric part of
\((\mathring{\tau}_{1})^{b\ppr b_{1}}
\mathring{\mscr{N}}_{\vec{y}\ppr;\vec{y}}^{b_{1}a}(\vartheta_{q_{\kappa}}^{(\kappa)\bprime}\boldsymbol{|}\vartheta_{q_{\kappa}}^{(\kappa)})\)
(\ref{s3_72}-\ref{s3_75}) only remains after the integration over the bilinear anti-commuting fields so that
the block diagonal self-energy densities
\(\delta\hat{\Sigma}_{D;\vec{y}_{2};\vec{y}_{1}}^{b_{2}=a_{1}}(\vartheta_{q_{\kappa}}^{(\kappa)\bprime})\) vanish in the
integral \(\mfrak{I}^{(\kappa)}[\mathring{\mscr{N}}_{\vec{y}\ppr;\vec{y}}^{(\kappa)ba}]\) (\ref{s3_76}) and act as 'hinge'
fields in a spontaneous symmetry breaking with a coset decomposition. Corresponding to the Weyl unitary trick,
we can factor out coset matrices \(\wt{T}_{\vec{y}\ppr;\vec{y}}^{ba}(\vartheta_{q_{\kappa}}^{(\kappa)})\)
\beq\lb{s3_80}
\mathring{I}\;\mathring{T}_{\vec{y}\ppr;\vec{y}_{2}}^{bb_{2}}\!(\vartheta_{q_{\kappa}}^{(\kappa)})\;\mathring{I}^{-1} &=&
\mathring{I}^{-1}\;\mathring{T}_{\vec{y}\ppr;\vec{y}_{2}}^{-1;bb_{2}}\!(\vartheta_{q_{\kappa}}^{(\kappa)})\;\mathring{I}=
\wt{T}_{\vec{y}\ppr;\vec{y}_{2}}^{bb_{2}}\!(\vartheta_{q_{\kappa}}^{(\kappa)}) \;;  \\   \lb{s3_81}
\mathring{S} &\rightarrow& \mbox{projection onto coset space !}\;; \\  \lb{s3_82}
\mathring{S}^{ba}\;\delta_{\vec{y}\ppr;\vec{y}}\;\mcal{N}_{x} &=& \sum_{\vec{y}_{2}}
\wt{T}_{\vec{y}\ppr;\vec{y}_{2}}^{bb_{2}}\!(\vartheta_{q_{\kappa}}^{(\kappa)})\;\mathring{S}^{b_{2}a_{1}}\;
\wt{T}_{\vec{y}_{2};\vec{y}}^{a_{1}a}\!(\vartheta_{q_{\kappa}}^{(\kappa)})  \;,
\eeq
and finally succeed in eq. (\ref{s3_83}) for the bilinear integration of Grassmann fields with matrix
\(\mathring{\mscr{M}}_{\vec{y}\ppr;\vec{y}}^{ba}\!(\vartheta_{q_{\kappa}}^{(\kappa)\bprime}\boldsymbol{|}
\vartheta_{q_{\kappa}}^{(\kappa)})\) (\ref{s3_84}) which solely contains the anomalous pairs as remaining degrees
of freedom in the coset matrices \(\mathring{T}_{\vec{y}\ppr;\vec{y}}^{-1;ba}\!(\vartheta_{q_{\kappa}}^{(\kappa)})\),
\(\mathring{T}_{\vec{y}\ppr;\vec{y}}^{ba}\!(\vartheta_{q_{\kappa}}^{(\kappa)})\)
\beq \lb{s3_83}
\mfrak{I}^{(\kappa)}[\mathring{\mscr{N}}_{\vec{y}\ppr;\vec{y}}^{(\kappa)ba}] &=&
\Big\{\mbox{DET}\Big(\mathring{\mscr{M}}_{\vec{y}\ppr;\vec{y}}^{ba}(\vartheta_{q_{\kappa}}^{(\kappa)\bprime}\boldsymbol{|}
\vartheta_{q_{\kappa}}^{(\kappa)})\Big)\Big\}^{1/2}\;\times \\ \no &\times&
\exp\bigg\{-\frac{1}{2}\int_{-\Delta\vartheta^{(\kappa)}}^{\mfrak{t}_{q_{\kappa}}^{(\kappa)}} \hspace*{-0.1cm}
\int_{-\Delta\vartheta^{(\kappa)}}^{\mfrak{t}_{q_{\kappa}}^{(\kappa)}} \sum_{\vec{y};\vec{y}\ppr}\mcal{N}_{x}\;
\mathring{\Xi}_{\vec{y}\ppr}^{T,b}\!(\vartheta_{q_{\kappa}}^{(\kappa)\bprime})\;\;
\mathring{\mscr{M}}_{\vec{y}\ppr;\vec{y}}^{-1;ba}\!
(\vartheta_{q_{\kappa}}^{(\kappa)\bprime}\boldsymbol{|}\vartheta_{q_{\kappa}}^{(\kappa)})\;\;
\mathring{\Xi}_{\vec{y}}^{a}\!(\vartheta_{q_{\kappa}}^{(\kappa)}) \bigg\}\;;  \\ \lb{s3_84}
\mathring{\mscr{M}}_{\vec{y}\ppr;\vec{y}}^{ba}\!
(\vartheta_{q_{\kappa}}^{(\kappa)\bprime}\boldsymbol{|}\vartheta_{q_{\kappa}}^{(\kappa)}) &=&
\Delta\vartheta^{(\kappa)}\;\mathring{\mscr{H}}_{\vec{y}\ppr;\vec{y}}^{(\kappa)ba}\!(\vartheta_{q_{\kappa}}^{(\kappa)\bprime}\boldsymbol{|}
\vartheta_{q_{\kappa}}^{(\kappa)}\boldsymbol{|}\hat{\mscr{S}}^{(\kappa;i)}\cdot\sigma_{D}^{(i)}) +
\Delta\vartheta^{(\kappa)}\;\delta\mathring{\mscr{H}}_{\vec{y}\ppr;\vec{y}}^{(\kappa)ba}\!(
\mathring{T}^{-1}\!(\vartheta_{q_{\kappa}}^{(\kappa)\bprime})\boldsymbol{|}
\mathring{T}\!(\vartheta_{q_{\kappa}}^{(\kappa)})\boldsymbol{|}\hat{\mscr{S}}^{(\kappa;i)}\cdot\sigma_{D}^{(i)}) \;.
\eeq
As we insert the integration term (\ref{s3_83}) with matrix (\ref{s3_84}) into the original path integral
(\ref{s3_68},\ref{s3_71}), we attain relation (\ref{s3_85}) for the delta functions of the
two-particle operators
\beq \lb{s3_85}
\lefteqn{\big\langle\xi^{(\kappa+1)}\big|\delta\big(\mfrak{v}^{(\kappa)}-
\boldsymbol{\hat{\mfrak{V}}^{(\kappa)}}(\hat{\psi}\pdag,\hat{\psi})\,\big)\big|\xi^{(\kappa)}\big\rangle =  }  \\ \no &=&
\sum_{q_{\kappa}=\pm} \;\lim_{|\mfrak{e}_{q_{\kappa}}^{(\kappa)}|\rightarrow 0} \;\lim_{\mcal{T}^{(\kappa)}\rightarrow+\infty}
\int_{0}^{\mcal{T}^{(\kappa)}}\frac{d\mfrak{t}_{q_{\kappa}}^{(\kappa)}}{2\pi\,\hbar}\;
\big\langle\xi^{(\kappa+1)}\big|\overleftarrow{\exp}\Big\{-\im\,\zeta_{q_{\kappa}}^{(\kappa)}\:\frac{\mfrak{t}_{q_{\kappa}}^{(\kappa)}}{\hbar}\;
\Big(\mfrak{v}^{(\kappa)}-\boldsymbol{\hat{\mfrak{V}}^{(\kappa)}}(\hat{\psi}\pdag,\hat{\psi})-\im\:
\mfrak{e}_{q_{\kappa}}^{(\kappa)}\Big)\Big\}\big|\xi^{(\kappa)}\big\rangle   \\ \no &=&\sum_{q_{\kappa}=\pm}\;
\lim_{|\mfrak{e}_{q_{\kappa}}^{(\kappa)}|\rightarrow 0} \;\lim_{\mcal{T}^{(\kappa)}\rightarrow+\infty}
\int_{0}^{\mcal{T}^{(\kappa)}}\frac{d\mfrak{t}_{q_{\kappa}}^{(\kappa)}}{2\pi\,\hbar}\;
\exp\Big\{-\im\,\zeta_{q_{\kappa}}^{(\kappa)}\:\frac{\mfrak{t}_{q_{\kappa}}^{(\kappa)}}{\hbar}\:
\big(\mfrak{v}^{(\kappa)}-\im\,\mfrak{e}_{q_{\kappa}}^{(\kappa)}\big)\Big\}\;\times
\\ \no &\times&\int d[\sigma_{D}^{(i)}(\vec{x},\mfrak{t}_{q_{\kappa}}^{(\kappa)})]\;
\exp\bigg\{-\frac{\im}{2\hbar}\int_{0}^{\mfrak{t}_{q_{\kappa}}^{(\kappa)}-\Delta\vartheta^{(\kappa)}} \hspace*{-0.6cm}
d\vartheta_{q_{\kappa}}^{(\kappa)}\;\zeta_{q_{\kappa}}^{(\kappa)}\sum_{i=i_{\kappa}}^{j_{\kappa}} \sum_{\vec{x},\vec{x}\ppr}
\sigma_{D}^{(i)}(\vec{x}\ppr,\vartheta_{q_{\kappa}}^{(\kappa)})\;\hat{V}_{|\vec{x}\ppr-\vec{x}|}^{(\kappa);\boldsymbol{-1}}
\;\sigma_{D}^{(i)}(\vec{x},\vartheta_{q_{\kappa}}^{(\kappa)})\bigg\} \\  \no &\times&
\int d[\hat{T}_{\vec{y}\ppr;\vec{y}_{1}}^{-1}\!(\vartheta_{q_{\kappa}}^{(\kappa)})\;
d\!\hat{T}_{\vec{y}_{1};\vec{y}}\!(\vartheta_{q_{\kappa}}^{(\kappa)})]\;\;
\boldsymbol{\Delta^{(\kappa)}}\Big(\hat{T}_{\vec{y}\ppr;\vec{y}}^{-1}\!(\vartheta_{q_{\kappa}}^{(\kappa)});
\hat{T}_{\vec{y}\ppr;\vec{y}}\!(\vartheta_{q_{\kappa}}^{(\kappa)})\Big)\;\times\;
\Big\{\mbox{DET}\Big(\mathring{\mscr{M}}_{\vec{y}\ppr;\vec{y}}^{ba}(\vartheta_{q_{\kappa}}^{(\kappa)\bprime}\boldsymbol{|}
\vartheta_{q_{\kappa}}^{(\kappa)})\Big)\Big\}^{1/2}\times \\ \no &\times&
\exp\bigg\{-\frac{1}{2}\int_{-\Delta\vartheta^{(\kappa)}}^{\mfrak{t}_{q_{\kappa}}^{(\kappa)}} \hspace*{-0.1cm}
\int_{-\Delta\vartheta^{(\kappa)}}^{\mfrak{t}_{q_{\kappa}}^{(\kappa)}} \sum_{\vec{y};\vec{y}\ppr}\mcal{N}_{x}\;
\mathring{\Xi}_{\vec{y}\ppr}^{T,b}\!(\vartheta_{q_{\kappa}}^{(\kappa)\bprime})\;\;
\mathring{\mscr{M}}_{\vec{y}\ppr;\vec{y}}^{-1;ba}\!
(\vartheta_{q_{\kappa}}^{(\kappa)\bprime}\boldsymbol{|}\vartheta_{q_{\kappa}}^{(\kappa)})\;\;
\mathring{\Xi}_{\vec{y}}^{a}\!(\vartheta_{q_{\kappa}}^{(\kappa)}) \bigg\}\;.
\eeq
The source field term with \(\mathring{\Xi}_{\vec{y}\ppr}^{T,b}\!(\vartheta_{q_{\kappa}}^{(\kappa)\bprime})\;
\mathring{\mscr{M}}_{\vec{y}\ppr;\vec{y}}^{-1;ba}\!
(\vartheta_{q_{\kappa}}^{(\kappa)\bprime}\boldsymbol{|}\vartheta_{q_{\kappa}}^{(\kappa)})\;
\mathring{\Xi}_{\vec{y}}^{a}\!(\vartheta_{q_{\kappa}}^{(\kappa)})\) can be further simplified with Pauli matrix
\((\mathring{\tau}_{1})^{ba}\) in anomalous doubled space to relation (\ref{s3_86}) with the originally
introduced fields \(\xi_{\vec{y}\ppr}^{(\kappa+1)*}\), \(\xi_{\vec{y}}^{(\kappa)}\)
\beq \lb{s3_86}
\lefteqn{\exp\bigg\{-\frac{1}{2}\int_{-\Delta\vartheta^{(\kappa)}}^{\mfrak{t}_{q_{\kappa}}^{(\kappa)}} \hspace*{-0.1cm}
\int_{-\Delta\vartheta^{(\kappa)}}^{\mfrak{t}_{q_{\kappa}}^{(\kappa)}} \sum_{\vec{y};\vec{y}\ppr}\mcal{N}_{x}\;
\mathring{\Xi}_{\vec{y}\ppr}^{T,b}\!(\vartheta_{q_{\kappa}}^{(\kappa)\bprime})\;
\mathring{\mscr{M}}_{\vec{y}\ppr;\vec{y}}^{-1;ba}\!
(\vartheta_{q_{\kappa}}^{(\kappa)\bprime}\boldsymbol{|}\vartheta_{q_{\kappa}}^{(\kappa)})\;
\mathring{\Xi}_{\vec{y}}^{a}\!(\vartheta_{q_{\kappa}}^{(\kappa)}) \bigg\} = }   \\   \no &=&
\exp\bigg\{-\frac{1}{2}\int_{-\Delta\vartheta^{(\kappa)}}^{\mfrak{t}_{q_{\kappa}}^{(\kappa)}} \hspace*{-0.1cm}
\int_{-\Delta\vartheta^{(\kappa)}}^{\mfrak{t}_{q_{\kappa}}^{(\kappa)}} \sum_{\vec{y};\vec{y}\ppr}\mcal{N}_{x}\;
\mathring{\Xi}_{\vec{y}\ppr}^{T,b}\!(\vartheta_{q_{\kappa}}^{(\kappa)\bprime})\;
\underbrace{\mathring{\mscr{M}}_{\vec{y}\ppr;\vec{y}}^{-1;ba\ppr}\!
\big(\vartheta_{q_{\kappa}}^{(\kappa)\bprime}\boldsymbol{|}\vartheta_{q_{\kappa}}^{(\kappa)}\big)\;(\mathring{\tau}_{1})^{a\ppr a}}_{
\mbox{\scz anti-symmetric}}\;
\mathring{\Xi}_{\vec{y}}^{a}\!(\vartheta_{q_{\kappa}}^{(\kappa)}) \bigg\} \\ \no &=&
\exp\bigg\{\sum_{\vec{y};\vec{y}\ppr}\mcal{N}_{x}\;\xi_{\vec{y}\ppr}^{(\kappa+1)*}\;\;
\mathring{\mscr{M}}_{\vec{y}\ppr;\vec{y}}^{-1;ba}\!
\big(\mfrak{t}_{q_{\kappa}}^{(\kappa)}\boldsymbol{|}-\Delta\vartheta^{(\kappa)}\big)\;\;
\xi_{\vec{y}}^{(\kappa)} \bigg\}  \;.
\eeq
The derived relations (\ref{s3_85}-\ref{s3_86}) for the delta functions
(\ref{s1_50}-\ref{s1_62}) of one-particle and two-particle operators
are combined and inserted into the trace relation (\ref{s1_61},\ref{s1_62})
of the set of maximal commuting operators which finally takes
the form (\ref{s3_87}) with self-energy density fields \(\sigma_{D}^{(i)}(\vec{x},\vartheta_{q_{\kappa}}^{(\kappa)})\)
and coset matrices \(\mathring{T}_{\vec{y}\ppr;\vec{y}}^{ba}(\vartheta_{q_{\kappa}}^{(\kappa)})\) for the anomalous
doubled degrees of freedom
\beq \lb{s3_87}
\lefteqn{\varrho(\mfrak{v}^{(\kappa=0,1,2)};\mfrak{o}^{(k=1,2,3)}) =
\prod_{k=1}^{3}\sum_{p_{k}=\pm}\;
\lim_{|\ve_{p_{k}}^{(k)}|\rightarrow0}\;\lim_{T^{(k)}\rightarrow+\infty}
\int_{0}^{T^{(k)}}\frac{dt_{p_{k}}^{(k)}}{2\pi\,\hbar}\;
\exp\Big\{-\frac{\im}{\hbar}\:t_{p_{k}}^{(k)}\:
\eta_{p_{k}}^{(k)}\:\big(\mfrak{o}^{(k)}-\im\:\ve_{p_{k}}^{(k)}\big)\Big\}\;\times   } \\ \no &\times&
\prod_{\kappa=0}^{2}\sum_{q_{\kappa}=\pm}\;
\lim_{|\mfrak{e}_{q_{\kappa}}^{(\kappa)}|\rightarrow 0} \;\lim_{\mcal{T}^{(\kappa)}\rightarrow+\infty}
\int_{0}^{\mcal{T}^{(\kappa)}}\frac{d\mfrak{t}_{q_{\kappa}}^{(\kappa)}}{2\pi\,\hbar}\;
\exp\Big\{-\im\,\zeta_{q_{\kappa}}^{(\kappa)}\:\frac{\mfrak{t}_{q_{\kappa}}^{(\kappa)}}{\hbar}\:
\big(\mfrak{v}^{(\kappa)}-\im\,\mfrak{e}_{q_{\kappa}}^{(\kappa)}\big)\Big\}\;\times
\\ \no &\times&\int d[\sigma_{D}^{(i)}(\vec{x},\mfrak{t}_{q_{\kappa}}^{(\kappa)})]\;
\exp\bigg\{-\frac{\im}{2\hbar}\int_{0}^{\mfrak{t}_{q_{\kappa}}^{(\kappa)}-\Delta\vartheta^{(\kappa)}} \hspace*{-0.6cm}
d\vartheta_{q_{\kappa}}^{(\kappa)}\;\zeta_{q_{\kappa}}^{(\kappa)}\sum_{i=i_{\kappa}}^{j_{\kappa}}\sum_{\vec{x},\vec{x}\ppr}
\sigma_{D}^{(i)}(\vec{x}\ppr,\vartheta_{q_{\kappa}}^{(\kappa)})\;\hat{V}_{|\vec{x}\ppr-\vec{x}|}^{(\kappa);\boldsymbol{-1}}
\;\sigma_{D}^{(i)}(\vec{x},\vartheta_{q_{\kappa}}^{(\kappa)})\bigg\} \\  \no &\times&
\int d[\hat{T}_{\vec{y}\ppr;\vec{y}_{1}}^{-1}\!(\vartheta_{q_{\kappa}}^{(\kappa)})\;
d\!\hat{T}_{\vec{y}_{1};\vec{y}}\!(\vartheta_{q_{\kappa}}^{(\kappa)})]\;\;
\boldsymbol{\Delta^{(\kappa)}}\Big(\hat{T}_{\vec{y}\ppr;\vec{y}}^{-1}\!(\vartheta_{q_{\kappa}}^{(\kappa)});
\hat{T}_{\vec{y}\ppr;\vec{y}}\!(\vartheta_{q_{\kappa}}^{(\kappa)})\Big)\;\times\;
\Big\{\mbox{DET}\Big(\mathring{\mscr{M}}_{\vec{y}\ppr;\vec{y}}^{ba}(\vartheta_{q_{\kappa}}^{(\kappa)\bprime}\boldsymbol{|}
\vartheta_{q_{\kappa}}^{(\kappa)})\Big)\Big\}^{1/2}\times \\ \no &\times&
\int d[\chi_{\vec{x},s}^{(k)*},d\chi_{\vec{x},s}^{(k)}]\;d[\xi_{\vec{x},s}^{(\kappa)*},d\xi_{\vec{x},s}^{(\kappa)}]\;
\exp\Big\{-\sum_{\vec{x},s}\Big(\chi_{\vec{x},s}^{(k)*}\;\chi_{\vec{x},s}^{(k)}+
\xi_{\vec{x},s}^{(\kappa)*}\;\xi_{\vec{x},s}^{(\kappa)}\Big)\Big\}  \;\times \\ \no &\times&
\exp\bigg\{\sum_{\vec{x}\ppr,s\ppr;\vec{x},s}\mcal{N}_{x}\:\chi_{\vec{x}\ppr,s\ppr}^{(k+1)*}\;\;
\bigg(\exp\bigg\{\frac{\im}{\hbar}\:\eta_{p_{k}}^{(k)}\;t_{p_{k}}^{(k)}\;\frac{1}{\mcal{N}_{x}}\,
\hat{\mfrak{O}}_{\vec{x}_{2}\ppr,s_{2}\ppr;\vec{x}_{1}\ppr,s_{1}\ppr}^{(k)}\bigg\}\bigg)_{\vec{x}\ppr,s\ppr;\vec{x},s}
\;\;\chi_{\vec{x},s}^{(k)}\bigg\} \\ \no &\times&
\exp\bigg\{\sum_{\vec{y};\vec{y}\ppr}\mcal{N}_{x}\;\xi_{\vec{y}\ppr}^{(\kappa+1)*}\;\;
\mathring{\mscr{M}}_{\vec{y}\ppr;\vec{y}}^{-1;ba}
\big(\mfrak{t}_{q_{\kappa}}^{(\kappa)}\boldsymbol{|}-\Delta\vartheta^{(\kappa)}\big)\;\;
\xi_{\vec{y}}^{(\kappa)} \bigg\}  \;; \\ \lb{s3_88}
&& \chi_{\vec{y}}^{(4)}=-\xi_{\vec{y}}^{(0)}\;;\;\;\;\;
\chi_{\vec{y}}^{(1)}=\xi_{\vec{y}}^{(3)}\;.
\eeq
In order to remove the source fields, we define the source vector field \(\Upsilon_{\vec{y}}\) (\ref{s3_89}) and the
\(6\times6\) matrix \(\hat{\mscr{B}}_{\vec{y}\ppr;\vec{y}}\) (\ref{s3_91}) with propagator terms of one-particle
(\ref{s3_92}) and two-particle parts (\ref{s3_93}) so that one obtains the determinant
\(\mbox{DET}(\hat{1}\:\delta_{\vec{y}\ppr;\vec{y}}-\hat{\mscr{B}}_{\vec{y}\ppr;\vec{y}})\) (\ref{s3_90})
after integration over the anti-commuting source fields
\beq \lb{s3_89}
\Upsilon_{\vec{y}} &=&\Big(\chi_{\vec{y}}^{(3)}\,,\,\chi_{\vec{y}}^{(2)}\,,\,\chi_{\vec{y}}^{(1)}\;;\;
\xi_{\vec{y}}^{(2)}\,,\,\xi_{\vec{y}}^{(1)}\,,\,\xi_{\vec{y}}^{(0)}\Big)^{T} \;; \\ \lb{s3_90}
\mfrak{I}[\chi_{\vec{y}}^{(k)}\,;\,\xi_{\vec{y}}^{(\kappa)}]  &=&
\prod_{k=1}^{3}\prod_{\kappa=0}^{2}\int d[\chi_{\vec{x},s}^{(k)*},d\chi_{\vec{x},s}^{(k)}]\;
d[\xi_{\vec{x},s}^{(\kappa)*},d\xi_{\vec{x},s}^{(\kappa)}]\;\;
\exp\bigg\{-\sum_{\vec{y}\ppr;\vec{y}}\mcal{N}_{x}\;\Upsilon_{\vec{y}\ppr}^{*}\;
\Big(\hat{1}\:\delta_{\vec{y}\ppr;\vec{y}}-\hat{\mscr{B}}_{\vec{y}\ppr;\vec{y}}\Big)\;
\Upsilon_{\vec{y}}\bigg\}  \\ \no &=&
\mbox{DET}\Big(\hat{1}\:\delta_{\vec{y}\ppr;\vec{y}}-\hat{\mscr{B}}_{\vec{y}\ppr;\vec{y}}\Big)\;; \\ \lb{s3_91}
\hat{\mscr{B}}_{\vec{y}\ppr;\vec{y}} &=&
\left(\bea{cccccc}
0 & {\scr(\exp\{\mfrak{O}^{(2)}\})}  \\ 0 & 0 & {\scr(\exp\{\mfrak{O}^{(1)}\})} \\
0 & & 0 & {\scr\mathring{\mscr{M}}^{-1}('2')} & \\
0 & & & 0 & {\scr\mathring{\mscr{M}}^{-1}('1')} & \\
0 & & & & 0 & {\scr\mathring{\mscr{M}}^{-1}('0')}  \\
{\scr-(\exp\{\mfrak{O}^{(3)}\})}  &&&&& 0
\eea\right)_{\mbox{;}} \\  \lb{s3_92}
(\exp\{\mfrak{O}^{(k)}\}) &=&
\bigg(\exp\bigg\{\frac{\im}{\hbar}\:\eta_{p_{k}}^{(k)}\;t_{p_{k}}^{(k)}\;\frac{1}{\mcal{N}_{x}}\,
\hat{\mfrak{O}}_{\vec{x}_{2}\ppr,s_{2}\ppr;
\vec{x}_{1}\ppr,s_{1}\ppr}^{(k)}\bigg\}\bigg)_{\vec{x}\ppr,s\ppr;\vec{x},s} \;; \\  \lb{s3_93}
\mathring{\mscr{M}}^{-1}('\kappa') &=&\mathring{\mscr{M}}_{\vec{y}\ppr;\vec{y}}^{-1;ba}
(\mfrak{t}_{q_{\kappa}}^{(\kappa)}\boldsymbol{|}-\Delta\vartheta^{(\kappa)}) \;;\;\;\;\;
{\scr(b=a\equiv1)}\;.
\eeq
The determinant (\ref{s3_90}) with its one-particle and two-particle entries (\ref{s3_92},\ref{s3_93}) can be
further simplified into a trace-logarithm expansion of period '6' so that the integral relation
\(\mfrak{I}[\chi_{\vec{y}}^{(k)}\,;\,\xi_{\vec{y}}^{(\kappa)}]\) (\ref{s3_90}) reduces to the trace-logarithm
term with weight \(1/6\) over the trace-log operation acting onto the product of the three one-particle
elements (\ref{s3_92}) and of the three two-particle elements (\ref{s3_93})
\beq\lb{s3_94}
\lefteqn{\mfrak{I}[\chi_{\vec{y}}^{(k)}\,;\,\xi_{\vec{y}}^{(\kappa)}]  =
\mbox{DET}\Big(\hat{1}\:\delta_{\vec{y}\ppr;\vec{y}}-\hat{\mscr{B}}_{\vec{y}\ppr;\vec{y}}\Big) =
\exp\Big\{\trxs\ln\Big(\hat{1}\:\delta_{\vec{y}\ppr;\vec{y}}-\hat{\mscr{B}}_{\vec{y}\ppr;\vec{y}}\Big)\Big\} =\exp\Big\{-\sum_{n=1}^{\infty}\frac{1}{n}\;\trxs\big[\hat{\mscr{B}}^{n}\big]\Big\} }
\\ \no &\stackrel{n\rightarrow 6n}{=}& \sum_{n=1}^{\infty}\frac{(-1)^{n+1}}{6n}\;
\trxs\bigg[\bigg(
\Big(\exp\Big\{\frac{\im}{\hbar}\:\eta_{p_{3}}^{(3)}\;t_{p_{3}}^{(3)}\;\frac{1}{\mcal{N}_{x}}\,
\hat{\mfrak{O}}_{\vec{y}_{7}\ppr;\vec{y}_{6}\ppr}^{(3)}\Big\}\Big)_{\vec{y}_{7};\vec{y}_{6}}\;
\Big(\exp\Big\{\frac{\im}{\hbar}\:\eta_{p_{2}}^{(2)}\;t_{p_{2}}^{(2)}\;\frac{1}{\mcal{N}_{x}}\,
\hat{\mfrak{O}}_{\vec{y}_{6}\ppr;\vec{y}_{5}\ppr}^{(2)}\Big\}\Big)_{\vec{y}_{6};\vec{y}_{5}}\;\times \\ \no &\times&
\hspace*{-0.55cm}\Big(\exp\Big\{\frac{\im}{\hbar}\:\eta_{p_{1}}^{(1)}\;t_{p_{1}}^{(1)}\;\frac{1}{\mcal{N}_{x}}\,
\hat{\mfrak{O}}_{\vec{y}_{5}\ppr;\vec{y}_{4}\ppr}^{(1)}\Big\}\Big)_{\vec{y}_{5};\vec{y}_{4}}\;
\;\mathring{\mscr{M}}_{\vec{y}_{4};\vec{y}_{3}}^{-1;ba}
(\mfrak{t}_{q_{2}}^{(2)}\boldsymbol{|}-\Delta\vartheta^{(2)})\;
\;\mathring{\mscr{M}}_{\vec{y}_{3};\vec{y}_{2}}^{-1;ba}
(\mfrak{t}_{q_{1}}^{(1)}\boldsymbol{|}-\Delta\vartheta^{(1)})\;
\mathring{\mscr{M}}_{\vec{y}_{2};\vec{y}_{1}}^{-1;ba}
(\mfrak{t}_{q_{0}}^{(0)}\boldsymbol{|}-\Delta\vartheta^{(0)})\bigg)_{\vec{y}_{7};\vec{y}_{1}}^{n}\bigg] \\ \no &=&
\hspace*{-0.55cm}\exp\Bigg\{\frac{1}{6}\trxs\ln\Bigg(\hat{1}\:\delta_{\vec{y}_{7};\vec{y}_{1}}+
\bigg(
\Big(\exp\Big\{\frac{\im}{\hbar}\:\eta_{p_{3}}^{(3)}\;t_{p_{3}}^{(3)}\;\frac{1}{\mcal{N}_{x}}\,
\hat{\mfrak{O}}_{\vec{y}_{7}\ppr;\vec{y}_{6}\ppr}^{(3)}\Big\}\Big)_{\vec{y}_{7};\vec{y}_{6}}\;
\Big(\exp\Big\{\frac{\im}{\hbar}\:\eta_{p_{2}}^{(2)}\;t_{p_{2}}^{(2)}\;\frac{1}{\mcal{N}_{x}}\,
\hat{\mfrak{O}}_{\vec{y}_{6}\ppr;\vec{y}_{5}\ppr}^{(2)}\Big\}\Big)_{\vec{y}_{6};\vec{y}_{5}}\;\times \\ \no
\lefteqn{\times
\Big(\exp\Big\{\frac{\im}{\hbar}\:\eta_{p_{1}}^{(1)}\;t_{p_{1}}^{(1)}\;\frac{1}{\mcal{N}_{x}}\,
\hat{\mfrak{O}}_{\vec{y}_{5}\ppr;\vec{y}_{4}\ppr}^{(1)}\Big\}\Big)_{\vec{y}_{5};\vec{y}_{4}}\;
\;\mathring{\mscr{M}}_{\vec{y}_{4};\vec{y}_{3}}^{-1;ba}
\big(\mfrak{t}_{q_{2}}^{(2)}\boldsymbol{|}-\Delta\vartheta^{(2)}\big)\;
\;\mathring{\mscr{M}}_{\vec{y}_{3};\vec{y}_{2}}^{-1;ba}
\big(\mfrak{t}_{q_{1}}^{(1)}\boldsymbol{|}-\Delta\vartheta^{(1)}\big)\;
\mathring{\mscr{M}}_{\vec{y}_{2};\vec{y}_{1}}^{-1;ba}
\big(\mfrak{t}_{q_{0}}^{(0)}\boldsymbol{|}-\Delta\vartheta^{(0)}\big)\bigg)_{\vec{y}_{7};\vec{y}_{1}}^{b=a\equiv1}
\Bigg)\Bigg\}_{\mbox{.}} }
\eeq
As we replace the Grassmann integration of source fields in (\ref{s3_87}) by (\ref{s3_94}), the trace over the
delta functions of the maximal commuting set of symmetry operators finally achieves the form (\ref{s3_95}) with the
coset matrices as the remaining field degrees of freedom. The self-energy density fields
\(\sigma_{D}^{(i)}(\vec{x},\vartheta_{q_{\kappa}}^{(\kappa)})\)  are to be determined from a saddle point
approximation (\ref{s3_96}) and stay as coefficients in a gradient expansion (\ref{s3_60}-\ref{s3_67},\ref{s3_97})
with the coset matrices for the anomalous doubled pairs of fields
\beq \lb{s3_95}
\lefteqn{\varrho(\mfrak{v}^{(\kappa=0,1,2)};\mfrak{o}^{(k=1,2,3)}) =
\prod_{k=1}^{3}\sum_{p_{k}=\pm}\;
\lim_{|\ve_{p_{k}}^{(k)}|\rightarrow0}\;\lim_{T^{(k)}\rightarrow+\infty}
\int_{0}^{T^{(k)}}\frac{dt_{p_{k}}^{(k)}}{2\pi\,\hbar}\;
\exp\Big\{-\frac{\im}{\hbar}\:t_{p_{k}}^{(k)}\:
\eta_{p_{k}}^{(k)}\:\big(\mfrak{o}^{(k)}-\im\:\ve_{p_{k}}^{(k)}\big)\Big\}\;\times   } \\ \no &\times&
\prod_{\kappa=0}^{2}\sum_{q_{\kappa}=\pm}\;
\lim_{|\mfrak{e}_{q_{\kappa}}^{(\kappa)}|\rightarrow 0} \;\lim_{\mcal{T}^{(\kappa)}\rightarrow+\infty}
\int_{0}^{\mcal{T}^{(\kappa)}}\frac{d\mfrak{t}_{q_{\kappa}}^{(\kappa)}}{2\pi\,\hbar}\;
\exp\Big\{-\im\,\zeta_{q_{\kappa}}^{(\kappa)}\:\frac{\mfrak{t}_{q_{\kappa}}^{(\kappa)}}{\hbar}\:
\big(\mfrak{v}^{(\kappa)}-\im\,\mfrak{e}_{q_{\kappa}}^{(\kappa)}\big)\Big\}\;\times
\\ \no &\times&\int d[\sigma_{D}^{(i)}(\vec{x},\mfrak{t}_{q_{\kappa}}^{(\kappa)})]\;
\exp\bigg\{-\frac{\im}{2\hbar}\int_{0}^{\mfrak{t}_{q_{\kappa}}^{(\kappa)}-\Delta\vartheta^{(\kappa)}} \hspace*{-0.6cm}
d\vartheta_{q_{\kappa}}^{(\kappa)}\;\zeta_{q_{\kappa}}^{(\kappa)}\sum_{i=i_{\kappa}}^{j_{\kappa}}\sum_{\vec{x},\vec{x}\ppr}
\sigma_{D}^{(i)}(\vec{x}\ppr,\vartheta_{q_{\kappa}}^{(\kappa)})\;\hat{V}_{|\vec{x}\ppr-\vec{x}|}^{(\kappa);\boldsymbol{-1}}
\;\sigma_{D}^{(i)}(\vec{x},\vartheta_{q_{\kappa}}^{(\kappa)})\bigg\} \\  \no &\times&
\int d[\hat{T}_{\vec{y}\ppr;\vec{y}_{1}}^{-1}\!(\vartheta_{q_{\kappa}}^{(\kappa)})\;
d\!\hat{T}_{\vec{y}_{1};\vec{y}}\!(\vartheta_{q_{\kappa}}^{(\kappa)})]\;\;
\boldsymbol{\Delta^{(\kappa)}}\Big(\hat{T}_{\vec{y}\ppr;\vec{y}}^{-1}\!(\vartheta_{q_{\kappa}}^{(\kappa)});
\hat{T}_{\vec{y}\ppr;\vec{y}}\!(\vartheta_{q_{\kappa}}^{(\kappa)})\Big)\;\times\;
\Big\{\mbox{DET}\Big(\mathring{\mscr{M}}_{\vec{y}\ppr;\vec{y}}^{ba}(\vartheta_{q_{\kappa}}^{(\kappa)\bprime}\boldsymbol{|}
\vartheta_{q_{\kappa}}^{(\kappa)})\Big)\Big\}^{1/2}\times \\ \no &\times&
\exp\Bigg\{\frac{1}{6}\trxs\ln\Bigg(\hat{1}\:\delta_{\vec{y}_{7};\vec{y}_{1}}+
\bigg(\Big(\exp\Big\{\frac{\im}{\hbar}\:\eta_{p_{3}}^{(3)}\;t_{p_{3}}^{(3)}\;\frac{1}{\mcal{N}_{x}}\,
\hat{\mfrak{O}}_{\vec{y}_{7}\ppr;\vec{y}_{6}\ppr}^{(3)}\Big\}\Big)_{\vec{y}_{7};\vec{y}_{6}}\;
\Big(\exp\Big\{\frac{\im}{\hbar}\:\eta_{p_{2}}^{(2)}\;t_{p_{2}}^{(2)}\;\frac{1}{\mcal{N}_{x}}\,
\hat{\mfrak{O}}_{\vec{y}_{6}\ppr;\vec{y}_{5}\ppr}^{(2)}\Big\}\Big)_{\vec{y}_{6};\vec{y}_{5}}\;\times \\ \no
\lefteqn{\hspace*{-0.6cm}\times
\Big(\exp\Big\{\frac{\im}{\hbar}\:\eta_{p_{1}}^{(1)}\;t_{p_{1}}^{(1)}\;\frac{1}{\mcal{N}_{x}}\,
\hat{\mfrak{O}}_{\vec{y}_{5}\ppr;\vec{y}_{4}\ppr}^{(1)}\Big\}\Big)_{\vec{y}_{5};\vec{y}_{4}}\;
\;\mathring{\mscr{M}}_{\vec{y}_{4};\vec{y}_{3}}^{-1;ba}
\big(\mfrak{t}_{q_{2}}^{(2)}\boldsymbol{|}-\Delta\vartheta^{(2)}\big)\;
\;\mathring{\mscr{M}}_{\vec{y}_{3};\vec{y}_{2}}^{-1;ba}
\big(\mfrak{t}_{q_{1}}^{(1)}\boldsymbol{|}-\Delta\vartheta^{(1)}\big)\;
\mathring{\mscr{M}}_{\vec{y}_{2};\vec{y}_{1}}^{-1;ba}
\big(\mfrak{t}_{q_{0}}^{(0)}\boldsymbol{|}-\Delta\vartheta^{(0)}\big)\bigg)_{\vec{y}_{7};\vec{y}_{1}}^{b=a\equiv1}
\Bigg)\Bigg\}_{\mbox{;}}  }
\eeq
\beq   \lb{s3_96}
\sigma_{D}^{(i)}(\vec{x},\vartheta_{q_{\kappa}}^{(\kappa)}) &\rightarrow&
\Re\langle\sigma_{D}^{(i)}(\vec{x},\vartheta_{q_{\kappa}}^{(\kappa)})\rangle +\im\;
\Im\langle\sigma_{D}^{(i)}(\vec{x},\vartheta_{q_{\kappa}}^{(\kappa)})\rangle\;\;;\;\;
(\mbox{Sign of imaginary part from the saddle point}  \\ \no \lefteqn{\mbox{approximation
has to comply with the corresponding part within delta functional for convergence !})\;;}  \\ \no
\sigma_{D}^{(i)}(\vec{x},\vartheta_{q_{\kappa=1,2}}^{(\kappa=1,2)}) &=&
\sigma_{D}^{(i)}(\vartheta_{q_{\kappa=1,2}}^{(\kappa=1,2)})\;;\;\;\;(\mbox{angular momentum cases '$\kappa=1,2$'
are independent of the spatial vector $\vec{x}$ !});
\\  \lb{s3_97} \mathring{\mscr{M}}_{\vec{y}\ppr;\vec{y}}^{ba}\!\!
\left(\vartheta_{q_{\kappa}}^{(\kappa)\bprime}\boldsymbol{|}\vartheta_{q_{\kappa}}^{(\kappa)}\right) &=&
\Delta\vartheta^{(\kappa)}\;\mathring{\mscr{H}}_{\vec{y}\ppr;\vec{y}}^{(\kappa)ba}\!\!\left(\vartheta_{q_{\kappa}}^{(\kappa)\bprime}\boldsymbol{|}
\vartheta_{q_{\kappa}}^{(\kappa)}\boldsymbol{|}\hat{\mscr{S}}^{(\kappa;i)}\cdot\sigma_{D}^{(i)}\right) +
\Delta\vartheta^{(\kappa)}\;\delta\mathring{\mscr{H}}_{\vec{y}\ppr;\vec{y}}^{(\kappa)ba}\!\!\left(
\mathring{T}^{-1}(\vartheta_{q_{\kappa}}^{(\kappa)\bprime})\boldsymbol{|}
\mathring{T}(\vartheta_{q_{\kappa}}^{(\kappa)})\boldsymbol{|}\hat{\mscr{S}}^{(\kappa;i)}\cdot\sigma_{D}^{(i)}\right) \;.
\eeq

\section{Summary and conclusions} \lb{s4}

Despite of a probably intricate appearance of various transformations, it is the aim of this article to verify
the exactness of coherent state path integrals with precise discrete time grids for the various
limit processes of many particle theory. Important steps are the introduction of two different anomalous doubled
fields in (\ref{s3_1}-\ref{s3_4}) with further extension in (\ref{s3_28},\ref{s3_29}). One has also to adapt
the 'Nambu' doubled one-particle Hamiltonians
\(\mathring{\mfrak{H}}_{\vec{x}\ppr,s\ppr;\vec{x},s}^{(\kappa)ba}(\vartheta_{q_{\kappa}}^{(\kappa)\bprime}
\boldsymbol{|}\vartheta_{q_{\kappa}}^{(\kappa)})\) (\ref{s3_30}) and metric tensors \((\mathring{\tau}_{1})\),
\((\mathring{S})\) (\ref{s3_32}) in order to preserve the exact generalized 'time steps' from insertion
of overcomplete sets of anti-commuting fields. Furthermore, we have proved the applicability of coherent states
and coherent state path integrals, which are usually anticipated with varying particle numbers, to problems
with definite, specific symmetry quantum numbers. Instead of a coordinate space-spin representation
\(\vec{y}=(\vec{x}\,,\,s=\uparrow,\downarrow)\) (\ref{s1_4},\ref{s1_5}), one can also choose the angular
momentum representation with various abbreviations of angular momentum labels in order to regard
the inherent rotational symmetries of the chosen problem of a many electron atom in a central potential
and an external magnetic field. Details of the coset decomposition for fermionic field degrees of freedom
are already described in Refs. \cite{physica6,pop1,pop2} and especially in \cite{precisecoh1} so that
the generator \(\hat{Y}_{\vec{y};\vec{y}\ppr}^{a\neq b}(\vartheta_{q_{\kappa}}^{(\kappa)})\) of the coset matrices
\(\hat{T}_{\vec{y};\vec{y}\ppr}^{ab}(\vartheta_{q_{\kappa}}^{(\kappa)})=
(\,\exp\{-\hat{Y}_{\vec{y}_{1};\vec{y}_{2}}^{a_{1}\neq b_{1}}(\vartheta_{q_{\kappa}}^{(\kappa)})\,)_{\vec{y};\vec{y}\ppr}^{ab}\)
is composed of the anti-symmetric, complex-valued sub-generators
\(\hat{X}_{\vec{y};\vec{y}\ppr}(\vartheta_{q_{\kappa}}^{(\kappa)})\),
\(\hat{X}_{\vec{y};\vec{y}\ppr}\pdag(\vartheta_{q_{\kappa}}^{(\kappa)})\) in the off-diagonal blocks \(a\neq b\) of the
'Nambu' doubled space for the anomalous doubled pairs of fields. The given approach of a coherent-state-trace
representation of delta functions with second quantized field operators can also be extended to configurations
of molecules with various nucleons where one achieves inclusion of all relativistic corrections by choosing
the total set of relativistically invariant one-particle and two-particle symmetry quantum numbers and corresponding
operators. Since the various precise time steps of normal ordered operators with field combinations
\(\psi_{\vec{y}\ppr}^{*}(\vartheta_{q_{\kappa}}^{(\kappa)}\!+\!\Delta\vartheta^{(\kappa)})\,\ldots\,
\psi_{\vec{y}}(\vartheta_{q_{\kappa}}^{(\kappa)})\) may involve intricate appearance of equations, it is of common use to
simplify relations to a hermitian kind \(\psi_{\vec{y}\ppr}^{*}(\vartheta_{q_{\kappa}}^{(\kappa)})\,\ldots\,
\psi_{\vec{y}}(\vartheta_{q_{\kappa}}^{(\kappa)})\) without explicit emphasis of the limit process
\(\Delta\vartheta^{(\kappa)}\rightarrow0\) for normal ordered, second quantized operators in many-body theory
(compare \cite{Negele}-\cite{Nagaosa2}). However, as one restricts to a purely classical field theory
from variation of anti-commuting fields or coset matrices within exponentials of the corresponding functional,
it is possible to neglect the various limit processes \(\Delta\vartheta^{(\kappa)}\rightarrow0\) for the classical
field equations and to regard the fields \(\psi_{\vec{y}\ppr}^{*}(\vartheta_{q_{\kappa}}^{(\kappa)}\!+\!\Delta\vartheta^{(\kappa)})\)
and \(\psi_{\vec{y}\ppr}^{*}(\vartheta_{q_{\kappa}}^{(\kappa)})\) within a Hamiltonian density as equivalent
\cite{precisecoh1}.


\end{document}